\documentclass[12pt]{article}
\usepackage{latexsym}
\usepackage{epsf}
\begin{document}
\bibliographystyle{unsrt}
\def\question#1{{{\marginpar{\small \sc #1}}}}

\title{\small \rm
\begin{flushright}
\small{OUTP-97-62P}\\
\end{flushright}
\vspace{2cm}
\LARGE \bf Physics from the lattice: glueballs in QCD; topology;
SU(N) for all N.}
\date{}

\author{Michael J Teper\thanks{Lectures at the Isaac Newton Institute 
NATO-ASI School, June 1997} \\ 
{\it Department of Physics }\\
{\it University of Oxford  }\\
{\it Oxford, OX1 3NP, U.K.}}
\maketitle
\vfill
\hfill
hep-lat/9711011

\newpage

\section {Introduction} 

In these lectures I will show, through three examples,
how current lattice calculations are able to tell us
interesting things about the continuum physics of
non-Abelian gauge theories.

My first topic concerns the glueball spectrum.   
The physics question here is: where, in the experimentally
determined hadron spectrum, are the glueballs hiding? 
I will first summarise what lattice calculations tell
us about the continuum glueball spectrum of the
$SU(3)$ gauge theory. I will then discuss what this tells 
us about the masses of the corresponding `bare' glueballs
in QCD. I will then turn to the experimental spectrum
with some discussion of the interpretation of the
observed states in the quark model. Finally I will pinpoint
the experimental states most likely to have large
glueball components.

The second topic concerns topological fluctuations
in the $SU(3)$ gauge theory. Here the simplest physics
question is: are these fluctuations large enough
to be consistent with the observed large
$\eta^\prime$ mass? Thanks to Witten and Veneziano this
is a question that can be posed in the pure gauge theory.
We shall see that the fluctuations of the topological
charge do indeed have the required magnitude. Along the way 
I will discuss the problems with topology on a lattice.
I then move onto the much less straightforward question
concerning the structure of these vacuum fluctuations: e.g. what
is the size distribution of instantons? I will discuss some
preliminary calculations that show the mean size to be about
${\bar \rho} \sim 0.5 fm$. I will also point to some
intriguing evidence for a long distance polarisation of
the topological fluctuations.

My third topic concerns the physics of $SU(N_c)$ gauge theories 
as a function of the number of colours, $N_c$. This will
be mainly in 2+1 dimensions, since that is where we have good
calculations. I will describe calculations of the mass spectrum 
for $N_c \le 4$ which explicitly show that for $N_c \ge 2$
mass ratios are independent of $N_c$ up to a modest
$\sim 1/N_c^2$ correction. This is a very elegant result:
it tells us that all the apparently different $SU(N_c)$
theories are actually one single theory, $SU(\infty)$,
to a reasonable first approximation. There is some 
very preliminary evidence, as I will show, that the same 
is true for $D=3+1$.

\section {Glueballs in QCD}

Ideally I should be telling you what happens when you
simulate QCD with realistically light quarks. But it is
going to be a few years yet before I can do that. What
current lattice Monte Carlo calculations are able to provide
is predictions for the low-lying  mass spectrum of the 
continuum $SU(3)$ gauge theory without quarks. These states
are glueballs - there being nothing other than gluons in the 
theory. If you want hadrons with quarks then you can propagate
quarks in this gluonic vacuum and then tie such propagators
together so that the object propagating has the appropriate
hadronic quantum numbers. That is to say, you calculate
hadron masses in the relativistic valence quark approximation.
(This is usually referred to as the `quenched approximation' to QCD.)
The spectrum one obtains this way is a remarkably good
approximation to the observed
hadron spectrum. This is not too surprising: one
reason we were able to learn of the existence of quarks in the 
first place is because the low-lying hadrons are in fact well 
described by a simple valence quark picture. 

Suppose that we begin with our SU(3) gauge theory and then
couple to it 3 flavours of very heavy quarks. Initially
the spectrum will contain the usual light glueball
spectrum, supplemented by a spectrum of very heavy
quarkonia that can be well accounted for in terms
of a valence quark potential model. Let us now gradually 
reduce the quark masses towards their physical values.
In principle the glueball and quarkonia states
might entirely change their character once their
masses become comparable. However, as we remarked above,
the experimental light quark spectrum still seems to retain the
essential features of valence quark physics. If the
quarkonia are not qualitatively alterred, it seems
reasonable to think that neither will the glueballs be.
Of course if a quarkonium state and a glueball are
close enough in mass they will mix. However there is
reason to believe that this mixing is weak. The reason is 
the Zweig (OZI) rule: hadron decays where the initial quarks all
have to annihilate are strongly suppressed. The classic
example is the $\phi$ meson. Such a decay may be thought
of as $quarks \to glue \to quarks$. Glueball mixing with quarks
should therefore be $\surd$OZI suppressed. As should 
glueball decays into hadrons composed of quarks. The
existence of such a suppression is supported by a recent 
lattice calculation
\cite{GF11-decay}.

The picture we have in mind is therefore as follows.
The glueballs will only be mildly affected by the 
presence of light quarks. They will decay into, say, pions
but their decay width will be relatively small; and there
will be a correspondingly small mass shift. Only if there
happens to be a flavour singlet quarkonium state close by
in mass will things be very different, because of the mixing
of these nearly degenerate states. In this context we
expect `close by' to mean within $\sim 100 MeV$.
So we view the glueballs in the pure SU(3) gauge theory
as being the `bare' glueballs of QCD which may mix with
nearby quarkonia to produce the hadrons that are observed
in experiments. All this is an assumption of course, albeit
a reasonable one. If true it tells us that the glueballs,
whether mixed with quarkonia or not, should lie close to
the masses they have in the gauge theory. So we now turn to 
the calculation of those masses.

I shall begin by reviewing the available lattice calculations, 
placing a particular emphasis on exposing the sources
of systematic error in arriving at a final mass
prediction in $MeV$ units. I do this in some detail,
so that you are able to judge for yourselves the 
credibility of lattice mass estimates.

As we shall soon see, it is only for three states that
the lattice calculations are reliable enough
that we can extract continuum predictions: the
lightest scalar, $0^{++}$, tensor,  $2^{++}$, and
pseudoscalar, $0^{-+}$, glueballs. For other states
we do have calculations for one or two values of the lattice
spacing $a$, but that is not enough to extrapolate to $a=0$.
Nonetheless the lattice results strongly suggest that glueballs
with other $J^{PC}$ are heavier
\cite{ukqcd}. 

The lightest glueball is the $0^{++}$, and it is for
this state that we have the most accurate lattice predictions.
Although there have been recent estimates for the mass
that appear to differ, e.g. $1.55 \pm 0.05$GeV
\cite{ukqcd} 
and $1.74 \pm 0.07$GeV
\cite{GF11-G1,GF11-decay},
we shall see that this difference is illusory. In fact
the apparent difference reflects different ways of
extrapolating to the continuum limit and  different ways 
of introducing physical $MeV$ units into the pure gauge
theory. That is to say, it reflects particular systematic 
errors which we need to estimate and this we try to do.
Since we find that the various lattice
calculations are consistent, we are able to
carry out a global analysis that provides the
best available glueball mass estimate. We find
$m_{0^{++}}=1606\pm 73 \pm 130 MeV$ 
where the first error is statistical and the
second is systematic. Performing a similar analysis
for the lightest tensor and pseudoscalar glueballs we find
$m_{2^{++}}=2200\pm 73 \pm 130 MeV$ and
$m_{0^{-+}}=2100\pm 73 \pm 130 MeV$.

From the experimental and phenomenological point of view
the scalar sector is complex, and I will review
the current state of play. As we shall see, there
appear to be too many scalar states to be explained
as quarkonia, and the strongest candidates for
states with large gluonic components are 
the $f_0(1500)$ and the $f_{(J=0?)}(1710)$.
This possibility is strongly reinforced by the
fact that these are the only experimental states
that are compatible with the lattice mass
estimate in the previous paragraph.

In order to assess which states have the largest gluonic
components, it is important to understand the mixing between
nearby quarkonia and glueballs. I shall briefly discuss
what happens in the case of
mixing between a glueball and the lightest
singlet and octet scalar quarkonium multiplets.   

For this review I have drawn very heavily on
\cite{fec-mt}.
\subsection{Calculating glueball masses}

I begin by briefly reminding you of some general aspects
of lattice calculations. (For more detail see the lectures
by Chris Michael at this School and the books by
Creutz 
\cite{book-C}
and Montvay and Munster
\cite{book-MM}.)
In a lattice calculation (Euclidean) space-time is discretised, usually
onto a hypercubic lattice.
What we calculate is the mass spectrum of the discretised
theory, but what we actually want is the corresponding
spectrum of the continuum theory.
Because the theory is renormalisable, the
effects of the lattice spacing, $a$, on physical length scales 
will vanish as $a \to 0$. Since QCD has effectively one
length scale, $\sim 1 fm$, one expects the effects of the 
discretisation to become negligible once $a << 1 fm$.
The same is true of the gauge theory without quarks. 
(By `$1 fm$' I really mean some characteristic physical length
scale. How one introduces actual fermi units into the pure
gauge theory is something we shall return to below.)
The lattice spacing is varied by changing the value of the 
bare (inverse) coupling which appears in the lattice action:
$\beta = 6/g^2$. 
Since the theory is asymptotically free, we know that
in order to approach the continuum limit, $a \to 0$, we need 
to take $g^2 \to 0$ and so $\beta \to \infty$. 
Indeed for sufficiently small $g^2$ one can determine
the relationship between $a$ and $g^2$ in 
low-order perturbation theory.
(Although in practice the latter does not work well for the range
of couplings currently accessible.) Reducing $a$ makes the
calculation numerically more intensive for various reasons.
An obvious one is that if we wish to maintain a constant
volume, the number of lattice sites grows $\propto 1/a^4$.
It is only in recent years that calculations for very small
values of $a$ have become practical. As we shall see, the mass
calculations I shall use here have been performed
over a range of lattice spacings $0.18 fm \geq a \geq 0.06 fm$,
corresponding to couplings $5.7 \le \beta \le 6.4$.

We now outline the main steps in a lattice Monte Carlo 
calculation of a glueball mass with a view to exposing
the main different sources of systematic error. (So we
ignore various unilluminating technicalities and
skate over various qualifications that are irrelevant
for this purpose.) 

$\bullet$ {\sl Extracting lattice masses.}
The first step is to calculate the glueball
spectrum on a lattice, for a given space-time volume $V$
and for a particular lattice spacing $a$. The Euclidean time
translation operator is $e^{-Ht}$, where $H$ is the Hamiltonian
of the theory. Thus the correlation function of an operator with
some particular $J^{PC}$ quantum numbers will, for large
enough values of $t$, vary as $e^{-m_0t}$ where $m_0$ is
the lightest mass with those quantum numbers. We
calculate such propagators numerically for $t$ large
enough that we see this asymptotic exponential decay.
From the exponent we extract the mass. Since $t$ is
given in lattice units ($t=a n$ where $n$ is the
number of lattice spacings) the exponent is
$m_0t = m_0an$ and so what we actually
obtain is the mass in lattice units, i.e. $am_0$.
The statistical errors are straightforward
to estimate. However there is also a systematic
error that has to do with determining the range of
$t$ where the asymptotic exponential dominates.
We do not attempt to quantify this error but simply note
that it will become increasingly important for the
heavier glueball states (such as the tensor and
pseudoscalar) where the exponential decrease with $t$
of the `signal', and hence its immersion
into the statistical noise, is more rapid.

$\bullet$ {\sl Finite $V$ corrections.}
The second step is to determine the corrections
due to the fact that the volume is finite. For most
quantities the functional forms of the leading 
large-$V$ corrections are known theoretically;
typically they will be of the form
$\delta m \propto \exp\{-c am L\}$ where
$am$ and $L$ are the mass-gap and lattice size in
lattice units
\cite{Luscher}.
By doing calculations for a variety of volumes at some
chosen value of $a$, any
unknown constants in these expressions can be fitted and
the resulting formulae can be used to apply corrections
at other values of $a$. Since the calculations we use
here have been performed on periodic volumes of
typical sizes 1.5 to 2.0 fm, these corrections are
smaller than our typical statistical errors; but it is 
hard to determine them more accurately than that.
This provides another source of systematic error.

$\bullet$  {\sl Finite $a$ corrections.}
The calculated glueball mass will depend 
on $a$; and it will do so in two ways. Firstly, it is
obtained in  lattice units, $am$. This trivial dependence
is removed when we take the ratio of two masses:
$ am_1/am_2 \equiv m_1/m_2$. The non-trivial
$a$ dependence is due to the distortion of the
dynamics by the discretisation. For the lattice
action we use and for quantities
such as glueball mass ratios it is known that the leading
small-$a$ correction is $O(a^2)$
\cite{Sym}. 
So  for small enough $a$
we can extrapolate our calculated mass ratios to $a=0$ using
\begin{equation}
{{m_1(a)} \over {m_2(a)}} =
{{m_1(a=0)} \over {m_2(a=0)}} + c (a m)^2 
\label{A1}
\end{equation}
where $m$ may be chosen to be $m_1$ or $m_2$ or some
other physical mass: the difference between these choices
is clearly higher order in $a^2$. Such neglected higher
order terms in the extrapolation are another source of
systematic error.

$\bullet$  {\sl Introducing the MeV scale.}
Having obtained the continuum mass spectrum in the
form of mass ratios, we want to express the masses
in usual $MeV$ units. This can be done if
at least one of the masses corresponds to a quantity
whose value is known in $MeV$. For example the
potential between heavy quarks is linear for
large separations: $V(r) \simeq \sigma r$ where
$\sigma$ is called the string tension. 
In simple string pictures for high $J$ hadrons 
$\sigma$ is related to the slope, $\alpha^{\prime}$, of 
Regge trajectories by $\alpha^{\prime}=1/2\pi\sigma$
and this provides the conventional estimate
$\surd\sigma \simeq 400-450 MeV$
\cite{book-Perkins}.
Knowing the continuum value of $m_G/\surd\sigma$ we can now
express $m_G$ in $MeV$ units. Of course one may
distrust this particular argument for the value
of $\sigma$. An alternative is to calculate
quark propagators in the pure gauge theory
and from these to form hadron propagators.
From the asymptotic exponential decay of the 
latter we obtain quarkonium masses in
the (relativistic) valence quark approximation.   
For example, we can obtain the continuum limit of the
mass ratio $m_{\rho}/\surd\sigma$. Setting
$m_{\rho}=770 MeV$ we obtain a value for
$\surd\sigma$ and hence for the glueball mass $m_G$.
Or we might set the scale using the $\phi$-meson or
the nucleon instead of the $\rho$. While the mass spectrum 
one obtains in the quenched approximation is remarkably close
to that which is experimentally observed, it is
not exactly the same and so these different ways of
setting the physical scale will lead to slightly
different glueball masses. This is a source of 
systematic error.

The first three types of systematic error can be 
made arbitrarily small by sufficiently improving 
the numerical calculations (in obvious ways).
The error in setting the $MeV$ scale is qualitatively
different. It is intrinsic to working within the
quenched approximation. How do we estimate it?
There are some quantities which we know are going to 
be sensitive to the absence of vacuum $q\bar{q}$ fluctuations,
e.g. the $\eta^{\prime}$ mass or the topological
susceptibility. These should obviously not be used
to set the $MeV$ scale in the pure gauge theory. 
There are other quantities which we expect to be no 
more sensitive to the absence of vacuum $q\bar{q}$ fluctuations
than the glueball masses themselves and which
are therefore suitable quantities with which
to attempt to set the $MeV$ scale. These include the masses of
typical quarkonia such as the $\rho$ meson, the
$\phi$ meson, the nucleon, certain matrix elements,
etc. Of course the ratios of these quantities cannot be
exactly the same in full and quenched QCD and so
the scale we extract will vary according to which
of these quantities we choose to use. The extent of
this variation can be used as an estimate of the
systematic error.

\subsection{The lightest glueballs}

The values of the glueball masses that we shall use are
from 
\cite{deF-G,cm-mt,ukqcd,GF11-G1} 
and those of the string tension are 
taken from 
\cite{cm-mt,ukqcd,string-tension}. 
As we have mentionned
already, only the $0^{++}$, $2^{++}$ and $0^{-+}$
glueballs are determined accurately enough that a
continuum extrapolation is possible. In 
\cite{ukqcd}
you can find mass estimates for glueballs of widely
varying $J^{PC}$, obtained for a very small lattice
spacing. These do suggest that the three masses
we shall obtain are in fact the lightest ones. 
There is clearly an urgent need for a new generation
of lattice glueball calculations which will provide
information on a much larger part of the continuum
mass spectrum. For a preview of what these are likely
to look like, see
\cite{P-M}.

\begin	{figure}[p]
\begin	{center}
\leavevmode
\setlength{\unitlength}{0.240900pt}
\ifx\plotpoint\undefined\newsavebox{\plotpoint}\fi
\sbox{\plotpoint}{\rule[-0.200pt]{0.400pt}{0.400pt}}%
\begin{picture}(1500,1800)(0,0)
\font\gnuplot=cmr10 at 12pt
\gnuplot
\sbox{\plotpoint}{\rule[-0.200pt]{0.400pt}{0.400pt}}%
\put(120.0,31.0){\rule[-0.200pt]{4.818pt}{0.400pt}}
\put(108,31){\makebox(0,0)[r]{{$0$}}}
\put(1436.0,31.0){\rule[-0.200pt]{4.818pt}{0.400pt}}
\put(120.0,383.0){\rule[-0.200pt]{4.818pt}{0.400pt}}
\put(108,383){\makebox(0,0)[r]{{$1$}}}
\put(1436.0,383.0){\rule[-0.200pt]{4.818pt}{0.400pt}}
\put(120.0,736.0){\rule[-0.200pt]{4.818pt}{0.400pt}}
\put(108,736){\makebox(0,0)[r]{{$2$}}}
\put(1436.0,736.0){\rule[-0.200pt]{4.818pt}{0.400pt}}
\put(120.0,1088.0){\rule[-0.200pt]{4.818pt}{0.400pt}}
\put(108,1088){\makebox(0,0)[r]{{$3$}}}
\put(1436.0,1088.0){\rule[-0.200pt]{4.818pt}{0.400pt}}
\put(120.0,1441.0){\rule[-0.200pt]{4.818pt}{0.400pt}}
\put(108,1441){\makebox(0,0)[r]{{$4$}}}
\put(1436.0,1441.0){\rule[-0.200pt]{4.818pt}{0.400pt}}
\put(120.0,1793.0){\rule[-0.200pt]{4.818pt}{0.400pt}}
\put(108,1793){\makebox(0,0)[r]{{$5$}}}
\put(1436.0,1793.0){\rule[-0.200pt]{4.818pt}{0.400pt}}
\put(491.0,31.0){\rule[-0.200pt]{0.400pt}{4.818pt}}
\put(491,19){\makebox(0,0){\shortstack{\\ \\ \\ {$0.05$}}}}
\put(491.0,1773.0){\rule[-0.200pt]{0.400pt}{4.818pt}}
\put(862.0,31.0){\rule[-0.200pt]{0.400pt}{4.818pt}}
\put(862,19){\makebox(0,0){\shortstack{\\ \\ \\ {$0.1$}}}}
\put(862.0,1773.0){\rule[-0.200pt]{0.400pt}{4.818pt}}
\put(1233.0,31.0){\rule[-0.200pt]{0.400pt}{4.818pt}}
\put(1233,19){\makebox(0,0){\shortstack{\\ \\ \\ {$0.15$}}}}
\put(1233.0,1773.0){\rule[-0.200pt]{0.400pt}{4.818pt}}
\put(120.0,31.0){\rule[-0.200pt]{321.842pt}{0.400pt}}
\put(1456.0,31.0){\rule[-0.200pt]{0.400pt}{424.466pt}}
\put(120.0,1793.0){\rule[-0.200pt]{321.842pt}{0.400pt}}
\put(-48,912){\makebox(0,0){{\Large{${m_{0^{++}} \over {\surd\sigma}}$}}}}
\put(788,-89){\makebox(0,0){{\large{$a^2\sigma$}}}}
\put(120.0,31.0){\rule[-0.200pt]{0.400pt}{424.466pt}}
\put(231,1233){\circle*{12}}
\put(312,1257){\circle*{12}}
\put(475,1176){\circle*{12}}
\put(627,1123){\circle*{12}}
\put(1129,894){\circle*{12}}
\put(231.0,1190.0){\rule[-0.200pt]{0.400pt}{20.476pt}}
\put(221.0,1190.0){\rule[-0.200pt]{4.818pt}{0.400pt}}
\put(221.0,1275.0){\rule[-0.200pt]{4.818pt}{0.400pt}}
\put(312.0,1212.0){\rule[-0.200pt]{0.400pt}{21.922pt}}
\put(302.0,1212.0){\rule[-0.200pt]{4.818pt}{0.400pt}}
\put(302.0,1303.0){\rule[-0.200pt]{4.818pt}{0.400pt}}
\put(475.0,1138.0){\rule[-0.200pt]{0.400pt}{18.549pt}}
\put(465.0,1138.0){\rule[-0.200pt]{4.818pt}{0.400pt}}
\put(465.0,1215.0){\rule[-0.200pt]{4.818pt}{0.400pt}}
\put(627.0,1088.0){\rule[-0.200pt]{0.400pt}{17.104pt}}
\put(617.0,1088.0){\rule[-0.200pt]{4.818pt}{0.400pt}}
\put(617.0,1159.0){\rule[-0.200pt]{4.818pt}{0.400pt}}
\put(1129.0,856.0){\rule[-0.200pt]{0.400pt}{18.549pt}}
\put(1119.0,856.0){\rule[-0.200pt]{4.818pt}{0.400pt}}
\put(1119.0,933.0){\rule[-0.200pt]{4.818pt}{0.400pt}}
\put(229,1286){\makebox(0,0){$\times$}}
\put(330,1250){\makebox(0,0){$\times$}}
\put(575,1138){\makebox(0,0){$\times$}}
\put(763,1056){\makebox(0,0){$\times$}}
\put(1115,965){\makebox(0,0){$\times$}}
\put(229.0,1250.0){\rule[-0.200pt]{0.400pt}{17.104pt}}
\put(219.0,1250.0){\rule[-0.200pt]{4.818pt}{0.400pt}}
\put(219.0,1321.0){\rule[-0.200pt]{4.818pt}{0.400pt}}
\put(330.0,1226.0){\rule[-0.200pt]{0.400pt}{11.804pt}}
\put(320.0,1226.0){\rule[-0.200pt]{4.818pt}{0.400pt}}
\put(320.0,1275.0){\rule[-0.200pt]{4.818pt}{0.400pt}}
\put(575.0,1123.0){\rule[-0.200pt]{0.400pt}{6.986pt}}
\put(565.0,1123.0){\rule[-0.200pt]{4.818pt}{0.400pt}}
\put(565.0,1152.0){\rule[-0.200pt]{4.818pt}{0.400pt}}
\put(763.0,1004.0){\rule[-0.200pt]{0.400pt}{25.294pt}}
\put(753.0,1004.0){\rule[-0.200pt]{4.818pt}{0.400pt}}
\put(753.0,1109.0){\rule[-0.200pt]{4.818pt}{0.400pt}}
\put(1115.0,937.0){\rule[-0.200pt]{0.400pt}{13.490pt}}
\put(1105.0,937.0){\rule[-0.200pt]{4.818pt}{0.400pt}}
\put(1105.0,993.0){\rule[-0.200pt]{4.818pt}{0.400pt}}
\sbox{\plotpoint}{\rule[-0.500pt]{1.000pt}{1.000pt}}%
\put(120,1316){\usebox{\plotpoint}}
\put(120.00,1316.00){\usebox{\plotpoint}}
\put(139.43,1308.70){\usebox{\plotpoint}}
\put(158.87,1301.44){\usebox{\plotpoint}}
\multiput(160,1301)(19.077,-8.176){0}{\usebox{\plotpoint}}
\put(178.02,1293.45){\usebox{\plotpoint}}
\put(197.49,1286.25){\usebox{\plotpoint}}
\multiput(201,1285)(19.372,-7.451){0}{\usebox{\plotpoint}}
\put(216.92,1278.96){\usebox{\plotpoint}}
\put(236.39,1271.77){\usebox{\plotpoint}}
\multiput(241,1270)(19.546,-6.981){0}{\usebox{\plotpoint}}
\put(255.89,1264.66){\usebox{\plotpoint}}
\put(275.32,1257.38){\usebox{\plotpoint}}
\put(294.41,1249.27){\usebox{\plotpoint}}
\multiput(295,1249)(19.546,-6.981){0}{\usebox{\plotpoint}}
\put(313.89,1242.12){\usebox{\plotpoint}}
\put(333.36,1234.94){\usebox{\plotpoint}}
\multiput(336,1234)(19.372,-7.451){0}{\usebox{\plotpoint}}
\put(352.79,1227.65){\usebox{\plotpoint}}
\put(372.26,1220.44){\usebox{\plotpoint}}
\multiput(376,1219)(19.546,-6.981){0}{\usebox{\plotpoint}}
\put(391.75,1213.33){\usebox{\plotpoint}}
\put(411.00,1205.57){\usebox{\plotpoint}}
\multiput(417,1203)(19.372,-7.451){0}{\usebox{\plotpoint}}
\put(430.28,1197.90){\usebox{\plotpoint}}
\put(449.78,1190.78){\usebox{\plotpoint}}
\put(469.26,1183.62){\usebox{\plotpoint}}
\multiput(471,1183)(19.372,-7.451){0}{\usebox{\plotpoint}}
\put(488.69,1176.33){\usebox{\plotpoint}}
\put(508.14,1169.10){\usebox{\plotpoint}}
\multiput(511,1168)(19.077,-8.176){0}{\usebox{\plotpoint}}
\put(527.30,1161.12){\usebox{\plotpoint}}
\put(546.75,1153.88){\usebox{\plotpoint}}
\multiput(552,1152)(19.372,-7.451){0}{\usebox{\plotpoint}}
\put(566.18,1146.58){\usebox{\plotpoint}}
\put(585.66,1139.44){\usebox{\plotpoint}}
\put(605.15,1132.30){\usebox{\plotpoint}}
\multiput(606,1132)(19.372,-7.451){0}{\usebox{\plotpoint}}
\put(624.58,1125.01){\usebox{\plotpoint}}
\put(643.73,1117.05){\usebox{\plotpoint}}
\multiput(646,1116)(19.546,-6.981){0}{\usebox{\plotpoint}}
\put(663.16,1109.78){\usebox{\plotpoint}}
\put(682.62,1102.56){\usebox{\plotpoint}}
\multiput(687,1101)(19.372,-7.451){0}{\usebox{\plotpoint}}
\put(702.05,1095.27){\usebox{\plotpoint}}
\put(721.53,1088.10){\usebox{\plotpoint}}
\multiput(727,1086)(19.546,-6.981){0}{\usebox{\plotpoint}}
\put(741.03,1080.99){\usebox{\plotpoint}}
\put(760.09,1072.83){\usebox{\plotpoint}}
\put(779.53,1065.56){\usebox{\plotpoint}}
\multiput(781,1065)(19.546,-6.981){0}{\usebox{\plotpoint}}
\put(799.03,1058.45){\usebox{\plotpoint}}
\put(818.49,1051.25){\usebox{\plotpoint}}
\multiput(822,1050)(19.372,-7.451){0}{\usebox{\plotpoint}}
\put(837.92,1043.96){\usebox{\plotpoint}}
\put(857.39,1036.77){\usebox{\plotpoint}}
\multiput(862,1035)(19.077,-8.176){0}{\usebox{\plotpoint}}
\put(876.55,1028.79){\usebox{\plotpoint}}
\put(895.98,1021.51){\usebox{\plotpoint}}
\put(915.42,1014.22){\usebox{\plotpoint}}
\multiput(916,1014)(19.546,-6.981){0}{\usebox{\plotpoint}}
\put(934.92,1007.11){\usebox{\plotpoint}}
\put(954.39,999.93){\usebox{\plotpoint}}
\multiput(957,999)(19.372,-7.451){0}{\usebox{\plotpoint}}
\put(973.82,992.64){\usebox{\plotpoint}}
\put(993.03,984.83){\usebox{\plotpoint}}
\multiput(997,983)(19.546,-6.981){0}{\usebox{\plotpoint}}
\put(1012.42,977.46){\usebox{\plotpoint}}
\put(1031.86,970.19){\usebox{\plotpoint}}
\multiput(1038,968)(19.372,-7.451){0}{\usebox{\plotpoint}}
\put(1051.29,962.90){\usebox{\plotpoint}}
\put(1070.78,955.78){\usebox{\plotpoint}}
\put(1090.26,948.62){\usebox{\plotpoint}}
\multiput(1092,948)(18.845,-8.698){0}{\usebox{\plotpoint}}
\put(1109.32,940.46){\usebox{\plotpoint}}
\put(1128.78,933.24){\usebox{\plotpoint}}
\multiput(1132,932)(19.546,-6.981){0}{\usebox{\plotpoint}}
\put(1148.28,926.12){\usebox{\plotpoint}}
\put(1167.73,918.88){\usebox{\plotpoint}}
\multiput(1173,917)(19.372,-7.451){0}{\usebox{\plotpoint}}
\put(1187.16,911.59){\usebox{\plotpoint}}
\put(1206.65,904.44){\usebox{\plotpoint}}
\put(1225.82,896.51){\usebox{\plotpoint}}
\multiput(1227,896)(19.372,-7.451){0}{\usebox{\plotpoint}}
\put(1245.22,889.14){\usebox{\plotpoint}}
\put(1264.67,881.90){\usebox{\plotpoint}}
\multiput(1267,881)(19.546,-6.981){0}{\usebox{\plotpoint}}
\put(1284.17,874.78){\usebox{\plotpoint}}
\put(1303.63,867.56){\usebox{\plotpoint}}
\multiput(1308,866)(19.372,-7.451){0}{\usebox{\plotpoint}}
\put(1323.01,860.14){\usebox{\plotpoint}}
\put(1342.19,852.23){\usebox{\plotpoint}}
\put(1361.69,845.11){\usebox{\plotpoint}}
\multiput(1362,845)(19.372,-7.451){0}{\usebox{\plotpoint}}
\put(1381.12,837.82){\usebox{\plotpoint}}
\put(1400.56,830.55){\usebox{\plotpoint}}
\multiput(1402,830)(19.546,-6.981){0}{\usebox{\plotpoint}}
\put(1420.05,823.44){\usebox{\plotpoint}}
\put(1439.52,816.24){\usebox{\plotpoint}}
\multiput(1443,815)(18.845,-8.698){0}{\usebox{\plotpoint}}
\put(1456,809){\usebox{\plotpoint}}
\end{picture}
\end	{center}
\vskip 0.15in
\caption{The scalar glueball mass: the GF11 values ($\times$)
and the rest ($\bullet$).  
The best linear extrapolation to the continuum limit
is shown.}
\label	{fig-glue-scalar}
\end 	{figure}
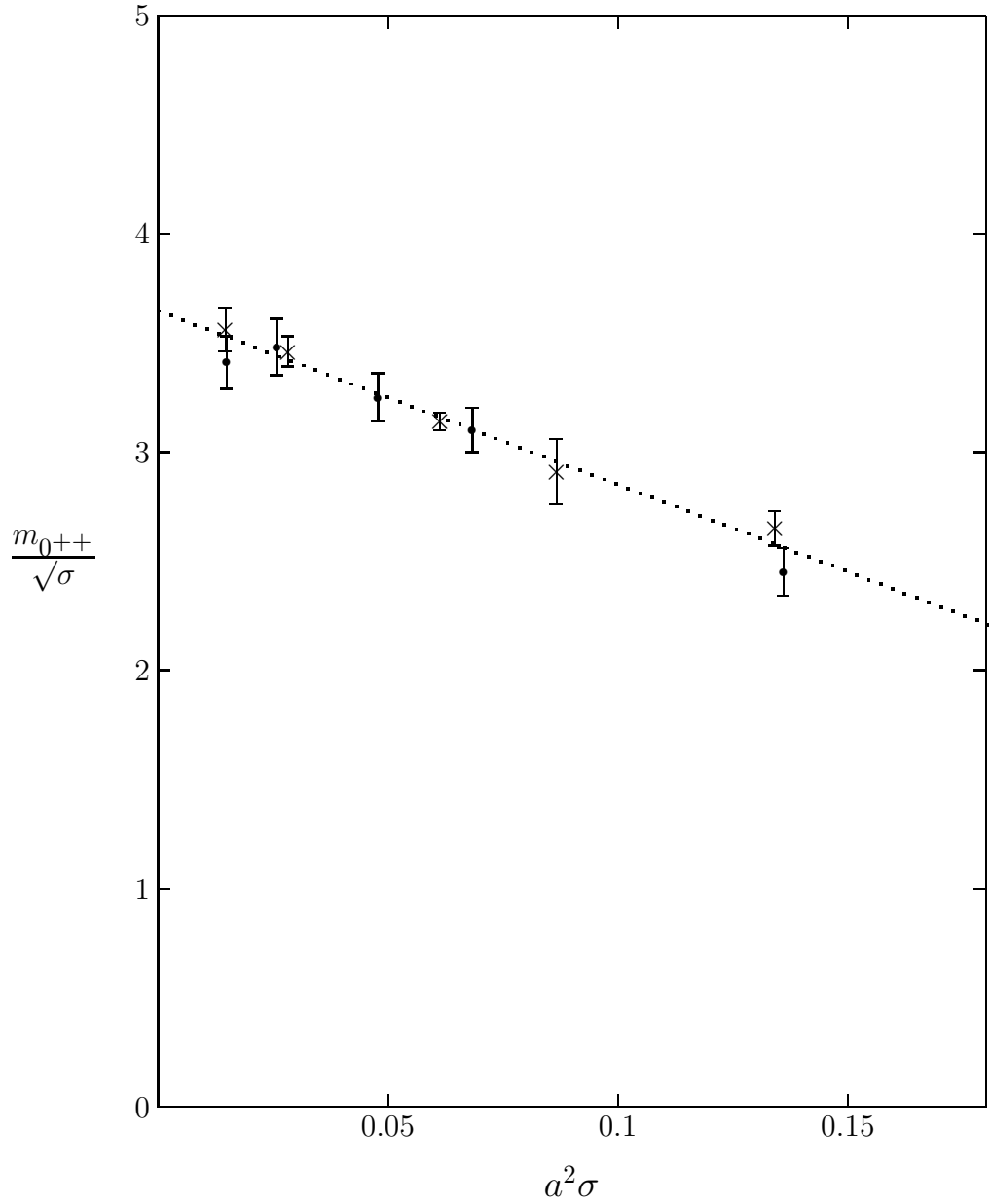

The first step is to take ratios of masses so that the 
scale, $a$, in which they are expressed cancels.
We choose to take ratios of the glueball masses, $am_G$,
to $a\surd\sigma$ since this latter quantity has
been very accurately calculated. Now we remarked above
that for small enough $a$ the leading discretisation 
effects in such mass ratios are $0(a^2)$. So for small enough $a$
we expect
\begin{equation}
{m_G(a) \over \surd\sigma(a)} = {m_G(a=0) \over \surd\sigma(a=0)} 
+ c a^2\sigma    \label{A2}
\end{equation}
We take all the available mass values and try to fit them 
using 
eqn~\ref{A2}. 
If a good fit is not possible we assume
that this is because the largest values of $a$ used is 
too large for the $O(a^2)$ correction to be adequate.
So we drop the mass corresponding to the largest 
value of $a$ and try again. We keep doing this until we
get a good fit. We find that the $0^{++}$ glueball
can be well fitted in this way over 
a range of lattice spacings $0.18 fm \geq a \geq 0.06 fm$,
corresponding to couplings $5.7 \le \beta \le 6.4$.
(Note that where we employ fermi units, these have been introduced
using $\surd\sigma=440MeV$; a value that will be made
plausible later on.)
Such lattice spacings are small enough that we are not
surprised that higher order terms in $a^2$ should be small.
We show the mass ratios in 
Fig~\ref{fig-glue-scalar}
together with the best
fit of the form in 
eqn~\ref{A2}. 
We obtain the $continuum$ mass ratio:
\begin{equation}
{{m_{0^{++}}}\over \surd\sigma} = 3.65 \pm 0.11 
\label{A3}
\end{equation}       
The fit is obviously a very good one (with a
confidence level of $85\%$) and it is clear that
the calculations of the different groups are entirely
consistent with each other.

\begin	{figure}[p]
\begin	{center}
\leavevmode
\setlength{\unitlength}{0.240900pt}
\ifx\plotpoint\undefined\newsavebox{\plotpoint}\fi
\sbox{\plotpoint}{\rule[-0.200pt]{0.400pt}{0.400pt}}%
\begin{picture}(1500,1800)(0,0)
\font\gnuplot=cmr10 at 12pt
\gnuplot
\sbox{\plotpoint}{\rule[-0.200pt]{0.400pt}{0.400pt}}%
\put(120.0,31.0){\rule[-0.200pt]{4.818pt}{0.400pt}}
\put(108,31){\makebox(0,0)[r]{{$2$}}}
\put(1436.0,31.0){\rule[-0.200pt]{4.818pt}{0.400pt}}
\put(120.0,383.0){\rule[-0.200pt]{4.818pt}{0.400pt}}
\put(108,383){\makebox(0,0)[r]{{$3$}}}
\put(1436.0,383.0){\rule[-0.200pt]{4.818pt}{0.400pt}}
\put(120.0,736.0){\rule[-0.200pt]{4.818pt}{0.400pt}}
\put(108,736){\makebox(0,0)[r]{{$4$}}}
\put(1436.0,736.0){\rule[-0.200pt]{4.818pt}{0.400pt}}
\put(120.0,1088.0){\rule[-0.200pt]{4.818pt}{0.400pt}}
\put(108,1088){\makebox(0,0)[r]{{$5$}}}
\put(1436.0,1088.0){\rule[-0.200pt]{4.818pt}{0.400pt}}
\put(120.0,1441.0){\rule[-0.200pt]{4.818pt}{0.400pt}}
\put(108,1441){\makebox(0,0)[r]{{$6$}}}
\put(1436.0,1441.0){\rule[-0.200pt]{4.818pt}{0.400pt}}
\put(387.0,31.0){\rule[-0.200pt]{0.400pt}{4.818pt}}
\put(387,19){\makebox(0,0){\shortstack{\\ \\ \\ {$0.02$}}}}
\put(387.0,1773.0){\rule[-0.200pt]{0.400pt}{4.818pt}}
\put(654.0,31.0){\rule[-0.200pt]{0.400pt}{4.818pt}}
\put(654,19){\makebox(0,0){\shortstack{\\ \\ \\ {$0.04$}}}}
\put(654.0,1773.0){\rule[-0.200pt]{0.400pt}{4.818pt}}
\put(922.0,31.0){\rule[-0.200pt]{0.400pt}{4.818pt}}
\put(922,19){\makebox(0,0){\shortstack{\\ \\ \\ {$0.06$}}}}
\put(922.0,1773.0){\rule[-0.200pt]{0.400pt}{4.818pt}}
\put(1189.0,31.0){\rule[-0.200pt]{0.400pt}{4.818pt}}
\put(1189,19){\makebox(0,0){\shortstack{\\ \\ \\ {$0.08$}}}}
\put(1189.0,1773.0){\rule[-0.200pt]{0.400pt}{4.818pt}}
\put(120.0,31.0){\rule[-0.200pt]{321.842pt}{0.400pt}}
\put(1456.0,31.0){\rule[-0.200pt]{0.400pt}{424.466pt}}
\put(120.0,1793.0){\rule[-0.200pt]{321.842pt}{0.400pt}}
\put(-48,912){\makebox(0,0){{\Large{${m_{2^{++}} \over {\surd\sigma}}$}}}}
\put(788,-89){\makebox(0,0){{\large{$a^2\sigma$}}}}
\put(120.0,31.0){\rule[-0.200pt]{0.400pt}{424.466pt}}
\put(319,1092){\circle*{12}}
\put(465,1176){\circle*{12}}
\put(759,1109){\circle*{12}}
\put(1032,1138){\circle*{12}}
\put(319.0,1042.0){\rule[-0.200pt]{0.400pt}{23.849pt}}
\put(309.0,1042.0){\rule[-0.200pt]{4.818pt}{0.400pt}}
\put(309.0,1141.0){\rule[-0.200pt]{4.818pt}{0.400pt}}
\put(465.0,1130.0){\rule[-0.200pt]{0.400pt}{22.163pt}}
\put(455.0,1130.0){\rule[-0.200pt]{4.818pt}{0.400pt}}
\put(455.0,1222.0){\rule[-0.200pt]{4.818pt}{0.400pt}}
\put(759.0,1067.0){\rule[-0.200pt]{0.400pt}{20.476pt}}
\put(749.0,1067.0){\rule[-0.200pt]{4.818pt}{0.400pt}}
\put(749.0,1152.0){\rule[-0.200pt]{4.818pt}{0.400pt}}
\put(1032.0,1085.0){\rule[-0.200pt]{0.400pt}{25.294pt}}
\put(1022.0,1085.0){\rule[-0.200pt]{4.818pt}{0.400pt}}
\put(1022.0,1190.0){\rule[-0.200pt]{4.818pt}{0.400pt}}
\put(316,1169){\makebox(0,0){$\times$}}
\put(498,1060){\makebox(0,0){$\times$}}
\put(940,1067){\makebox(0,0){$\times$}}
\put(316.0,1102.0){\rule[-0.200pt]{0.400pt}{32.281pt}}
\put(306.0,1102.0){\rule[-0.200pt]{4.818pt}{0.400pt}}
\put(306.0,1236.0){\rule[-0.200pt]{4.818pt}{0.400pt}}
\put(498.0,1004.0){\rule[-0.200pt]{0.400pt}{26.981pt}}
\put(488.0,1004.0){\rule[-0.200pt]{4.818pt}{0.400pt}}
\put(488.0,1116.0){\rule[-0.200pt]{4.818pt}{0.400pt}}
\put(940.0,1039.0){\rule[-0.200pt]{0.400pt}{13.490pt}}
\put(930.0,1039.0){\rule[-0.200pt]{4.818pt}{0.400pt}}
\put(930.0,1095.0){\rule[-0.200pt]{4.818pt}{0.400pt}}
\sbox{\plotpoint}{\rule[-0.500pt]{1.000pt}{1.000pt}}%
\put(120,1141){\usebox{\plotpoint}}
\put(120.00,1141.00){\usebox{\plotpoint}}
\put(140.70,1139.45){\usebox{\plotpoint}}
\multiput(147,1139)(20.756,0.000){0}{\usebox{\plotpoint}}
\put(161.43,1138.90){\usebox{\plotpoint}}
\put(182.13,1137.37){\usebox{\plotpoint}}
\multiput(187,1137)(20.703,-1.479){0}{\usebox{\plotpoint}}
\put(202.83,1135.86){\usebox{\plotpoint}}
\put(223.56,1135.00){\usebox{\plotpoint}}
\multiput(228,1135)(20.694,-1.592){0}{\usebox{\plotpoint}}
\put(244.26,1133.77){\usebox{\plotpoint}}
\put(264.96,1132.23){\usebox{\plotpoint}}
\multiput(268,1132)(20.703,-1.479){0}{\usebox{\plotpoint}}
\put(285.66,1130.72){\usebox{\plotpoint}}
\put(306.39,1130.00){\usebox{\plotpoint}}
\multiput(309,1130)(20.694,-1.592){0}{\usebox{\plotpoint}}
\put(327.09,1128.64){\usebox{\plotpoint}}
\put(347.79,1127.09){\usebox{\plotpoint}}
\multiput(349,1127)(20.703,-1.479){0}{\usebox{\plotpoint}}
\put(368.51,1126.00){\usebox{\plotpoint}}
\put(389.23,1125.05){\usebox{\plotpoint}}
\multiput(390,1125)(20.694,-1.592){0}{\usebox{\plotpoint}}
\put(409.93,1123.51){\usebox{\plotpoint}}
\multiput(417,1123)(20.694,-1.592){0}{\usebox{\plotpoint}}
\put(430.63,1122.00){\usebox{\plotpoint}}
\put(451.36,1121.43){\usebox{\plotpoint}}
\multiput(457,1121)(20.703,-1.479){0}{\usebox{\plotpoint}}
\put(472.06,1119.92){\usebox{\plotpoint}}
\put(492.76,1118.37){\usebox{\plotpoint}}
\multiput(498,1118)(20.694,-1.592){0}{\usebox{\plotpoint}}
\put(513.46,1117.00){\usebox{\plotpoint}}
\put(534.19,1116.29){\usebox{\plotpoint}}
\multiput(538,1116)(20.703,-1.479){0}{\usebox{\plotpoint}}
\put(554.89,1114.78){\usebox{\plotpoint}}
\put(575.59,1113.24){\usebox{\plotpoint}}
\multiput(579,1113)(20.756,0.000){0}{\usebox{\plotpoint}}
\put(596.33,1112.69){\usebox{\plotpoint}}
\put(617.02,1111.15){\usebox{\plotpoint}}
\multiput(619,1111)(20.703,-1.479){0}{\usebox{\plotpoint}}
\put(637.72,1109.64){\usebox{\plotpoint}}
\put(658.42,1108.11){\usebox{\plotpoint}}
\multiput(660,1108)(20.756,0.000){0}{\usebox{\plotpoint}}
\put(679.16,1107.56){\usebox{\plotpoint}}
\put(699.86,1106.01){\usebox{\plotpoint}}
\multiput(700,1106)(20.703,-1.479){0}{\usebox{\plotpoint}}
\put(720.56,1104.50){\usebox{\plotpoint}}
\multiput(727,1104)(20.756,0.000){0}{\usebox{\plotpoint}}
\put(741.29,1103.98){\usebox{\plotpoint}}
\put(761.99,1102.43){\usebox{\plotpoint}}
\multiput(768,1102)(20.694,-1.592){0}{\usebox{\plotpoint}}
\put(782.69,1100.88){\usebox{\plotpoint}}
\put(803.41,1100.00){\usebox{\plotpoint}}
\multiput(808,1100)(20.703,-1.479){0}{\usebox{\plotpoint}}
\put(824.13,1098.84){\usebox{\plotpoint}}
\put(844.82,1097.30){\usebox{\plotpoint}}
\multiput(849,1097)(20.694,-1.592){0}{\usebox{\plotpoint}}
\put(865.52,1095.75){\usebox{\plotpoint}}
\put(886.25,1095.00){\usebox{\plotpoint}}
\multiput(889,1095)(20.703,-1.479){0}{\usebox{\plotpoint}}
\put(906.96,1093.70){\usebox{\plotpoint}}
\put(927.66,1092.17){\usebox{\plotpoint}}
\multiput(930,1092)(20.694,-1.592){0}{\usebox{\plotpoint}}
\put(948.37,1091.00){\usebox{\plotpoint}}
\put(969.09,1090.07){\usebox{\plotpoint}}
\multiput(970,1090)(20.703,-1.479){0}{\usebox{\plotpoint}}
\put(989.79,1088.55){\usebox{\plotpoint}}
\put(1010.49,1087.04){\usebox{\plotpoint}}
\multiput(1011,1087)(20.694,-1.592){0}{\usebox{\plotpoint}}
\put(1031.20,1086.00){\usebox{\plotpoint}}
\multiput(1038,1086)(20.694,-1.592){0}{\usebox{\plotpoint}}
\put(1051.92,1084.93){\usebox{\plotpoint}}
\put(1072.62,1083.41){\usebox{\plotpoint}}
\multiput(1078,1083)(20.703,-1.479){0}{\usebox{\plotpoint}}
\put(1093.32,1082.00){\usebox{\plotpoint}}
\put(1114.05,1081.35){\usebox{\plotpoint}}
\multiput(1119,1081)(20.694,-1.592){0}{\usebox{\plotpoint}}
\put(1134.75,1079.80){\usebox{\plotpoint}}
\put(1155.45,1078.27){\usebox{\plotpoint}}
\multiput(1159,1078)(20.756,0.000){0}{\usebox{\plotpoint}}
\put(1176.19,1077.75){\usebox{\plotpoint}}
\put(1196.89,1076.22){\usebox{\plotpoint}}
\multiput(1200,1076)(20.694,-1.592){0}{\usebox{\plotpoint}}
\put(1217.58,1074.67){\usebox{\plotpoint}}
\put(1238.28,1073.13){\usebox{\plotpoint}}
\multiput(1240,1073)(20.756,0.000){0}{\usebox{\plotpoint}}
\put(1259.02,1072.61){\usebox{\plotpoint}}
\put(1279.72,1071.09){\usebox{\plotpoint}}
\multiput(1281,1071)(20.694,-1.592){0}{\usebox{\plotpoint}}
\put(1300.41,1069.54){\usebox{\plotpoint}}
\multiput(1308,1069)(20.756,0.000){0}{\usebox{\plotpoint}}
\put(1321.15,1068.99){\usebox{\plotpoint}}
\put(1341.85,1067.47){\usebox{\plotpoint}}
\multiput(1348,1067)(20.703,-1.479){0}{\usebox{\plotpoint}}
\put(1362.55,1065.96){\usebox{\plotpoint}}
\put(1383.25,1064.41){\usebox{\plotpoint}}
\multiput(1389,1064)(20.756,0.000){0}{\usebox{\plotpoint}}
\put(1403.98,1063.86){\usebox{\plotpoint}}
\put(1424.68,1062.33){\usebox{\plotpoint}}
\multiput(1429,1062)(20.703,-1.479){0}{\usebox{\plotpoint}}
\put(1445.38,1060.82){\usebox{\plotpoint}}
\put(1456,1060){\usebox{\plotpoint}}
\end{picture}
\end	{center}
\vskip 0.15in
\caption{The tensor glueball mass: the GF11 values ($\times$)
and the rest ($\bullet$).  
The best linear extrapolation to the continuum limit
is shown.}
\label	{fig-glue-tensor}
\end 	{figure}
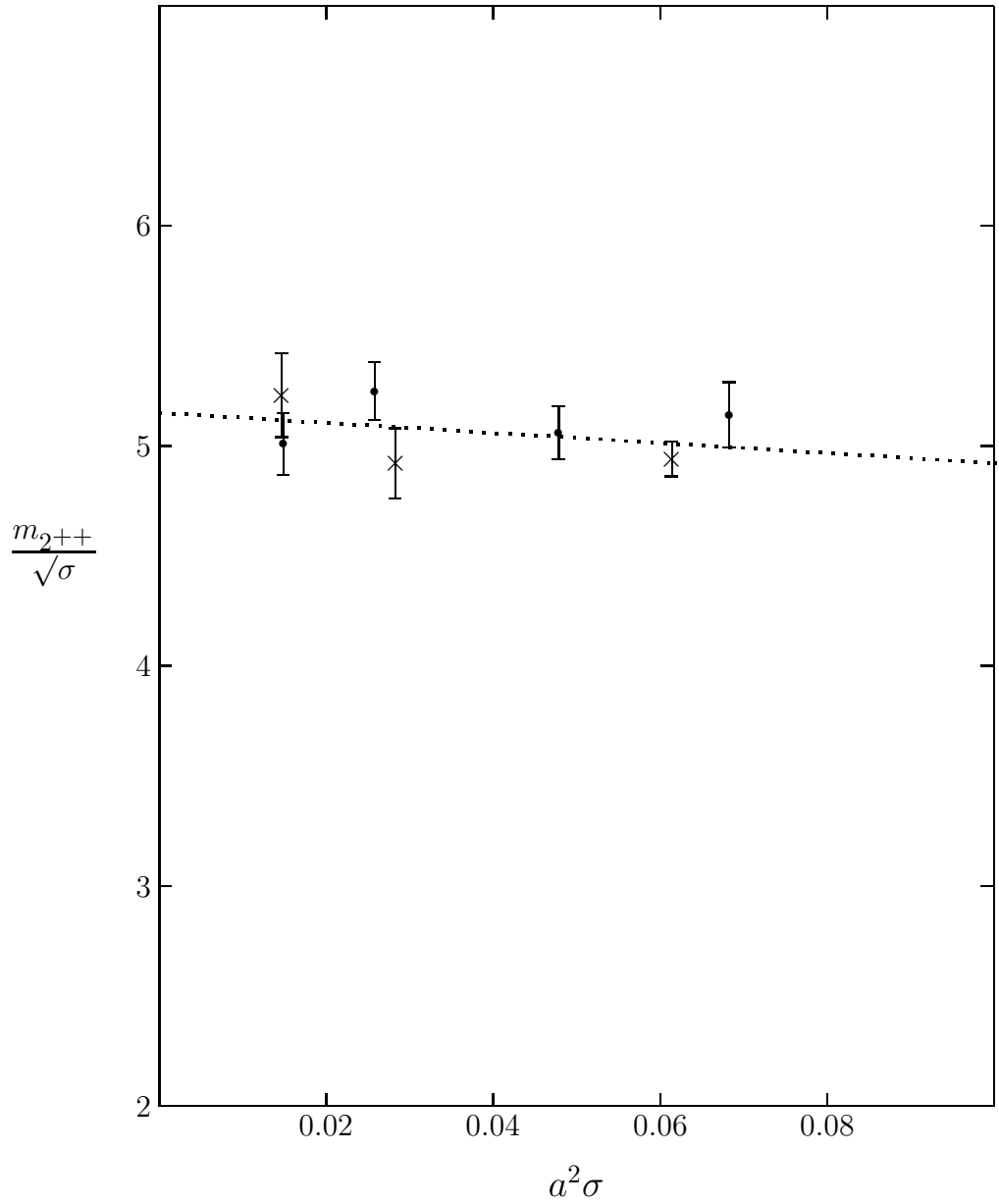

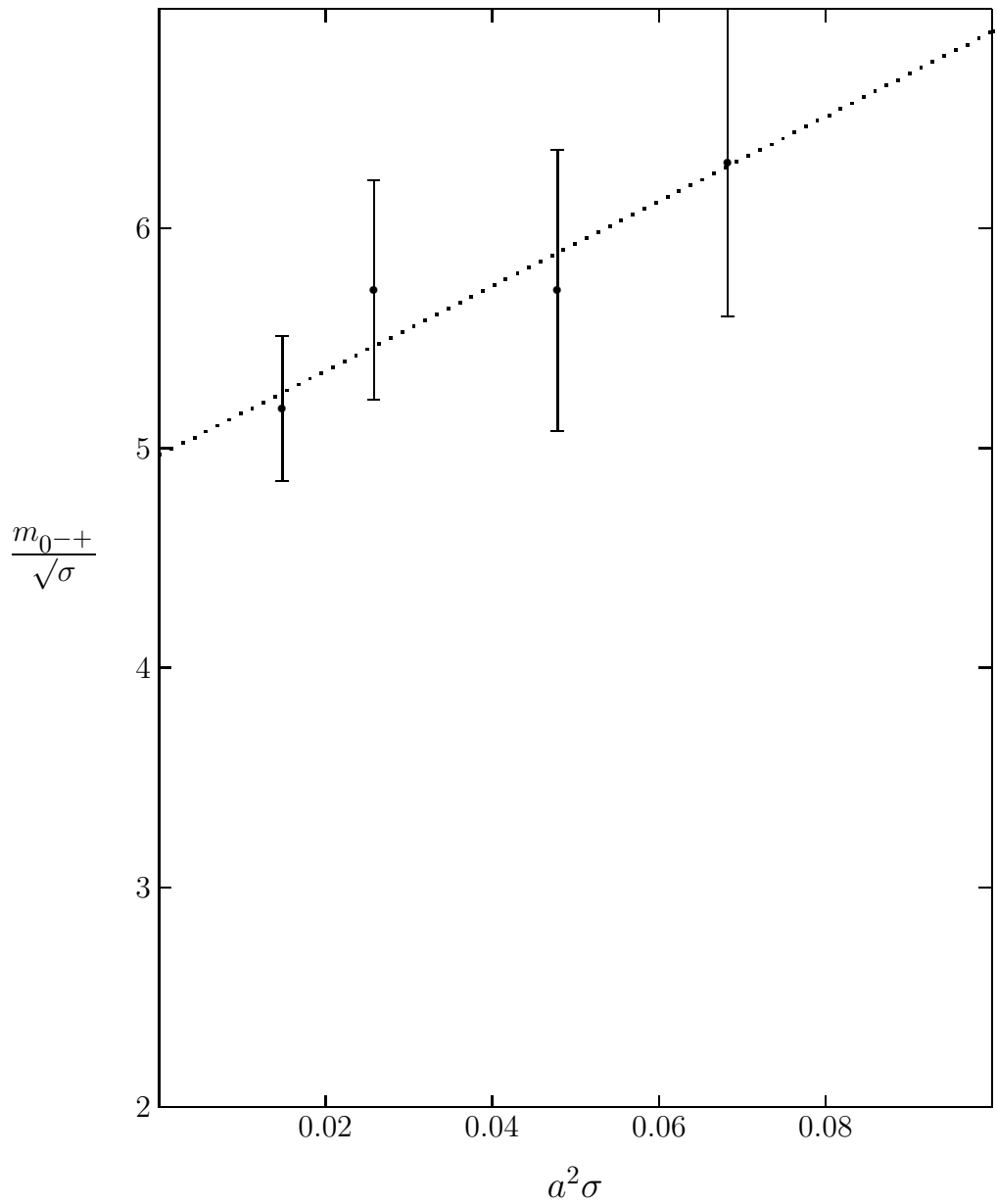
\begin	{figure}[p]
\begin	{center}
\leavevmode
\setlength{\unitlength}{0.240900pt}
\ifx\plotpoint\undefined\newsavebox{\plotpoint}\fi
\sbox{\plotpoint}{\rule[-0.200pt]{0.400pt}{0.400pt}}%
\begin{picture}(1500,1800)(0,0)
\font\gnuplot=cmr10 at 12pt
\gnuplot
\sbox{\plotpoint}{\rule[-0.200pt]{0.400pt}{0.400pt}}%
\put(120.0,31.0){\rule[-0.200pt]{4.818pt}{0.400pt}}
\put(108,31){\makebox(0,0)[r]{{$2$}}}
\put(1436.0,31.0){\rule[-0.200pt]{4.818pt}{0.400pt}}
\put(120.0,383.0){\rule[-0.200pt]{4.818pt}{0.400pt}}
\put(108,383){\makebox(0,0)[r]{{$3$}}}
\put(1436.0,383.0){\rule[-0.200pt]{4.818pt}{0.400pt}}
\put(120.0,736.0){\rule[-0.200pt]{4.818pt}{0.400pt}}
\put(108,736){\makebox(0,0)[r]{{$4$}}}
\put(1436.0,736.0){\rule[-0.200pt]{4.818pt}{0.400pt}}
\put(120.0,1088.0){\rule[-0.200pt]{4.818pt}{0.400pt}}
\put(108,1088){\makebox(0,0)[r]{{$5$}}}
\put(1436.0,1088.0){\rule[-0.200pt]{4.818pt}{0.400pt}}
\put(120.0,1441.0){\rule[-0.200pt]{4.818pt}{0.400pt}}
\put(108,1441){\makebox(0,0)[r]{{$6$}}}
\put(1436.0,1441.0){\rule[-0.200pt]{4.818pt}{0.400pt}}
\put(387.0,31.0){\rule[-0.200pt]{0.400pt}{4.818pt}}
\put(387,19){\makebox(0,0){\shortstack{\\ \\ \\ {$0.02$}}}}
\put(387.0,1773.0){\rule[-0.200pt]{0.400pt}{4.818pt}}
\put(654.0,31.0){\rule[-0.200pt]{0.400pt}{4.818pt}}
\put(654,19){\makebox(0,0){\shortstack{\\ \\ \\ {$0.04$}}}}
\put(654.0,1773.0){\rule[-0.200pt]{0.400pt}{4.818pt}}
\put(922.0,31.0){\rule[-0.200pt]{0.400pt}{4.818pt}}
\put(922,19){\makebox(0,0){\shortstack{\\ \\ \\ {$0.06$}}}}
\put(922.0,1773.0){\rule[-0.200pt]{0.400pt}{4.818pt}}
\put(1189.0,31.0){\rule[-0.200pt]{0.400pt}{4.818pt}}
\put(1189,19){\makebox(0,0){\shortstack{\\ \\ \\ {$0.08$}}}}
\put(1189.0,1773.0){\rule[-0.200pt]{0.400pt}{4.818pt}}
\put(120.0,31.0){\rule[-0.200pt]{321.842pt}{0.400pt}}
\put(1456.0,31.0){\rule[-0.200pt]{0.400pt}{424.466pt}}
\put(120.0,1793.0){\rule[-0.200pt]{321.842pt}{0.400pt}}
\put(-48,912){\makebox(0,0){{\Large{${m_{0^{-+}} \over {\surd\sigma}}$}}}}
\put(788,-89){\makebox(0,0){{\large{$a^2\sigma$}}}}
\put(120.0,31.0){\rule[-0.200pt]{0.400pt}{424.466pt}}
\put(318,1152){\circle*{12}}
\put(465,1342){\circle*{12}}
\put(759,1342){\circle*{12}}
\put(1032,1546){\circle*{12}}
\put(318.0,1035.0){\rule[-0.200pt]{0.400pt}{56.130pt}}
\put(308.0,1035.0){\rule[-0.200pt]{4.818pt}{0.400pt}}
\put(308.0,1268.0){\rule[-0.200pt]{4.818pt}{0.400pt}}
\put(465.0,1166.0){\rule[-0.200pt]{0.400pt}{84.797pt}}
\put(455.0,1166.0){\rule[-0.200pt]{4.818pt}{0.400pt}}
\put(455.0,1518.0){\rule[-0.200pt]{4.818pt}{0.400pt}}
\put(759.0,1116.0){\rule[-0.200pt]{0.400pt}{108.646pt}}
\put(749.0,1116.0){\rule[-0.200pt]{4.818pt}{0.400pt}}
\put(749.0,1567.0){\rule[-0.200pt]{4.818pt}{0.400pt}}
\put(1032.0,1300.0){\rule[-0.200pt]{0.400pt}{118.764pt}}
\put(1022.0,1300.0){\rule[-0.200pt]{4.818pt}{0.400pt}}
\put(1022.0,1793.0){\rule[-0.200pt]{4.818pt}{0.400pt}}
\sbox{\plotpoint}{\rule[-0.500pt]{1.000pt}{1.000pt}}%
\put(120,1077){\usebox{\plotpoint}}
\put(120.00,1077.00){\usebox{\plotpoint}}
\put(138.36,1086.68){\usebox{\plotpoint}}
\put(156.77,1096.26){\usebox{\plotpoint}}
\multiput(160,1098)(19.077,8.176){0}{\usebox{\plotpoint}}
\put(175.63,1104.88){\usebox{\plotpoint}}
\put(194.02,1114.51){\usebox{\plotpoint}}
\put(212.40,1124.14){\usebox{\plotpoint}}
\multiput(214,1125)(18.564,9.282){0}{\usebox{\plotpoint}}
\put(230.89,1133.56){\usebox{\plotpoint}}
\put(249.30,1143.15){\usebox{\plotpoint}}
\multiput(255,1146)(18.845,8.698){0}{\usebox{\plotpoint}}
\put(268.05,1152.03){\usebox{\plotpoint}}
\put(286.55,1161.45){\usebox{\plotpoint}}
\put(304.98,1170.99){\usebox{\plotpoint}}
\multiput(309,1173)(18.275,9.840){0}{\usebox{\plotpoint}}
\put(323.33,1180.67){\usebox{\plotpoint}}
\put(341.81,1190.13){\usebox{\plotpoint}}
\put(360.26,1199.63){\usebox{\plotpoint}}
\multiput(363,1201)(18.845,8.698){0}{\usebox{\plotpoint}}
\put(379.02,1208.51){\usebox{\plotpoint}}
\put(397.46,1218.02){\usebox{\plotpoint}}
\put(415.94,1227.47){\usebox{\plotpoint}}
\multiput(417,1228)(18.275,9.840){0}{\usebox{\plotpoint}}
\put(434.30,1237.15){\usebox{\plotpoint}}
\put(452.72,1246.70){\usebox{\plotpoint}}
\multiput(457,1249)(19.077,8.176){0}{\usebox{\plotpoint}}
\put(471.59,1255.32){\usebox{\plotpoint}}
\put(489.95,1264.98){\usebox{\plotpoint}}
\put(508.35,1274.58){\usebox{\plotpoint}}
\multiput(511,1276)(18.564,9.282){0}{\usebox{\plotpoint}}
\put(526.85,1283.99){\usebox{\plotpoint}}
\put(545.23,1293.62){\usebox{\plotpoint}}
\put(563.61,1303.25){\usebox{\plotpoint}}
\multiput(565,1304)(19.077,8.176){0}{\usebox{\plotpoint}}
\put(582.48,1311.87){\usebox{\plotpoint}}
\put(600.89,1321.45){\usebox{\plotpoint}}
\multiput(606,1324)(18.275,9.840){0}{\usebox{\plotpoint}}
\put(619.25,1331.12){\usebox{\plotpoint}}
\put(637.74,1340.55){\usebox{\plotpoint}}
\put(656.17,1350.09){\usebox{\plotpoint}}
\multiput(660,1352)(18.275,9.840){0}{\usebox{\plotpoint}}
\put(674.57,1359.67){\usebox{\plotpoint}}
\put(693.37,1368.43){\usebox{\plotpoint}}
\put(711.83,1377.91){\usebox{\plotpoint}}
\multiput(714,1379)(18.275,9.840){0}{\usebox{\plotpoint}}
\put(730.19,1387.59){\usebox{\plotpoint}}
\put(748.63,1397.11){\usebox{\plotpoint}}
\put(767.11,1406.56){\usebox{\plotpoint}}
\multiput(768,1407)(18.845,8.698){0}{\usebox{\plotpoint}}
\put(785.87,1415.43){\usebox{\plotpoint}}
\put(804.29,1425.00){\usebox{\plotpoint}}
\multiput(808,1427)(18.564,9.282){0}{\usebox{\plotpoint}}
\put(822.78,1434.42){\usebox{\plotpoint}}
\put(841.15,1444.07){\usebox{\plotpoint}}
\put(859.55,1453.68){\usebox{\plotpoint}}
\multiput(862,1455)(18.564,9.282){0}{\usebox{\plotpoint}}
\put(878.10,1462.97){\usebox{\plotpoint}}
\put(896.83,1471.91){\usebox{\plotpoint}}
\put(915.20,1481.57){\usebox{\plotpoint}}
\multiput(916,1482)(18.564,9.282){0}{\usebox{\plotpoint}}
\put(933.69,1490.99){\usebox{\plotpoint}}
\put(952.11,1500.56){\usebox{\plotpoint}}
\multiput(957,1503)(18.275,9.840){0}{\usebox{\plotpoint}}
\put(970.48,1510.21){\usebox{\plotpoint}}
\put(989.32,1518.87){\usebox{\plotpoint}}
\put(1007.77,1528.38){\usebox{\plotpoint}}
\multiput(1011,1530)(18.275,9.840){0}{\usebox{\plotpoint}}
\put(1026.13,1538.06){\usebox{\plotpoint}}
\put(1044.59,1547.55){\usebox{\plotpoint}}
\put(1063.05,1557.02){\usebox{\plotpoint}}
\multiput(1065,1558)(18.275,9.840){0}{\usebox{\plotpoint}}
\put(1081.50,1566.50){\usebox{\plotpoint}}
\put(1100.22,1575.42){\usebox{\plotpoint}}
\put(1118.71,1584.85){\usebox{\plotpoint}}
\multiput(1119,1585)(18.275,9.840){0}{\usebox{\plotpoint}}
\put(1137.06,1594.53){\usebox{\plotpoint}}
\put(1155.48,1604.10){\usebox{\plotpoint}}
\multiput(1159,1606)(18.564,9.282){0}{\usebox{\plotpoint}}
\put(1174.00,1613.46){\usebox{\plotpoint}}
\put(1192.74,1622.37){\usebox{\plotpoint}}
\put(1211.13,1631.99){\usebox{\plotpoint}}
\multiput(1213,1633)(18.564,9.282){0}{\usebox{\plotpoint}}
\put(1229.63,1641.41){\usebox{\plotpoint}}
\put(1248.02,1651.01){\usebox{\plotpoint}}
\put(1266.39,1660.67){\usebox{\plotpoint}}
\multiput(1267,1661)(18.564,9.282){0}{\usebox{\plotpoint}}
\put(1285.01,1669.85){\usebox{\plotpoint}}
\put(1303.71,1678.85){\usebox{\plotpoint}}
\multiput(1308,1681)(18.275,9.840){0}{\usebox{\plotpoint}}
\put(1322.06,1688.53){\usebox{\plotpoint}}
\put(1340.54,1697.98){\usebox{\plotpoint}}
\put(1358.99,1707.49){\usebox{\plotpoint}}
\multiput(1362,1709)(18.275,9.840){0}{\usebox{\plotpoint}}
\put(1377.34,1717.17){\usebox{\plotpoint}}
\put(1396.01,1726.24){\usebox{\plotpoint}}
\put(1414.67,1735.33){\usebox{\plotpoint}}
\multiput(1416,1736)(18.275,9.840){0}{\usebox{\plotpoint}}
\put(1433.03,1745.01){\usebox{\plotpoint}}
\put(1451.46,1754.55){\usebox{\plotpoint}}
\put(1456,1757){\usebox{\plotpoint}}
\end{picture}
\end	{center}
\vskip 0.15in
\caption{The pseudoscalar glueball mass.
The best linear extrapolation to the continuum limit
is shown.}
\label	{fig-glue-pseudo}
\end 	{figure}

In 
Fig~\ref{fig-glue-tensor}
and 
Fig~\ref{fig-glue-pseudo}
I show corresponding plots for the
$2^{++}$ and $0^{-+}$ glueballs. The former is
well determined, and we obtain the continuum ratio
\begin{equation}
{{m_{2^{++}}}\over \surd\sigma} = 5.15 \pm 0.21 .
\label{A4}
\end{equation}      
It is however clear that our
control over the $0^{-+}$ is marginal. This translates
into a very large error when we perform the continuum
extrapolation: 
\begin{equation}
{{m_{0^{-+}}}\over \surd\sigma} = 4.97 \pm 0.58.
\label{A5}
\end{equation}      

We now wish to transform the above continuum glueball
masses to physical $MeV$ units. There are several
reasonable ways to do this and how they differ
will give us an estimate of the systematic error
intrinsic to introducing physical units into a theory
that is not quite physical. 

\begin	{figure}[p]
\begin	{center}
\leavevmode
\input	{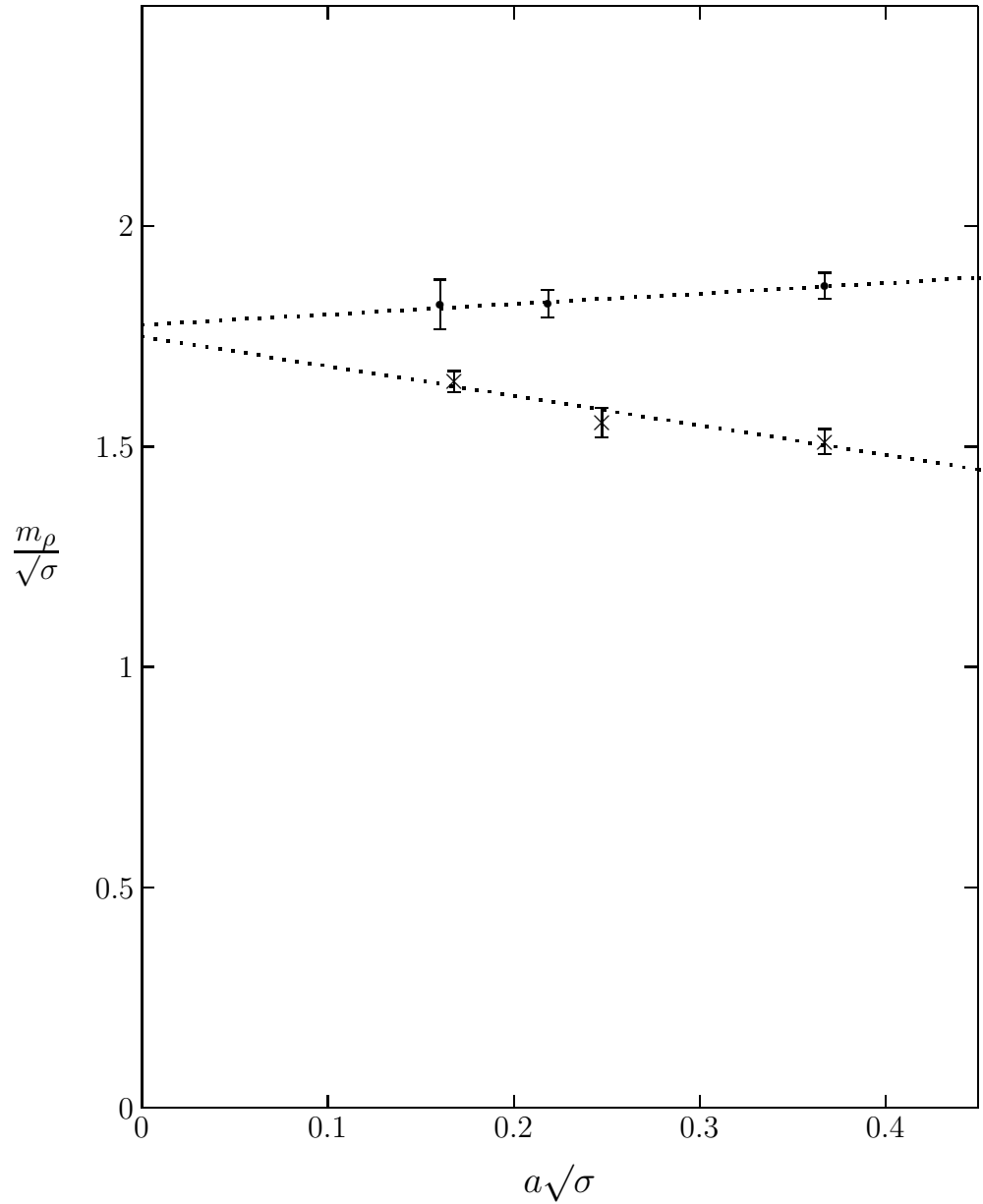}
\end	{center}
\vskip 0.15in
\caption{The $\rho$ mass: GF11 ($\times$)
and UKQCD ($\bullet$) values.  
The best linear extrapolations to the respective continuum limits
are shown.}
\label	{fig-rho}
\end 	{figure}

As we remarked earlier, one way is to infer 
from the observed Regge slopes that
$\surd\sigma \simeq 400-450MeV$. This estimate does however 
suffer from being somewhat model dependent. 
An alternative is to take lattice calculations of
the mass of the $\rho$ and extrapolate the ratio
$m_{\rho}/\surd\sigma$ to the continuum limit, just
as we did for the glueball. The only difference with 
eqn~\ref{A2}
is that the leading correction will be $O(a)$ rather
than $O(a^2)$. We do this for two recent `state-of-the-art' 
calculations: the GF11 collaboration
\cite{GF11-H} 
and the UKQCD collaboration 
\cite{UKQCD-H}
who use quite different discretisations for the quark action. 
UKQCD uses an improved action which should have smaller
discretisation errors. We plot the two sets of ratios
in 
Fig~\ref{fig-rho}
with their corresponding continuum extrapolations.
These give us 
\begin{equation}
 {{m_{\rho}}\over \surd\sigma} = 1.72 \pm 0.08 
\  \  \  \  \  GF11       \label{A6}
\end{equation}
and
\begin{equation}
 {{m_{\rho}}\over \surd\sigma} = 1.78 \pm 0.08 
\  \  \  \  \  UKQCD      \label{A7}
\end{equation}
respectively. It is reassuring that these two calculations are
entirely consistent in the continuum limit,
despite the fact that they have very different lattice 
discretisation corrections. (Indeed one can argue that the
dominant correction in the UKQCD calculation will be
$0(a^2)$ rather than $0(a)$.) To extract a value for
$\sigma$ in $MeV$ units we average the above results, 
set $m_{\rho}=770 MeV$ and so obtain
$\surd\sigma = 440 \pm 15 MeV$.
We observe that this is entirely consistent with
the scale we inferred from Regge slopes; but the argument
is much cleaner here.

The error on $\sigma$ is largely statistical. We now need
to estimate the systematic errors as well. These are
discussed in detail in 
\cite{fec-mt}. 
There we consider errors 
due to finite volume corrections, to extrapolations
in the quark mass, to uncertainties in our estimates of
the lattice values of $\sigma$ and to using the $K^{*}$
or nucleon to set the scale rather than the $\rho$.
We estimate a $\pm 8 \%$ systematic error in total.
This leads to our final estimate for the value of the
string tension as being: 
\begin{equation}
 \surd\sigma = 440 \pm 15 \pm 35 MeV 
\label{A8}
\end{equation}       
where the first error is statistical and the second
is systematic.

We can now use this value in eqns~\ref{A3}-~\ref{A5} to express our
glueball masses in $MeV$ units. We obtain
\begin{equation}
m_{0^{++}} = 1.61 \pm 0.07 \pm 0.13 GeV 
\label{A9}
\end{equation}         
\begin{equation}
m_{2^{++}} = 2.26 \pm 0.12 \pm 0.18 GeV 
\label{A10}
\end{equation}        
and
\begin{equation}
m_{0^{-+}} = 2.19 \pm 0.26 \pm 0.18 GeV. 
\label{A11}
\end{equation}        
The first error combines the statistical errors
on the glueball and $\rho$ continuum extrapolations, and 
the second is our estimate of the systematic error. 
This, then, is our best lattice prediction
for the lightest glueballs prior to any mixing with
nearby quarkonium states.

Where are the quarkonia? We are interested in flavour-singlet
states because those are the ones that can mix with glueballs.
It would clearly be very useful if lattice calculations
were to provide estimates for the masses of the 
$u\bar{u}+d\bar{d}$ and $s\bar{s}$ mesons prior to
their mixing with glue. One could then 
introduce the mixing  with the scalar glueball
using the (standard) formalism described below and compare
the resulting states with the experimental spectrum.  
There are some indications from recent lattice calculations
\cite{GF11-ss,CM-ss}
that there is a scalar $s\bar{s}$ state close to the
scalar glueball mass. However these quenched calculations 
do not try to incorporate the essential quark annihilation
contributions and so must be regarded as indicative at best.
We will therefore turn now to experiment and phenomenology.

\subsection {Experiment and phenomenology}

An illuminating (if arbitrary) starting point is provided
by first considering the lightest $2^{++}$ mesons.
In the quark model 
\cite{book-Close}
these are very similar to
the scalars: the quark spins are aligned in both cases 
and the only difference is that the net spin is
parallel to the unit orbital angular momentum
for the tensors and antiparallel for the
scalars - which, in quark potential models, leads to minor differences
in the masses
\cite{Isgur}.
However in the real world we expect 
the tensors to have narrower decay widths than
the scalars because their decays into light 
pseudoscalars require non-zero angular momentum
and corresponding near-threshold suppression factors.
Thus they should be easier to identify experimentally
- and that is their interest for us here.

Experimentally we find the following lightest tensor states
\cite{PDT}.
There is an isoscalar $f_2(1270)$ with
width, $\Gamma \sim 185 MeV$, a second isoscalar 
$f^{\prime}_2(1525)$ ($\Gamma \sim 76 MeV$),
an isovector $a_2(1320)$ ($\Gamma \sim 107 MeV$) 
and a strange isodoublet
$K^{\star}_2(1430)$ ($\Gamma \sim 100 MeV$). The
first isoscalar decays mainly into pions while 
the second decays mainly into strange mesons. Thus
it is natural to infer that the $f_2(1270)$
is mainly $u\bar{u} + d\bar{d}$ while the  
$f^{\prime}_2(1525)$ is mainly $s\bar{s}$. We have
a clear nonet of tensor mesons and we note that
the splittings are exactly what one would expect
from a mass-difference between strange and
non-strange (constituent) quarks of $\sim 100 MeV$.
Thus the lightest tensors provide a nice illustration 
of the quark model at its most successful: 
the mesons fall into $SU(3)$ multiplets with a modest
symmetry breaking driven by the $m_s - m_n$
mass difference. (For convenience we shall adopt the
shorthand notation $n\bar{n}$ for  
${1\over{\surd 2}}(u\bar{u}+d\bar{d})$ in the
following.)

We now turn to the $0^{++}$ mesons. The easiest 
such meson to see 
experimentally should be the strange isodoublet and 
the lightest such state turns out to be the 
$K^{\star}_0(1430)$ with $\Gamma \sim 290 MeV$.
Note that this is close in mass to the corresponding
tensor although, as expected, its decay width is
much larger. (This is what makes the scalars very 
much harder to identify experimentally.) There is
also a candidate isovector 
$a_0(1450)$ ($\Gamma \sim 270 MeV$). In addition
there are two isoscalars, the $f_0(1370)$ 
($\Gamma \sim 300-500 MeV$) and the $f_0(1500)$ 
($\Gamma \sim 120 MeV$). The former decays mainly
into pions and so one would suppose it to be composed
mainly of non-strange quarks. 

So far all this looks much like the tensor nonet. 
However there is a puzzle. The $f_0(1500)$
does $not$ have the obvious decays for a predominantly
$s\bar{s}$ state. Moreover
if we compare its width to the other scalars then
we see that it is remarkably narrow -- particularly for 
a state with, apparently, a large non-strange 
component. This motivates us to 
look for other nearby scalar states. There is
evidence that the $f_J(1710)$ is, or contains,
a $0^{++}$ state. Moreover the predominant 2-body
decays of this state involve strange quarks
($K\bar{K}$ or $\eta\eta$). The state is
relatively narrow $\Gamma \sim 175 MeV$. So if we
include this state into our discussion, we
no longer have an obvious $s\bar{s}$ problem.
However now we have three isoscalar states in the
$1370 - 1710 MeV$ mass region - too many for a
quark model nonet! Since, as we have seen, lattice 
calculations predict a scalar glueball in
the 1600 MeV mass region, with a width comparable
to that which we observe for the $f_0(1500)$ and $f_J(1710)$, 
it is natural to conjecture that the three
observed isoscalars are in fact the results of mixing between
two quarkonium isoscalars and the lightest scalar glueball.
This is the scenario explored in 
\cite{Genov,Amsler-Close,GF11-G3}
with some differing assumptions.

All this might be a convincing picture if it were
not for the existence of some 
lighter scalar states that we have so
far failed to mention. These are the isoscalar 
$f_0(980)$ and the isovector $a_0(980)$. Both are
narrow ($\Gamma \sim 40-100 MeV$). (There may also be
an extremely  wide iso-iscalar in the $400-1200 MeV$ mass
range.) Since there is no nearby strange isodoublet, 
these states certainly do not fit into the usual quark
picture where mesons should 
fall into approximate $SU(3)$ multiplets with modest
symmetry breakings driven by the $m_s - m_n$
mass difference. In fact these mysterious states were
interpreted some time ago
\cite{Jaffe} 
as being loosely-bound
$K\bar{K}$ molecules - recall that they are narrow
and occur very close to the $K\bar{K}$ threshold.
If we accept this interpretation - and it has
been widely accepted as being plausible - then
we can ignore these states for our purposes
and the interpretation of the previous paragraph
remains. I should stress that alternative
interpretations of this spectrum do exist, 
but I will ignore these here and instead
refer you to
\cite{fec-mt} 
for a detailed discussion.

The picture we are thus led to is one where the
$f_0(1370)$, the $f_0(1500)$ and the $f_0(1710)$
are the result of mixing between the glueball
and the would-be $s\bar{s}$ and $n\bar{n}$
scalar quarkonia. We argued above that
these quarkonia belong to the same nonet so
the $s\bar{s}$ will be heavier than the 
$n\bar{n}$. Moreover the relative strengths
of the glueball mixings will be determined on symmetry
grounds. 

What are the constraints on the mixing? 

First the output masses of the
mixing should correspond to the $f_0(1370)$, $f_0(1500)$ and 
$f_0(1710)$. The first state is very broad and so we
shall allow its mass to lie in the region
$M_1 \in [1.31,1.40] GeV$ with a preference for $1.37 GeV$.
The second state is both quite narrow and well-defined
and so we fix its mass to $M_2=1.5 GeV$. The third state
appears to be again quite narrow but the precise location of
the $0^{++}$ component is still quite uncertain. We shall
consider the range $M_3 \in [1.64,1.80] GeV$ with a preference for 
$1.71 GeV$.

There are also some constraints on the input parameters. Given
the lattice predictions, a generous range for the glueball
mass would be $m_G \in [1.40,1.80] GeV$. We also have a qualitative
constraint on the glueball-quarkonium matrix element of the
Hamiltonian: it should be small, $\sim O(100)MeV$ or less. 
Similarly we expect a glueball to be narrow, and so any
state with a large glueball component should be
narrower than one would otherwise expect.
As for the quarkonia, we expect $m_{s\bar{s}} > m_{n\bar{n}}$,
with a mass difference in the $\sim 150-200MeV$ ball-park. 
And the mixing matrix elements should be close
to their $SU(3)_{flavour}$ values. There are further, and important, 
constraints
that arise from the observed decays of the three output
states and again I refer to
\cite{fec-mt} 
for a more detailed discussion.

Here I shall give one example of what appears to be
an acceptable mixing scheme. We assume that prior to
mixing what we have are the `bare'
quarkonia with masses $m_{s\bar{s}} = 1.61 GeV$
and $m_{n\bar{n}} = 1.36 GeV$ and a bare glueball
with mass $m_G = 1.48 GeV$. These are not eigenstates
of $H$ and so the mass matrix will not be diagonal.
We can write the latter as 
$$
\left({\begin{array}{c c c c}
m_G & z & \sqrt{2} z \\
z & m_{s\bar{s}} & 0  \\
\sqrt{2} z & 0 & m_{n\bar{n}}\\
\end{array}
}\right)
$$
(The factor of $\surd2$ follows from the fact that
${1\over{\surd 2}}\{<u\bar{u}|+<d\bar{d}|\}H | G>
= \surd 2 <u\bar{u}| H | G>$.) The physical 
masses, after mixing, are the eigenvalues of this mass matrix.
With an acceptably weak mixing of $z=62 MeV$ we obtain 
as output masses $1.31, 1.50, 1.64GeV$ for the
$f_0(1370)$, $f_0(1500)$ and $f_0(1710)$, which is
perfectly acceptable. The overlaps of the output
states can be written in terms if the bare input
states as follows: 
$$
\begin{array}{c r r r}
i &f_{in} & f_{is} & f_{iG} \\
f_0(1370) & 0.87 &  0.10  & -0.49 \\
f_0(1500) & 0.48 & -0.43  & 0.77 \\
f_0(1710) & 0.13 &  0.90  & 0.42 \\
\end{array}
$$
We see that most of the glueball resides in the $f_0(1500)$
and the $f_0(1710)$ is mainly an $s\bar{s}$ quarkonium.
Such a mixing scheme is qualitatively compatible with
the observed decays of the output states. This and
other possible mixing scenarios are discussed in
\cite{fec-mt} 
For some other recent discussions of mixing in this
context see
\cite{Amsler-Close,GF11-G3}
\subsection {Conclusion}

We now know quite accurately the lightest scalar glueball
mass in the pure gauge theory in units of the string tension.
If we translate this into physical units, as described in this 
lecture, we find that the glueball should appear around
$\sim 1.60 \pm 0.15 GeV$. This is just the mass range where
one naively expects the two scalar flavour singlet quarkonia
to lie. If they do then all these states will inevitably
mix to some extent, even though we expect the dynamical
mixing parameter to be weak. It is therefore intriguing that
experimentally not only are there definitely states in this mass range, 
the $f_0(1370)$ and the $f_0(1500)$, but there is evidence for
a third, the  $f_0(1710)$. We presented a sample mixing scheme in
which the glueball mainly resides in the $f_0(1500)$. This
is not a new suggestion
\cite{Amsler-Close}.
While it is still a little too early to come to a convincing conclusion,
it certainly seems that it will not be very long before we are 
able to do so.

I have emphasised the scalar glueball because that is where most of
the recent interest has been. What about the tensor and pseudoscalar?
In looking for glueball candidates experimentalists 
naturally look for states
that appear in `gluon-rich' processes. For example in $J/\psi$
decays, or in Pomeron-Pomeron collisions, or states that have large
$\eta\eta$ decays. This picks out the $f_0(1500)$ and $f_0(1710)$
amongst the scalars. Amongst the tensors this picks out the
$f_2(1900)$ and the $G(2150)$. These are, of course, in the right
ball-park for the $2^{++}$ glueball as predicted by lattice 
calculations: $\sim 2.26 \pm 0.22 GeV$. So here too, things
look interesting. 

With the pseudoscalar, on the other hand, we have little reason
to feel smug. First, the lattice calculations are poor: a mass
estimate of $\sim 2.16 \pm 0.32$ barely qualifies as an
estimate at all. Moreover the obvious experimental candidate
is the $\iota(1490)$ seen in $J/\psi$ decays. We should however
remember that this state is special: it has the quantum numbers
of the vacuum topological charge. Which brings me smoothly to my 
next topic.

\section {Topological fluctuations}

As you know, SU(N) gauge fields in 3+1 dimensions possess 
a topological charge. 
\cite{U1-Cole}.
And it is the topological fluctuations 
of the gauge fields that are the reason why the $\eta^\prime$ 
has a mass $\sim 1 GeV$  rather than being almost a Goldstone boson
\cite{U1-H}.
Moreover there is good reason to think that these fluctuations
lead to the spontaneous breaking of chiral symmetry. 
The reason is that isolated instantons produce zero modes
in the Dirac operator;
these mix with each other, and shift away from zero, when 
the instantons are not isolated (as in the real vacuum). 
It is not hard to imagine that this might leave
a non-zero density of modes close to zero, and this would
suffice to break chiral symmetry (via the Banks-Casher formula
\cite{Banks-Casher}).
This occurs
explicitly in some instanton models (see Shuryak's lectures at
this School and 
\cite{Shuryak,dowrick-teper})
and has been observed in some lattice calculations 
\cite{SH-MT}.
It is also possible that instantons may affect the properties
of some hadrons, as the near-zero modes may lead to
large contributions to the valence-quark propagators.
This is more speculative. All this to say that
topology is interesting. In addition it is intrinsically
non-perturbative. Here I will tell you some things that we
have learned by studying topology in lattice gauge theory.

The first thing I need to address is the fact that when we
discretise space-time we lose topology in a formal sense,
since any field on a discrete set of points can be
smoothly deformed to a trivial field. One should not get 
too excited about this; the same happens with dimensional
regularisation. If the space-time dimension is not exactly
4 we have no topological winding. Since the theory is
renormalisable we expect that, as the cut-off is removed, 
$a\to 0$, we recover all the properties of the continuum theory.
We shall see how this occurs for topology in the lattice 
gauge theory.

So in this lecture I will address the following topics.
First I discuss the basic ambiguity with defining 
topology on a lattice; and show why it does not really
matter. Then I discuss the practical problem of calculating
the topological charge of fields that have fluctuations on all
length scales. This is a large subject and I will focus
simply on the one technique that goes by the name of
`cooling'. I will then move onto the first bit of
physics: the Witten-Veneziano fomula that relates
the strength of the topological fluctuations in the
pure gauge theory to the mass of the $\eta^\prime$
in QCD. Finally I will attempt to give some insight
into the structure of the topological fluctuations 
in the vacuum.

\subsection {A basic ambiguity and why it does not matter}

Let me start by giving an explicit example. A lattice gauge
field is defined by a set of SU(2) matrices on the links, $l$,
of the lattice: $\{U_l\}$. (I will stick here to SU(2) for simplicity.)
Consider now the following continuum gauge potential for an
instanton of size $\rho$ centered at $x=0$:
\begin{equation}
A_\mu^I(x) = {{x^2}\over{x^2+\rho^2}}g^{-1}(x)\partial_\mu g(x)
\label{B1}
\end{equation}       
with
\begin{equation}
g(x) = {{x_0+ix_j\sigma_j}\over{{(x_{\mu}x_{\mu})}^{1/2}}}
\label{B2}
\end{equation}       
Let us now translate this field by $a/2$ in each direction
so that it is centered at $\bar{x}=(a/2,a/2,a/2,a/2)$, i.e
at the centre of a lattice hypercube.
Define a lattice field by:
\begin{equation}
U_\mu^I(x) = {\cal P} \exp \int\limits_x^{x+a{\hat\mu}}
A_\mu^I(x)dx         \label{B3}
\end{equation}
For $\rho \gg a$ the lattice discretisation is very fine compared
to the instanton core in which the instanton action and 
topological charge density, $Q(x)={1\over{32\pi^2}}
\epsilon_{\mu\nu\rho\sigma}
Tr\{F_{\mu\nu}(x)F_{\rho\sigma}(x)\}$, reside. In this case any 
reasonable definition of the topological charge (see below for an
example) will assign a topological charge of $Q = 1$ to this
lattice field. Suppose we now continuously reduce $\rho$.
The lattice field will also vary continuously. Eventually
we will have $\rho \ll a$. At this stage the instanton core,
which is at the centre of a lattice hypercube, will be
very far, in units of its size, from any of the lattice links.
Thus even the nearest link matrices on the lattice will be 
arbitrarily close to pure gauge.
That is to say, the field configuration will be the same as
that due to a gauge singularity located at $x=\bar{x}$.
In this case any reasonable definition of the topological charge 
will assign a topological charge of $Q = 0$ to this
lattice field. Thus we have passed continuously from a field
with $Q=1$ to one with $Q=0$. 

Does this raise a fundamental problem with topology on the lattice?
The answer is: not really. At least if we are interested, as here,
with the limit $a\to 0$. Let me argue why this is so.

Consider the density of topological fluctuations as a function of
their size $\rho$. This is only an unambiguous notion if
$\rho \ll \xi_d$ where $\xi_d$ is the typical dynamical length
scale of the theory (e.g. 1 fermi in QCD or the inverse
mass gap in the pure gauge theory). In that case
we know the density of these `instantons':
\begin{equation}
D(\rho) d\rho = {{d\rho}\over{\rho}}{1\over{\rho^4}}
e^{-{{8\pi^2}\over{g^2(\rho)}}} ....
\label{B4}
\end{equation}          
where the `...' represent factors varying weakly with $\rho$. You 
recognise in this equation the scale-invariant integration measure; 
also a factor to account for the fact that a ball of volume $\rho^4$
can be placed in $1/\rho^4$ different ways in a unit volume;
and finally a factor arising from the classical instanton
action, $S_I=8\pi^2/g^2$, with perturbative fluctuations
promoting the bare $g^2$ to a running $g^2(\rho)$ in the
usual way. This last sentence is the most important one for us 
here: if we substitute for $g^2(\rho)$ in 
eqn~\ref{B4}
we find that
\begin{equation}
D(\rho) d\rho \propto \rho^6 d\rho \ \ \ \ \ \ \ \  : SU(3)
\label{B5}
\end{equation}
with a power $\rho^{7/3}$ in the case of SU(2). 
So , because of the scale anomaly, the number
of instantons rapidly vanishes as $\rho\to 0$ rather
than diverging as $1/\rho^5$.

On the lattice this density will change as follows
if $a \ll \rho$ and $a \ll \xi_d$:
\begin{equation}
D(\rho) \to D(\rho) \times \{1 + 0({{a^2}\over{\rho^2}})\}
\label{B6}
\end{equation}
Now, suppose a lattice field configuration is to be smoothly
deformed from $Q=1$ to $Q=0$. This requires a topological fluctuation
to be squuezed out of the lattice, as described above.  
While we do not know much about the structure of the original 
fluctuation (it will typically be on a size scale $\sim \xi_d$
which is beyond the reach of our analytic techniques) we
do know that if the lattice spacing is sufficiently small
then to reach $\rho \sim a$ the `instanton' will have to pass
through sizes $\xi_d \gg \rho \gg a$. In this region the density
is calculable as we saw above, with a probability that is very strongly
suppressed; at least as $\sim (\rho/\xi_d)^6$ for SU(3). So
the changing of $Q$ is conditional upon 
the involvement of field configurations 
whose probability$\to 0$ as $a \to 0$. Thus, as we approach the
continuum limit this lattice ambiguity vanishes very rapidly.
We knew that this had to happen because the theory is 
renormalisable and the lattice is surely a good regulator.
It is, however, nice to see it happen explicitly.

\subsection {Cooling}

Having convinced you that it makes sense to discuss topology on
the lattice, I now want to discuss how you can calculate it.
The obvious ways are three. 

$\bullet$ Calculate the zero-modes of the Dirac operator
${\not{\rm D}}[A]$. $Q$ will equal the difference between the
number of left and right handed zero-modes.

$\bullet$ Interpolate a smooth gauge potential. If $Q \not= 0$
and if space-time is compact (as it is here: a hypertorus)
then we will need more than one patch. The net winding
of the transition functions between the patches equals $Q$. 

$\bullet$ We calculate the topological charge density
$Q(x)= {1\over{32\pi^2}} \epsilon_{\mu\nu\rho\sigma}
Tr\{F_{\mu\nu}(x)F_{\rho\sigma}(x)\}$.
Then $Q=\int Q(x) d^4 x$.

All these methods have been explored in lattice calculations.
I will focus on the last because it is the simplest and
because it immediately tells us something about the
size and location of the core of the `instanton' (since
that is where $Q(x)$ is localised).

We need a lattice operator that becomes $Q(x)$ in the
continuum limit. This is easy. Recall (see Chris
Michael's lectures) that if we define the plaquette
matrix, $U_{\mu\nu}(x)$, as the ordered product of
link matrices around the corresponding plaquette of
the lattice, then it is easy to see 
that $U_{\mu\nu}(x) = 1 + a^2 F_{\mu\nu}(x) + ....$
and hence that
\cite{diV}
\begin{equation}
Q_L(x) \equiv {1\over{32\pi^2}} \epsilon_{\mu\nu\rho\sigma}
Tr\{U_{\mu\nu}(x)U_{\rho\sigma}(x)\}
= a^4 Q(x) +0(a^6).        \label{B7}
\end{equation}
If we apply this formula to an instanton of size $\rho$
(discretised as described above) then we find, as expected, that 
$Q_L = \int Q_L(x)dx = 1 + 0(a^2/\rho^2)$.

Applied to the real vacuum however $Q_L(x)$ has problems.
The operator is dimensionless and so $0(a^6)$ actually
means terms like $\sim a^6 F^3$, $\sim a^6 FD^2F$ etc.
For smooth fields these are indeed  $0(a^6)$. However
realistic fields (those that contribute to the path integral)
have fluctuations all the way up to frequencies of $0(1/a)$.
So if we take matrix elements of $Q_L(x)$ in the vacuum
the contribution of the high frequency modes to the 
$0(a^6)$ terms will be
$\delta Q_L(x) \sim a^6 \times 1/a^6 \sim 0(a^0)$. 
In practice this contribution is suppressed by some
powers of $\beta$ that can be calculated in perturbation 
theory (since these contributions are short-distance).
Thus in the real world $Q_L(x)$ possesses interesting topological
contributions that are of order $a^4 \propto e^{-c\beta}$
(since $g^2(a) \propto 1/\log(a\Lambda)$)
and uninteresting ultraviolet contributions that are 
$\propto 1/\beta^n$. So as we approach the continuum limit,
$\beta \to \infty$, the latter dominate and we are in trouble.

Actually things are a little worse than this. Like other
composite lattice operators, $Q_L(x)$ possesses a 
multiplicative lattice renormalisation factor:
$Z_Q \simeq 1+c^{\prime}/\beta + ...$
\cite{PISA}.
This looks innocuous, and indeed in the continuum limit
it obviously is. However it turns out that the value of
$c^{\prime}$ is such that $Z_Q \ll 1$ precisely in
the range of values of $\beta$ where current lattice
calculations are performed.

There are different ways to deal with these problems.
I will describe a particularly simple technique
\cite{cool-MT}.
The idea rests on the observation that the
problems are all caused by the ultraviolet fluctuations 
on wavelengths $\sim a$. By contrast, 
if we are close to the continuum limit, the topology 
is on wavelengths $\rho \gg a$. One can therefore imagine
taking the lattice fields and locally smoothing them
over distances $\gg a$ but $\ll \rho$. Such a
smoothing would  erase the unwanted ultraviolet
fluctuations while not significantly disturbing the
physical topological charge fluctuations. One could then
apply the operator $Q_L(x)$ to these `cooled' fields
to reveal the topological charge distribution of the
vacuum.

How do we cool a lattice gauge field? The simplest
procedure is to take the field and generate from it
a new field by the standard Monte Carlo heat bath
algorithm subject to one crucial modification: we always
choose the new link matrix to locally minimise the plaquette
action. Since $Tr U_{\mu\nu}(x)$ measures the variations of
the link matrices over a distance $a$, minimising the plaquette 
action is a very efficient way to erase the ultraviolet 
fluctuations. (Obviously there are many possible variations 
on this theme.) 

Thus the notion is that we take our ensemble of $N$ gauge 
fields, $\{U^{I=1,..,N}\}$, perform a suitable number of cooling 
sweeps on each one of these, so obtaining a corresponding
ensemble $\{U_c^{I=1,..,N}\}$ of cooled fields, and then
extract the desired topological properties from these
cooled fields. As one might expect there are 
ambiguities for realistic values of $a$. As we
cool, topological charges of opposite sign will gradually
annihilate. This changes the topological charge density
but not the total value of $Q$. Eventually this leads to a  
very dilute gas of instantons.
As we cool even further these isolated
instantons will gradually shrink and will eventually
shrink within a hypercube and at this point even $Q$ will change.
(This is for a plaquette action on a large enough volume:
other actions may have other effects.) Of course when
an instanton becomes narrow it has a very peaked charge
density and is impossible to miss. So we certainly know 
when it disappears out of the lattice
and can, if we think it appropriate,
correct for that. All this to say that
cooling is a good way to calculate the total topological charge, 
but is not necessarily a reliable way to learn about the
topological charge density.

\begin	{figure}[p]
\begin	{center}
\leavevmode
\setlength{\unitlength}{0.240900pt}
\ifx\plotpoint\undefined\newsavebox{\plotpoint}\fi
\sbox{\plotpoint}{\rule[-0.200pt]{0.400pt}{0.400pt}}%
\begin{picture}(1500,900)(0,0)
\font\gnuplot=cmr10 at 12pt
\gnuplot
\sbox{\plotpoint}{\rule[-0.200pt]{0.400pt}{0.400pt}}%
\put(120.0,203.0){\rule[-0.200pt]{4.818pt}{0.400pt}}
\put(108,203){\makebox(0,0)[r]{{$-2$}}}
\put(1436.0,203.0){\rule[-0.200pt]{4.818pt}{0.400pt}}
\put(120.0,376.0){\rule[-0.200pt]{4.818pt}{0.400pt}}
\put(108,376){\makebox(0,0)[r]{{$-1$}}}
\put(1436.0,376.0){\rule[-0.200pt]{4.818pt}{0.400pt}}
\put(120.0,548.0){\rule[-0.200pt]{4.818pt}{0.400pt}}
\put(108,548){\makebox(0,0)[r]{{$0$}}}
\put(1436.0,548.0){\rule[-0.200pt]{4.818pt}{0.400pt}}
\put(120.0,721.0){\rule[-0.200pt]{4.818pt}{0.400pt}}
\put(108,721){\makebox(0,0)[r]{{$1$}}}
\put(1436.0,721.0){\rule[-0.200pt]{4.818pt}{0.400pt}}
\put(120.0,893.0){\rule[-0.200pt]{4.818pt}{0.400pt}}
\put(108,893){\makebox(0,0)[r]{{$2$}}}
\put(1436.0,893.0){\rule[-0.200pt]{4.818pt}{0.400pt}}
\put(417.0,31.0){\rule[-0.200pt]{0.400pt}{4.818pt}}
\put(417,19){\makebox(0,0){\shortstack{\\ \\ \\ {$20$}}}}
\put(417.0,873.0){\rule[-0.200pt]{0.400pt}{4.818pt}}
\put(714.0,31.0){\rule[-0.200pt]{0.400pt}{4.818pt}}
\put(714,19){\makebox(0,0){\shortstack{\\ \\ \\ {$40$}}}}
\put(714.0,873.0){\rule[-0.200pt]{0.400pt}{4.818pt}}
\put(1011.0,31.0){\rule[-0.200pt]{0.400pt}{4.818pt}}
\put(1011,19){\makebox(0,0){\shortstack{\\ \\ \\ {$60$}}}}
\put(1011.0,873.0){\rule[-0.200pt]{0.400pt}{4.818pt}}
\put(1308.0,31.0){\rule[-0.200pt]{0.400pt}{4.818pt}}
\put(1308,19){\makebox(0,0){\shortstack{\\ \\ \\ {$80$}}}}
\put(1308.0,873.0){\rule[-0.200pt]{0.400pt}{4.818pt}}
\put(120.0,31.0){\rule[-0.200pt]{321.842pt}{0.400pt}}
\put(1456.0,31.0){\rule[-0.200pt]{0.400pt}{207.656pt}}
\put(120.0,893.0){\rule[-0.200pt]{321.842pt}{0.400pt}}
\put(-48,462){\makebox(0,0){{\Large{$Q_L$}}}}
\put(788,-89){\makebox(0,0){{\large{sweep}}}}
\put(120.0,31.0){\rule[-0.200pt]{0.400pt}{207.656pt}}
\put(135,543){\circle*{12}}
\put(150,540){\circle*{12}}
\put(165,541){\circle*{12}}
\put(179,545){\circle*{12}}
\put(194,390){\circle*{12}}
\put(209,390){\circle*{12}}
\put(224,391){\circle*{12}}
\put(239,388){\circle*{12}}
\put(254,390){\circle*{12}}
\put(268,390){\circle*{12}}
\put(283,388){\circle*{12}}
\put(298,386){\circle*{12}}
\put(313,388){\circle*{12}}
\put(328,388){\circle*{12}}
\put(343,390){\circle*{12}}
\put(358,388){\circle*{12}}
\put(372,390){\circle*{12}}
\put(387,388){\circle*{12}}
\put(402,393){\circle*{12}}
\put(417,386){\circle*{12}}
\put(432,388){\circle*{12}}
\put(447,390){\circle*{12}}
\put(461,390){\circle*{12}}
\put(476,388){\circle*{12}}
\put(491,388){\circle*{12}}
\put(506,390){\circle*{12}}
\put(521,388){\circle*{12}}
\put(536,390){\circle*{12}}
\put(550,398){\circle*{12}}
\put(565,252){\circle*{12}}
\put(580,250){\circle*{12}}
\put(595,260){\circle*{12}}
\put(610,115){\circle*{12}}
\put(625,122){\circle*{12}}
\put(640,229){\circle*{12}}
\put(654,238){\circle*{12}}
\put(669,243){\circle*{12}}
\put(684,247){\circle*{12}}
\put(699,267){\circle*{12}}
\put(714,381){\circle*{12}}
\put(729,386){\circle*{12}}
\put(743,391){\circle*{12}}
\put(758,396){\circle*{12}}
\put(773,393){\circle*{12}}
\put(788,390){\circle*{12}}
\put(803,391){\circle*{12}}
\put(818,390){\circle*{12}}
\put(833,391){\circle*{12}}
\put(847,543){\circle*{12}}
\put(862,391){\circle*{12}}
\put(877,388){\circle*{12}}
\put(892,391){\circle*{12}}
\put(907,396){\circle*{12}}
\put(922,393){\circle*{12}}
\put(936,395){\circle*{12}}
\put(951,396){\circle*{12}}
\put(966,393){\circle*{12}}
\put(981,391){\circle*{12}}
\put(996,390){\circle*{12}}
\put(1011,388){\circle*{12}}
\put(1026,390){\circle*{12}}
\put(1040,384){\circle*{12}}
\put(1055,234){\circle*{12}}
\put(1070,233){\circle*{12}}
\put(1085,236){\circle*{12}}
\put(1100,378){\circle*{12}}
\put(1115,378){\circle*{12}}
\put(1129,381){\circle*{12}}
\put(1144,526){\circle*{12}}
\put(1159,381){\circle*{12}}
\put(1174,379){\circle*{12}}
\put(1189,372){\circle*{12}}
\put(1204,388){\circle*{12}}
\put(1218,390){\circle*{12}}
\put(1233,393){\circle*{12}}
\put(1248,540){\circle*{12}}
\put(1263,538){\circle*{12}}
\put(1278,405){\circle*{12}}
\put(1293,545){\circle*{12}}
\put(1308,398){\circle*{12}}
\put(1322,398){\circle*{12}}
\put(1337,395){\circle*{12}}
\put(1352,391){\circle*{12}}
\put(1367,538){\circle*{12}}
\put(1382,546){\circle*{12}}
\put(1397,545){\circle*{12}}
\put(1411,540){\circle*{12}}
\put(1426,545){\circle*{12}}
\put(1441,550){\circle*{12}}
\put(1456,555){\circle*{12}}
\put(135,496){\makebox(0,0){$\times$}}
\put(150,348){\makebox(0,0){$\times$}}
\put(165,436){\makebox(0,0){$\times$}}
\put(179,457){\makebox(0,0){$\times$}}
\put(194,509){\makebox(0,0){$\times$}}
\put(209,395){\makebox(0,0){$\times$}}
\put(224,833){\makebox(0,0){$\times$}}
\put(239,750){\makebox(0,0){$\times$}}
\put(254,495){\makebox(0,0){$\times$}}
\put(268,519){\makebox(0,0){$\times$}}
\put(283,402){\makebox(0,0){$\times$}}
\put(298,569){\makebox(0,0){$\times$}}
\put(313,341){\makebox(0,0){$\times$}}
\put(328,362){\makebox(0,0){$\times$}}
\put(343,462){\makebox(0,0){$\times$}}
\put(358,428){\makebox(0,0){$\times$}}
\put(372,286){\makebox(0,0){$\times$}}
\put(387,665){\makebox(0,0){$\times$}}
\put(402,610){\makebox(0,0){$\times$}}
\put(417,350){\makebox(0,0){$\times$}}
\put(432,495){\makebox(0,0){$\times$}}
\put(447,465){\makebox(0,0){$\times$}}
\put(461,559){\makebox(0,0){$\times$}}
\put(476,579){\makebox(0,0){$\times$}}
\put(491,717){\makebox(0,0){$\times$}}
\put(506,560){\makebox(0,0){$\times$}}
\put(521,529){\makebox(0,0){$\times$}}
\put(536,474){\makebox(0,0){$\times$}}
\put(550,521){\makebox(0,0){$\times$}}
\put(565,365){\makebox(0,0){$\times$}}
\put(580,405){\makebox(0,0){$\times$}}
\put(595,360){\makebox(0,0){$\times$}}
\put(610,512){\makebox(0,0){$\times$}}
\put(625,467){\makebox(0,0){$\times$}}
\put(640,350){\makebox(0,0){$\times$}}
\put(654,655){\makebox(0,0){$\times$}}
\put(669,493){\makebox(0,0){$\times$}}
\put(684,448){\makebox(0,0){$\times$}}
\put(699,541){\makebox(0,0){$\times$}}
\put(714,583){\makebox(0,0){$\times$}}
\put(729,340){\makebox(0,0){$\times$}}
\put(743,443){\makebox(0,0){$\times$}}
\put(758,240){\makebox(0,0){$\times$}}
\put(773,495){\makebox(0,0){$\times$}}
\put(788,621){\makebox(0,0){$\times$}}
\put(803,498){\makebox(0,0){$\times$}}
\put(818,650){\makebox(0,0){$\times$}}
\put(833,496){\makebox(0,0){$\times$}}
\put(847,431){\makebox(0,0){$\times$}}
\put(862,479){\makebox(0,0){$\times$}}
\put(877,362){\makebox(0,0){$\times$}}
\put(892,648){\makebox(0,0){$\times$}}
\put(907,617){\makebox(0,0){$\times$}}
\put(922,628){\makebox(0,0){$\times$}}
\put(936,433){\makebox(0,0){$\times$}}
\put(951,453){\makebox(0,0){$\times$}}
\put(966,284){\makebox(0,0){$\times$}}
\put(981,483){\makebox(0,0){$\times$}}
\put(996,560){\makebox(0,0){$\times$}}
\put(1011,455){\makebox(0,0){$\times$}}
\put(1026,557){\makebox(0,0){$\times$}}
\put(1040,471){\makebox(0,0){$\times$}}
\put(1055,481){\makebox(0,0){$\times$}}
\put(1070,505){\makebox(0,0){$\times$}}
\put(1085,652){\makebox(0,0){$\times$}}
\put(1100,464){\makebox(0,0){$\times$}}
\put(1115,452){\makebox(0,0){$\times$}}
\put(1129,821){\makebox(0,0){$\times$}}
\put(1144,691){\makebox(0,0){$\times$}}
\put(1159,522){\makebox(0,0){$\times$}}
\put(1174,424){\makebox(0,0){$\times$}}
\put(1189,412){\makebox(0,0){$\times$}}
\put(1204,428){\makebox(0,0){$\times$}}
\put(1218,491){\makebox(0,0){$\times$}}
\put(1233,662){\makebox(0,0){$\times$}}
\put(1248,495){\makebox(0,0){$\times$}}
\put(1263,510){\makebox(0,0){$\times$}}
\put(1278,426){\makebox(0,0){$\times$}}
\put(1293,557){\makebox(0,0){$\times$}}
\put(1308,743){\makebox(0,0){$\times$}}
\put(1322,286){\makebox(0,0){$\times$}}
\put(1337,664){\makebox(0,0){$\times$}}
\put(1352,631){\makebox(0,0){$\times$}}
\put(1367,381){\makebox(0,0){$\times$}}
\put(1382,650){\makebox(0,0){$\times$}}
\put(1397,496){\makebox(0,0){$\times$}}
\put(1411,524){\makebox(0,0){$\times$}}
\put(1426,391){\makebox(0,0){$\times$}}
\put(1441,690){\makebox(0,0){$\times$}}
\put(1456,560){\makebox(0,0){$\times$}}
\end{picture}
\end	{center}
\vskip 0.15in
\caption{The lattice topological charge: before cooling ($\times$)
and after cooling ($\bullet$) the fields. Calculated on a sequence
of fields separated by one Monte Carlo sweep.}  
\label	{fig-qseq}
\end 	{figure}
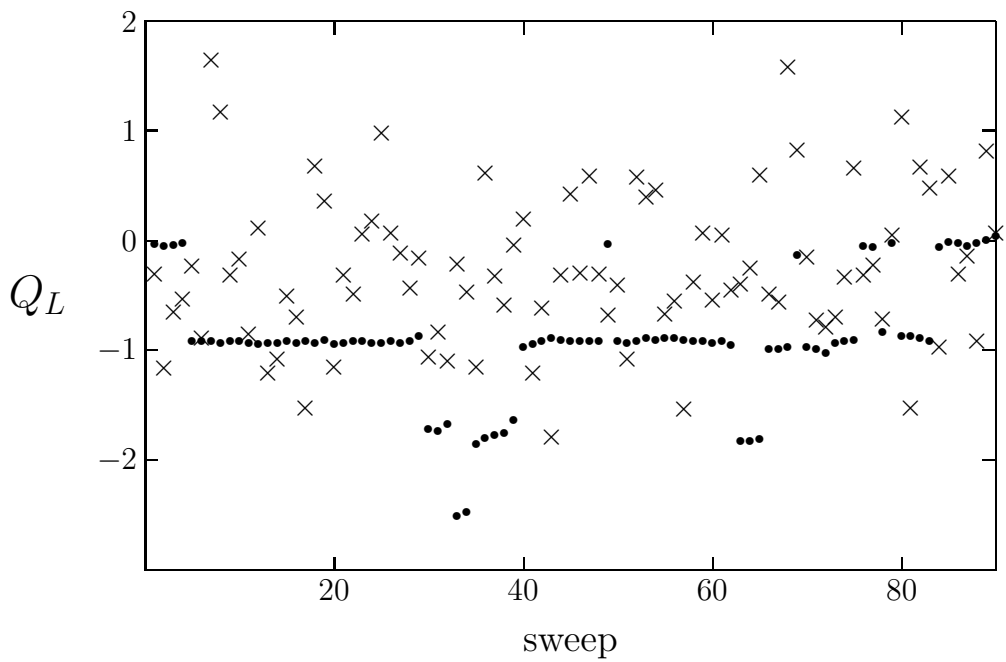

Let me give you an example of how it works. I have produced
a sequence of 90 Monte Carlo sweeps on a $12^4$ lattice
at $\beta=2.5$. These field configurations are separated
by just one heat bath sweep (i.e. each link matrix has been 
changed only once) so one expects the long distance physics on
neighbouring field configurations to be almost identical.
That is, the value of $Q$ should change little. In 
Fig.~\ref{fig-qseq}
I show you the values of $Q_L$ when calculated on these
fields. They jump all over the place and are nowhere near
the expected integer values. Now let us cool each of these
90 configurations with 25 cooling sweeps. Calculating
$Q_L$ on these one finds a dramatic difference - as shown in
Fig.~\ref{fig-qseq}. 
(The cooled charges differ from integers because there is
an $0(a^2/\rho^2)$ error which is substantial for our
not-so-small choice of $a$.)

We now have a technique for calculating the topological charge 
of a lattice field. As we have seen, its application requires
some care. Another reason to take care is the following.
The probability of instantons with $\rho \sim a$ 
depends on the lattice discretisation. If the lattice action
is much less than the continuum action then even when $a\to 0$
there may be a finite density (per unit physical volume) of
these lattice artifacts. They might survive a couple of cooling 
sweeps. For realistic values of $a$ the gap in scales between 
$a$ and $\xi_d$ is not very large and so such a narrow instanton,
if sitting on a background field due to a physical but 
moderately narrow anti-instanton, might expand under cooling 
rather than shrinking. So it might 
occasionally survive our 20, or whatever, 
cooling sweeps and bias our calculations. Obviously this latter
effect disappears as $a\to 0$. Another correction is due to the
fact that with finite $a$ one loses the 
tail of the continuum density that extends to
$\rho \leq a$. This correction also disappears as $a\to 0$.
To deal with these problems one needs to perform a suitable scaling
analysis. For example suppose we define a topological charge
$Q_L(\rho_c)$ which only includes topological charges
for which $\rho \geq \rho_c = an_c$. As $a\to 0$ $Q_L(\rho_c)$ should
become independent of $\rho_c$ for a growing range
$n_c^{min} \leq n_c \leq n_c^{max}$: with $an_c^{min}$
large enough to exclude any $\rho\sim a$ artifacts
and $n_c^{max}$ growing exponentially with $\beta$. We shall
demonstrate a simple calculation of this kind next.

\subsection {The topological susceptibility}

Since $<Q> = 0$ (we have no $\theta$-term) the simplest
quantity we can calculate is $<Q^2>$. However there is a 
much better reason for calculating it: it is directly related
\cite{U1-Wit,U1-Ven}
to the $\eta^{\prime}$ mass:
\begin{equation}
\chi_t \equiv {{<Q^2>}\over{volume}}
\simeq {{f_{\pi}^2}\over{2N_f}}
(m_{\eta^{\prime}}^2 + m_{\eta}^2 - 2 m_K^2)      \label{B8}
\end{equation}
If we put in the experimental numbers into the above formula, 
then we get 
\begin{equation}
\chi_t \sim (180 \ MeV)^4          \label{B9}
\end{equation}
In this relation the susceptibility, $\chi_t$, is that of the 
pure gauge theory - which is something that we can calculate. 
We now describe such a calculation and see what happens.

The values of $<Q^2>$ that I am going to use come from
calculations that I have done in the past; for SU(3) they
come from 
\cite{JH-MT,MT-SU3}
while for SU(2) they come from
\cite{CM-MT-Q,DP-MT,unpub-QSU2}.
All these calculations have been performed using 20 to 25
cooling sweeps. The charge $Q_L$ is then calculated on
these cooled lattice fields. This charge is non-integer
because the smallest instanton charges suffer significant
$0(a^2/\rho^2)$ corrections. However such small instantons
have very peaked densities and are easy to identify.
One can then estimate the corrections due to these small
instantons, and shift $Q_L$ to the appropriate integer topological
charge $Q$. This has been done in all these calculations.

In 
Fig.~\ref{fig-QSU3} 
I plot $\chi_t/\surd\sigma$ against the string
tension, $a^2 \sigma$ in lattice units. We expect the leading
lattice corrections to this dimensionless mass ratio to be
$0(a^2/\rho^2)$, as in
eqn~\ref{A1}. 
So we would attempt a continuum extrapolation
of the form
\begin{equation}
{{\chi_t(a)}\over{\surd\sigma(a)}}
= {{\chi_t(0)}\over{\surd\sigma(0)}} + ca^2\sigma      \label{B10}
\end{equation}
which is a simple straight line on the plot. As we see the 
calculated values are consistent with this functional form.
In addition to the susceptibility calculated from the
total charge we also calculate $Q$ with narrow instantons
removed. The particular cut we have used is to remove all
charges whose peak density is greater than $1/16\pi^2$.
This corresponds to removing instantons with $\rho \leq 3a$. 
In doing this we are largely removing lattice artifacts
with $\rho \sim a$ which, because of their environment,
survive the cooling. Of course we are also removing a
part of the small-$\rho$ tail of the continuum
density, $D(\rho)$. So we are not saying that this is 
necessarily a better 
measure of the continuum susceptibility. What we do wish to
check is that both these susceptibilities are consistent
when extrapolated to the continuum limit. And as we see in
Fig.~\ref{fig-QSU3} 
this is indeed so. We extract:
\begin{equation}
\lim_{a\to 0} {{\chi_t}\over{\surd\sigma}}
= 0.437 \pm 0.020 \pm 0.015
\label{B11}
\end{equation}           
Here the first error is statistical and 
the second is a systematic error estimated using the
two different extrapolations. If we now plug in our
favourite value for the string tension, $\surd\sigma = 440 MeV$,
we obtain  
\begin{equation}
\chi_t = (192 \pm 12 MeV)^4        \label{B12}
\end{equation}
Of course we should really incorporate the uncertainty in 
the value of  $\surd\sigma$, most of that coming from
setting MeV units, and this would increase the error
to $\pm 20 MeV$. Irrespective of such details
it is clear that the 
pure gauge theory susceptibility is indeed consistent
with being large enough to drive the large $\eta^\prime$
mass.

In 
Fig.~\ref{fig-QSU2} 
I show the corresponding plot for the 
SU(2) susceptibility. Again we see that the two
susceptibilities we calculate are consistent
when extrapolated to the continuum limit. Note that
the difference between the two at corresponding
values of $a$ is much greater for $SU(2)$ than for $SU(3)$.
This is because the $SU(N)$ running coupling, $g^2(\rho)$, runs 
much faster to zero as $N\uparrow$ and so any finite-$a$
ambiguities between physical and ultraviolet topological
charges rapidly decrease. We extract:
\begin{equation}
\lim_{a\to 0} {{\chi_t}\over{\surd\sigma}}
= 0.470 \pm 0.011 \pm 0.015 \ \ \ \ \ \ \ : \ SU(2)     \label{B13}
\end{equation}
which translates to
\begin{equation}
\chi_t = (207 \pm 9 MeV)^4 \ \ \ \ \ \ \ : \ SU(2)      \label{B14}
\end{equation}
if we use $\surd\sigma = 440 MeV$. Unfortunately we do 
not know the $\eta^\prime$ etc. masses for the SU(2) theory
so we cannot say if this is what is expected. And neither
does it make much sense to introduce MeV units by using 
the $SU(3)$ value of $\surd\sigma$ in this way. 
It does however allow me to
compare to two recent estimates of $200 \pm 15 MeV$
\cite{deF-QSU2}
and of $230 \pm 30 MeV$
\cite{DeG-SU2}
which have been obtained using quite different methods,
but a similar $MeV$ scale. As we see, the values are
all consistent - which is reassuring.

\subsection {Vacuum topological structure}

The calculation of $Q$ on the lattice is relatively straightforward.
If we continuously deform a continuum gauge field so as to
minimise the action then we necessarily get driven to the
semiclassical multi-instanton minimum. We do not change topological
charge sectors since these are separated by infinite action barriers.
Cooling is just a naive lattice version of this procedure
and so it should work in this way for $a\to 0$. As we have seen, 
it does indeed seem to work well.

However there are other things we would like to know about the instanton
density if we are interested in gauging the possible influence of
instantons on chiral symmetry breaking, the quarkonium spectrum
and related physics. For example a picture in which instantons
have a mean size of $\sim 1/3 fm$ and are relatively dilute 
produces a plethora of interesting physics
\cite{Shuryak}.
Can we say something about such more detailed features of the
vacuum topological structure?

This question involves both technical and conceptual ambiguities.
There are two main technical problems. The first is that as
we cool a gauge field its
topological structure changes: instantons will change their
sizes and nearby instantons and anti-instantons will annihilate. 
These quantities are not invariant under minimising the action;
they are not even quasi-stable. So we must do as little cooling 
as possible and trust only those features that we find to be
relatively insensitive to cooling. If the vacuum that one obtains 
is a dilute gas of instantons then there is no further difficulty.
However what one might expect, and indeed finds, is a relatively dense
gas of overlapping charges. This raises a difficult problem of
pattern recognition, particularly for the larger instantons
whose density, $Q_L(x)$ will be very small (it obviously
varies as $1/\rho^4$) and which are most likely to overlap with 
several other large instantons. Such large instantons will, 
in any case, not be completely smooth, and may possess multiple
peaks (`ripples') since we are trying to minimise the number of
cooling sweeps used. These problems have been addressed in
different ways by several recent calculations of this kind
\cite{CM-PS,deF-QSU2,DeG-SU2,Brower,DS-MT}.
I think it is fair to say that all these should be regarded as 
exploratory. However, to whet your appetite I am going to show you
some results from an analysis that I have been involved in
\cite{DS-MT}.
What I am not going to do here is to tell you anything at all
about the pattern recognition algorithms etc.

I mentionned a conceptual difficulty. Instantons are semiclassical
objects. The real vacuum has fluctuations on all scales and
there is no reason to think that the topological charge
resides in instanton-like objects. Does it even make sense
to talk of a distribution in $\rho$? This raises all kinds of
interesting questions, including: do we care? After all,
models such as that in
\cite{Shuryak}
would not claim to be more than simplifications that
are appropriate to the physics being considered. Perhaps we
should regard a minimal amount of smoothening as performing
such a simplification upon the fluctuating gauge fields?
I am not going to address these questions any further here,
but you should be aware of their existence.

Let me now move to some results of this kind of calculation
\cite{DS-MT}.
The calculations are in the pure SU(3) gauge theory. We have used
lattice fields that have been generated and stored (for other 
purposes) by the UKQCD Collaboration. We are performing 
calculations on $16^3 48$ and $32^3 64$ lattices at $\beta=6.0$, 
$24^3 48$ lattices at $\beta=6.2$,
and $32^3 64$ lattices at $\beta=6.4$.
The point of the 2 lattice sizes at $\beta = 6.0$ is to check
for finite volume effects - especially for the very large
instantons. The various lattice spacings enable us to check
whether the features we identify have the right scaling
properties to survive into
the continuum limit. In addition we do the calculations for
various different numbers of cooling sweeps so that we
can check whether these features are insensitive to cooling
or not.

\begin{figure}
\centerline{\epsfysize=16.0cm
            \epsfxsize=16.0cm
            \epsfbox{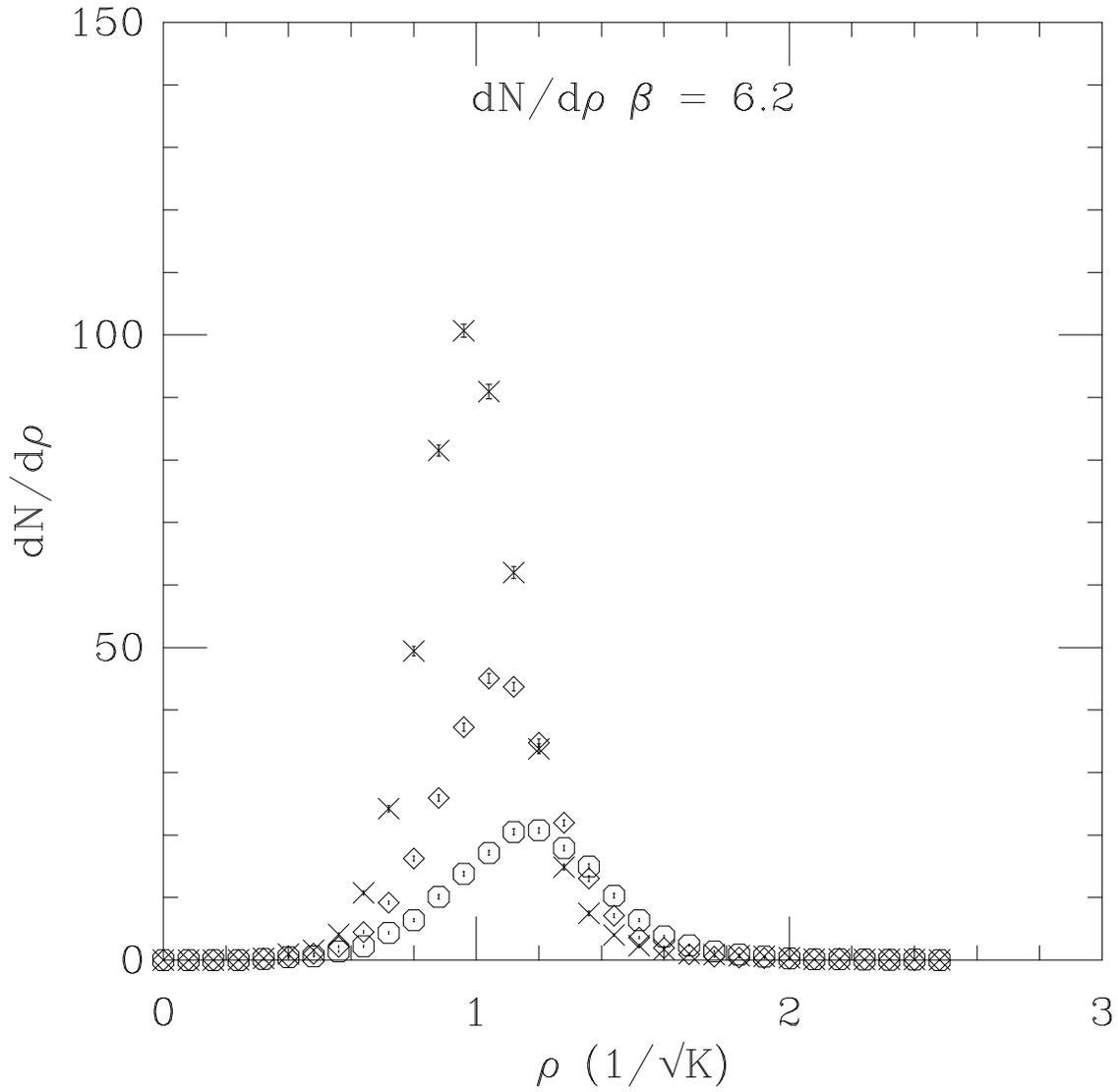}}
\caption{The number of topological charges
versus size at $\beta=6.2$: after 23 ($\times$), 32 ($\diamond$)
and 46 ($\circ$) (under-relaxed) cooling sweeps. The size is in
units of $1/\surd K$ where K is the string tension.}
\label{fig-Qsize}
\end{figure}

In 
Fig.~\ref{fig-Qsize} 
I show the size distribution, $D(\rho)$, with
the size $\rho$ expressed in units of $1/\surd\sigma \equiv
1/\surd K$. The peak  is around $\rho \sim 0.5 fm$. This is
not very sensitive to $\beta$ or to the lattice size or to the
number of cooling sweeps. The total number of charges is of 
course sensitive to the amount of cooling. We do, however,
note that unless we
go to a very large number of cools, the vacuum is dense
with a great deal of overlap between the charges.

It is not straightforward to test for scaling, since a
cooling sweep is intrinsically non-scaling. What we do 
find is that if we tune the number of cooling sweeps
so that the total number of charges is independent of
$\beta$, then the detailed densities,  $D(\rho)$, seem
to scale as well.

Two other aspects of  $D(\rho)$ are of particular interest:
the fall-offs at small and large $\rho$. The former
is interesting because it provides a check on our
calculations: we know  that 
$D(\rho) \propto \rho^6$ for $a \ll \rho \ll 1/\surd\sigma$.
Of course we can only approach these severe inequalities
by looking for the trend as we increase $\beta$. We do
indeed find a tendency to approach something close to this
functional behaviour. (Which, in any case, is modified by
powers of $\log\rho$.) By contrast
large $\rho$ is interesting because we do not know what
happens to large instantons in a confining vacuum.
If we used the semiclassical formula with a coupling
that froze at large $\rho$ then we would get
$D(\rho) \propto 1/\rho^5$ as $a \to \infty$. What we
find is a much stronger suppression: 
\begin{equation}
D(\rho) \propto {1\over{\rho^{\sim 11}}} \ \ \ \ \ \ \ : \ 
\rho \gg {1\over{\surd\sigma}}         \label{B15}
\end{equation}

\begin{figure}
\centerline{\epsfysize=16.0cm
            \epsfxsize=16.0cm
            \epsfbox{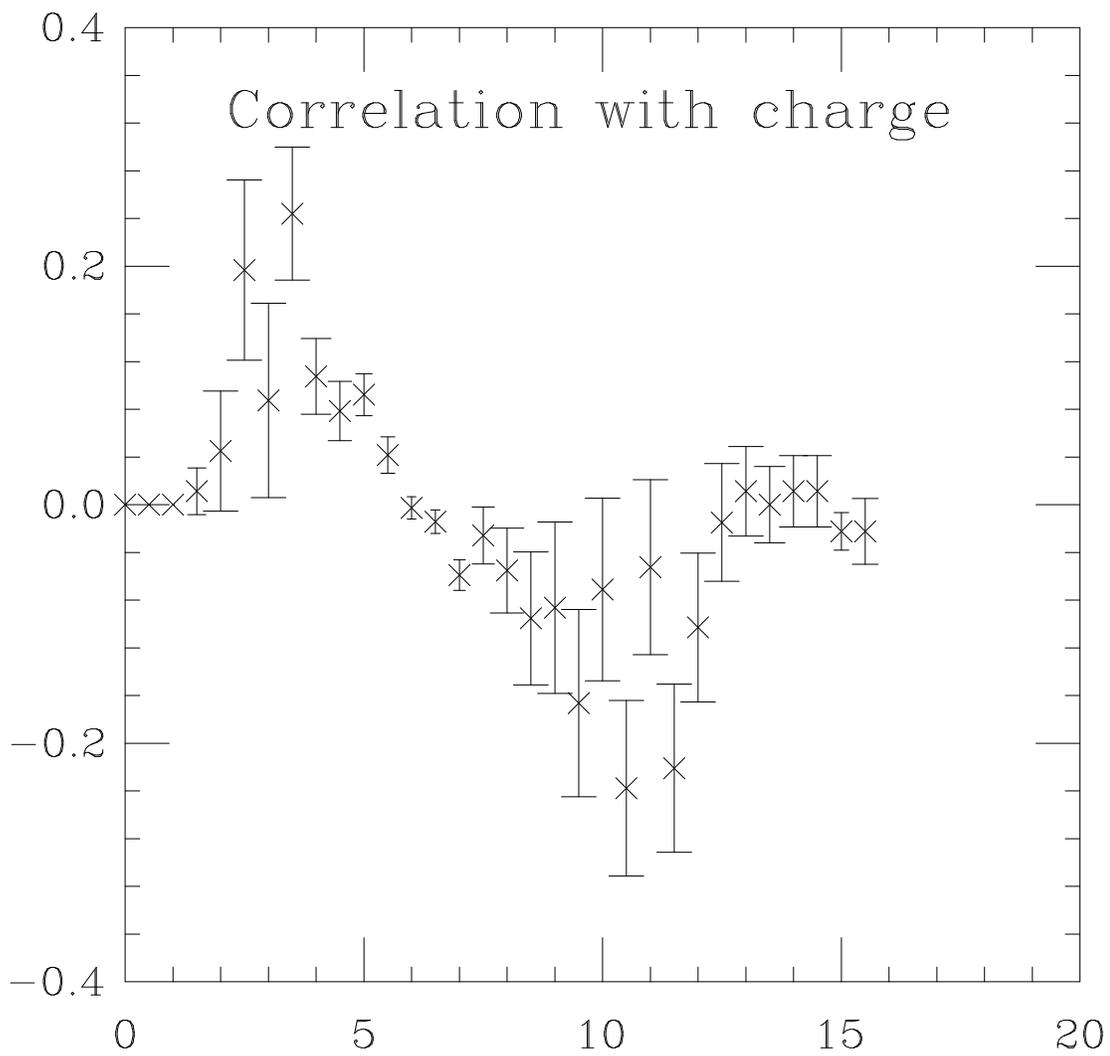}}
\caption{The average correlation between the sign of a charge of size 
$\rho$ and the sign of the total charge of the gauge field;
at $\beta=6.2$ after 23 cooling sweeps}
\label{fig-Qqcor}
\end{figure}

\begin{figure}
\centerline{\epsfysize=16.0cm
            \epsfxsize=16.0cm
            \epsfbox{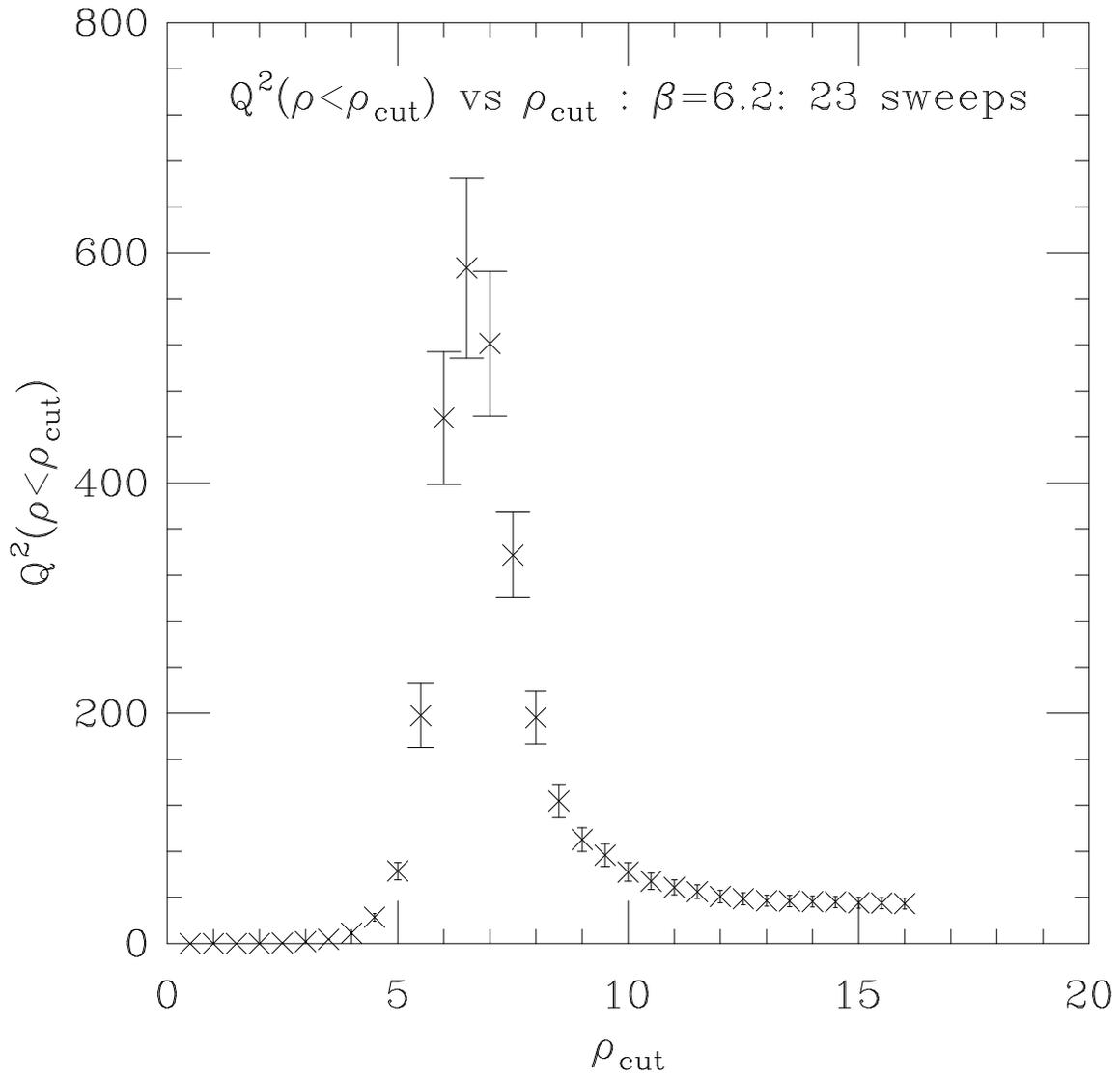}}
\caption{The average of $Q^2$ when only charges with sizes
$\rho<\rho_c$  are included : at $\beta=6.2$ after 23 cooling sweeps}
\label{fig-Qqcut}
\end{figure}

Before closing this subject let me move to a quite different
and more subtle aspect of the vacuum structure. In a dilute
gas the correlation between the sign of a charge and the sign
of the total charge will be independent of size
$\rho$. In 
Fig.~\ref{fig-Qqcor} 
I show the correlation we actually find. There 
is a striking size-dependence. Charges smaller than average tend
to have the same sign as $Q$, larger charges tend to have the
opposite charge. Since the sign of $Q$ simply tells you which
sign wins out on that configuration, this tells us that
net charge of the small instantons wins out over the net charge of 
the larger instantons. This suggests a picture where the very large
instantons are polarised, so that the small instantons sitting on
this background tend to have the opposite charge thoughout the
volume. Since the charge of the small instantons wins out, this
means that the large instantons are actually overscreened.
Of course this would be a lot to glean from the one plot; however
this intriguing picture is in fact supported by more
detailed calculations. It leads to quite striking effects.
Suppose for example we decide to calculate the total charge
that includes instantons smaller than some value $\rho_c$:
$Q_{\rho \leq \rho_c}$. The fluctuations are shown in 
Fig.~\ref{fig-Qqcut}. 
We note that if we were to limit ourselves
to instantons with $\rho \leq 1/\surd\sigma$ then we would
obtain a topological susceptibility that is $\sim 10$ times
greater than the total susceptibility! It is clearly 
important to investigate the effect of these structures
on quark propagators and hence on the observable physics. 

\subsection {Conclusions}

In this lecture I hope that I have convinced you that
simulations of lattice gauge theories can be both useful
and interesting in telling us something about the 
topological fluctuations of the vacuum. Some of what
I have shown you, particularly concerning the long distance
polarisation of the vacuum, must be regarded as preliminary
- to a potentially embarrassing degree. The calculations
of the topological susceptibility, on the other hand, are
now quite reliable. They show us that the topological
vacuum fluctuations are indeed large enough to drive
the large $\eta^{\prime}$ mass. The theoretical
expectations with which we compared our result
arise, most straightforwardly, from arguments about what happens
in $SU(N)$ QCD at large $N$. This takes me smoothly to the final of
my three topics in these lectures.

\section {${\bf SU(N_c)}$ gauge theories for all ${\bf N_c}$}

Quantum Chromodynamics is an $SU(3)$ gauge theory coupled to 
3 lightish quark colour triplets. If we change the gauge group 
to $SU(2), SU(4), SU(5),$ ... then we obtain an infinite set of,
a priori, quite different theories. However from an analysis
of Fenyman diagrams to all orders 
\cite{N-H}
one finds that the $N_c \to \infty$ limit of such $SU(N_c)$ theories
is smooth if we vary the coupling as $g^2 \propto 1/N_c$. This
suggests that it should be possible to describe $SU(N_c)$ gauge  
theories as perturbations in powers of $1/N_c$ 
around $SU(\infty)$ 
\cite{N-H},
at least for large enough $N_c$. Moreover if one assumes 
confinement for all $N_c$, then one can easily show 
\cite{N-H,N-W}
that the phenomenology of the $SU(\infty)$ quark-gluon theory 
is strikingly similar to that of (the non-baryonic sector of) QCD.
This makes it conceivable that the physically
interesting $SU(3)$ theory could be largely
understood by solving the much simpler $SU(\infty)$ theory.
If all the $SU(N_c)$ theories down to $SU(3)$ can be treated
in this way, then this represents an elegant and enormous
theoretical simplification.

I do not have the time to review 
\cite{N-C}
this subject here, but let me at least indicate something of
what is involved, albeit using arguments that lack any rigour.
The constraint that $g^2 \propto 1/N_c$ is easy to motivate. 
Consider inserting a gluon loop into a gluon propagator.  
We have added two triple-gluon vertices and this gives
a factor $g^2$. At the same time the sum over colour
in the loop gives a factor of $N_c$. (Not $N_c^2$ because
the colour of the incoming/outgoing gluon is fixed.)
So we have a total factor $\sim g^2 N_c$. Now such
loops can be inserted any number of times, so if we
want a smooth large-$N_c$ limit ( at least in all-order 
perturbation theory) then we clearly need
to impose $g^2 \propto 1/N_c$. Assume therefore that
we have done so. Consider a typical meson decay,
e.g. $\rho \to \pi\pi$. This requires the production
of a $q\bar{q}$ pair. So the decay width contains
a factor of $g^2 \propto 1/N_c$. But if we have confinement all the
hadrons are colour singlets and so we do not acquire any 
compensating factors from summing over the colours of the
decay products. Thus the decay width is suppressed
by a factor $1/N_c$. That is, at large $N_c$
hadrons do not decay. This is like a `narrow-width'
caricature of the real world. In fact this is a reasonable
first approximation to the hadrons we know; mostly
their widths are very much smaller than their masses.
A similar argument tells us that mixings, e.g. between 
mesons and glueballs, are suppressed. This is reminiscent
of the OZI rule discussed in my first lecture. 
This (and much more) suggests that the $SU(\infty)$
theory is indeed a first approximation to $SU(3)$.

All this motivates us to try and answer by explicit
calculation some basic questions:

$\bullet$ does a non-perturbative calculation support the
(all-orders) perturbative argument for a smooth 
$N_c \to \infty$ limit?

$\bullet$ is $N_c \to \infty$ confining?

$\bullet$ does such a limit really require $g^2 \propto 1/N_c$;
and if so what does this mean when we have a running coupling?

$\bullet$ is $SU(N_c) \simeq SU(\infty)$ only for $N_c \gg 1$
or is it the case down to $N_c = 3$, or even  $N_c = 2$?

$\bullet$ what is the $SU(\infty)$ mass spectrum?

There have been a number of interesting
computational explorations of the lattice $SU(\infty)$ theory
(for a review see
\cite{N-Das})
based on the fact that it can be re-expressed as a single plaquette
theory
\cite{N-EK}.
Unfortunately this scheme makes no statement
about the size of the leading corrections to the
$N_c=\infty$ limit, and so gives us no clue as to
how close $N_c=3$ is to $N_c=\infty$.

In this lecture I will describe an extremely straightforward
approach to this problem. I will simply calculate 
the properties of $SU(N_c)$ gauge theories for several values 
of $N_c$ and so determine explicitly how the physics varies 
as $N_c$ increases. I will only look at the pure gauge theory,
but there are good reasons for believing that the 
inclusion of quarks will not alter any of our conclusions
(except in some obvious ways). Ideally I would like to present
you with accurate calculations in 4 dimensions. Unfortunately,
at present what I can  provide you there is very rough
and tentative. But I will make up for that by describing the results of 
the corresponding, but much more precise, calculation in 3 dimensions.
Of course you will want to know what is the relevance of such a 
`substitution'. I will come to that shortly.

My study was originally motivated by the observation that
the $C=+$ sector of the light mass spectrum turned 
out to be quite similar in the $D=2+1$ $SU(2)$ 
\cite{MTD3SU2G,MTD3SU2K}
and $SU(3)$
\cite{MTD3SU3}
theories. (This also appears to be the case in $D=3+1$, 
although there the comparison is weakened by 
the much larger errors.) One reason for this might be
that both are close to the $N_c=\infty$ limit. In that case
we would have an economical understanding of the spectra
of $SU(N_c)$ gauge theories for all $N_c$ :
there is a common spectrum with small corrections.

A second, more practical reason for studying $N_c \to \infty$ 
was my interest in obtaining some model understanding of
the structure of glueballs. Models are of interest even
if they are very approximate in comparison with the
results of simulations. A good model will embody the
essential degrees of freedom in a problem and show how
this leads to the main features of the physics, within some 
transparent and plausible approximation scheme. A model could
provide the intuition necessary for an economical understanding
of the role of glueballs in a wide range of contexts. For example,
if one understands the structure of glueballs, one can make crude
but reliable estimates  of glueball-quarkonium mixing,
of glueball decay, of the effects of dynamical quarks etc.
In many cases the approximations made in the model include the 
neglect of decays and mixing. These are features of the $SU(\infty)$
rather than of the $SU(3)$ theory, and so it would be better to 
test the model against the spectrum of the former theory.
One example that has been of particular interest to me is the 
flux tube model of glueballs
\cite{ISG-PAT,TM-MT,RJ-MT}.
The formulation of this model is identical for all $N_c > 2$.
However, because the model does not incorporate the
effects of glueball decay, it should presumably be tested
against the $N_c\to\infty$ spectrum since it is only in that 
limit that there are no decays. It is also the case that many
theoretical approaches are simpler in that limit. 
An example is provided by the recent progress in calculating the
large $N_c$ mass spectrum using light-front quantisation
techniques
\cite{FA-SD}.

The $D=2+1$ analysis that I shall present here 
is based on my calculations over the last 
few years of the properties of $SU(N_c)$ gauge theories with
$N_c=2,3,4$ and 5. In $D=3+1$ what I have done is to perform
some $SU(4)$ calculations to supplement what is known about
$SU(2)$ and $SU(3)$. My strategy is the very simple one of
directly calculating the mass spectra of these theories and
seeing whether they are approximately independent of $N_c$.
The calculations are performed through the Monte Carlo
simulation of the corresponding lattice theories, using
the standard plaquette action. In the $D=2+1$ 
case the calculations are very accurate and we are able to 
extrapolate our mass ratios to the continuum limit prior to
the comparison. In the 3+1 dimensional case our $SU(4)$
calculations are not good enough for that, and our
comparisons with $SU(2)$ and $SU(3)$ are correspondingly
less precise. Some of the $SU(2)$ results have been published 
\cite{MTD3SU2G,MTD3SU2K}
as have brief summaries of the results discussed here
\cite{MTSUN,MTLAT96}.
A long paper is in preparation.

\subsection {D=2+1 $\sim$ D=3+1 ?}

While one might naively expect that the $D=2+1$ and $D=3+1$ 
gauge theories would be so different as to make a unified
treatment misleading, this is not in fact so. Theoretically
the $D=2+1$ theory shares with its $D=3+1$ homologue
four important properties.

$\bullet$  Both theories become free at short distances.
In 3 dimensions  the coupling, $g^2$, has dimensions
of mass so that the effective dimensionless expansion
parameter on a scale $l$ will be 
\begin{equation}
g_3^2(l) \equiv lg^2 \stackrel{l\to 0}{\longrightarrow} 0    \label{C1}
\end{equation}
In 4 dimensions the coupling is dimensionless and runs
in a way we are all familar with:
\begin{equation}
g_4^2(l) \simeq {c\over{\ln(l\Lambda)}}
 \stackrel{l\to 0}{\longrightarrow} 0        \label{C2}
\end{equation}
In both cases the interactions vanish as $l \to 0$,
although they do so much faster in the super-renormalisable
$D=2+1$ case than in the merely asymptotically free
$D=3+1$ case.

$\bullet$  Both theories become strongly coupled at large distances.
This we see immediately by letting $l\uparrow$ in the above formulae.
Thus in both cases the interesting physics is nonperturbative.

$\bullet$ In both theories the coupling sets the mass scale.
In 3 dimensions it does so explicitly:
\begin{equation}
m_i = c_i g^2           \label{C3}
\end{equation}
In 4 dimensions it does so through the phenomenon of 
dimensional transmutation: the classical scale invariance is 
anomalous, the coupling runs and this introduces 
a mass scale through the rate at which it runs:
\begin{equation}
m_i = c_i \Lambda       \label{C4}
\end{equation}

$\bullet$ Both theories confine with a linear potential.
This is not something that we can prove by a simple argument.
However lattice simulations provide convincing evidence
that this is indeed the case. Note that although the
$D=2+1$ Coulomb potential is already confining, this
is a weak logarithmic confinement,
$V_C(r) \sim g^2 \ln(r)$, which has nothing to do with
the nonperturbative linear potential, $V(r) \simeq \sigma r$,
that one finds at large $r$.

In addition to these theoretical similarities, 
the calculated spectra also
show some striking similarities.

$\bullet$ In both theories the lightest glueball is the
scalar $0^{++}$ with a similar mass 
$m_{0^{++}} \sim 4\surd\sigma$. In the $C=+$ sector, the 
$2^{++}$ is the next lightest glueball (ignoring any
excited scalars) with $m_{2^{++}}/m_{0^{++}} \sim 3/2$
in both cases.

All this motivates us to believe that a unified treatment makes sense.

Of course there are significant differences as well. For example:

$\circ$ There are no instantons in $D=2+1$ non-Abelian gauge theories.
This probably implies quite different quark physics.

$\circ$ The rotation group is Abelian.

$\circ$ The details of the mass spectrum are very different in
3 and 4 dimensions.
A particularly striking difference is that in 3 dimensions
the spectrum exhibits parity doubling for states with
non-zero angular momentum:
\begin{equation}
m_{J^{+}} = m_{J^{-}} \ \ \ \ if \ J\not=0      \label{C5}
\end{equation}
This follows from the fact that angular momentum flips sign
under parity - which can be defined in $D=3$ as 
$(x,y)\to(-x,y)$. The proof is elementary and I leave it as an 
exercise. Actually to show this you need the continuum rotation
group. If you only have $\pi/2$ rotations then you find
that you lose parity doubling for the $J=2$ state. This can
happen either because you are too close to the strong coupling
limit or because the volume is too small and the rotation
symmetry is broken by the boundary conditions. In either 
case this is a useful check: that we are effectively in the
continuum limit and in an effectively infinite volume.

\subsection {2+1 dimensions.}

\label	{3dim}

The calculations in 3 dimensions are performed in the same way as 
in 4 dimensions except that the computational problem is much
more manageable. (Lattices grow as $L^3$ rather than as $L^4$.)
The basic steps are just as outlined in my first lecture, and
you will have to trust me that all the checks have been performed
sufficiently carefully! After extrapolating to the continuum limit
I obtain the string tension and a mass spectrum in units of $g^2$.
Only the lightest portion of the mass spectrum is calculated,
but that includes $J^{PC}$ states for $J=0,1,2$ and $P=\pm$
and $C=\pm$ as well as one or two further excited states
in many cases. I do this for $SU(2)$, $SU(3)$, $SU(4)$ and $SU(5)$.
In the case of $SU(2)$ there is no $C=-$ sector. In fact the main 
point of the $SU(5)$ calculation was to have 3 values of $N_c$
for the $C=-$ states, so as to provide some control over their
$N_c$ dependence. The $SU(5)$ calculation is recent and has not
been incorporated into the analysis that follows.

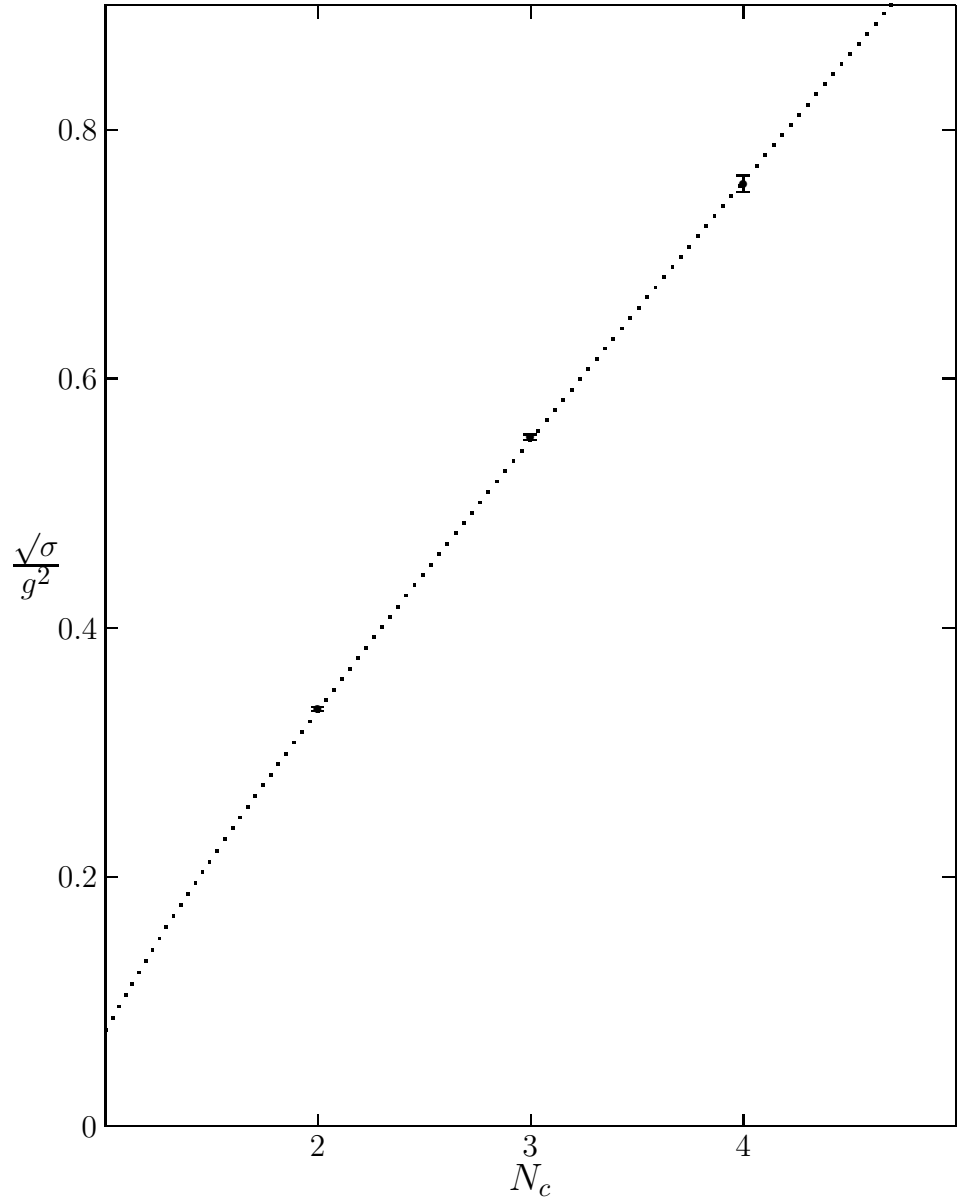
\begin	{figure}[p]
\begin	{center}
\leavevmode
\setlength{\unitlength}{0.240900pt}
\ifx\plotpoint\undefined\newsavebox{\plotpoint}\fi
\sbox{\plotpoint}{\rule[-0.200pt]{0.400pt}{0.400pt}}%
\begin{picture}(1500,1800)(0,0)
\font\gnuplot=cmr10 at 12pt
\gnuplot
\sbox{\plotpoint}{\rule[-0.200pt]{0.400pt}{0.400pt}}%
\put(120.0,31.0){\rule[-0.200pt]{4.818pt}{0.400pt}}
\put(108,31){\makebox(0,0)[r]{{$0$}}}
\put(1436.0,31.0){\rule[-0.200pt]{4.818pt}{0.400pt}}
\put(120.0,423.0){\rule[-0.200pt]{4.818pt}{0.400pt}}
\put(108,423){\makebox(0,0)[r]{{$0.2$}}}
\put(1436.0,423.0){\rule[-0.200pt]{4.818pt}{0.400pt}}
\put(120.0,814.0){\rule[-0.200pt]{4.818pt}{0.400pt}}
\put(108,814){\makebox(0,0)[r]{{$0.4$}}}
\put(1436.0,814.0){\rule[-0.200pt]{4.818pt}{0.400pt}}
\put(120.0,1206.0){\rule[-0.200pt]{4.818pt}{0.400pt}}
\put(108,1206){\makebox(0,0)[r]{{$0.6$}}}
\put(1436.0,1206.0){\rule[-0.200pt]{4.818pt}{0.400pt}}
\put(120.0,1597.0){\rule[-0.200pt]{4.818pt}{0.400pt}}
\put(108,1597){\makebox(0,0)[r]{{$0.8$}}}
\put(1436.0,1597.0){\rule[-0.200pt]{4.818pt}{0.400pt}}
\put(454.0,31.0){\rule[-0.200pt]{0.400pt}{4.818pt}}
\put(454,19){\makebox(0,0){\shortstack{\\ \\ \\ {$2$}}}}
\put(454.0,1773.0){\rule[-0.200pt]{0.400pt}{4.818pt}}
\put(788.0,31.0){\rule[-0.200pt]{0.400pt}{4.818pt}}
\put(788,19){\makebox(0,0){\shortstack{\\ \\ \\ {$3$}}}}
\put(788.0,1773.0){\rule[-0.200pt]{0.400pt}{4.818pt}}
\put(1122.0,31.0){\rule[-0.200pt]{0.400pt}{4.818pt}}
\put(1122,19){\makebox(0,0){\shortstack{\\ \\ \\ {$4$}}}}
\put(1122.0,1773.0){\rule[-0.200pt]{0.400pt}{4.818pt}}
\put(120.0,31.0){\rule[-0.200pt]{321.842pt}{0.400pt}}
\put(1456.0,31.0){\rule[-0.200pt]{0.400pt}{424.466pt}}
\put(120.0,1793.0){\rule[-0.200pt]{321.842pt}{0.400pt}}
\put(12,912){\makebox(0,0){{\Large{${{\surd\sigma} \over g^2}$}}}}
\put(788,-53){\makebox(0,0){{\large{$N_c$}}}}
\put(120.0,31.0){\rule[-0.200pt]{0.400pt}{424.466pt}}
\put(454,687){\circle*{12}}
\put(788,1114){\circle*{12}}
\put(1122,1512){\circle*{12}}
\put(454.0,684.0){\rule[-0.200pt]{0.400pt}{1.445pt}}
\put(444.0,684.0){\rule[-0.200pt]{4.818pt}{0.400pt}}
\put(444.0,690.0){\rule[-0.200pt]{4.818pt}{0.400pt}}
\put(788.0,1110.0){\rule[-0.200pt]{0.400pt}{1.927pt}}
\put(778.0,1110.0){\rule[-0.200pt]{4.818pt}{0.400pt}}
\put(778.0,1118.0){\rule[-0.200pt]{4.818pt}{0.400pt}}
\put(1122.0,1499.0){\rule[-0.200pt]{0.400pt}{6.263pt}}
\put(1112.0,1499.0){\rule[-0.200pt]{4.818pt}{0.400pt}}
\put(1112.0,1525.0){\rule[-0.200pt]{4.818pt}{0.400pt}}
\sbox{\plotpoint}{\rule[-0.500pt]{1.000pt}{1.000pt}}%
\put(120,183){\usebox{\plotpoint}}
\multiput(120,183)(9.885,18.250){2}{\usebox{\plotpoint}}
\put(140.16,219.28){\usebox{\plotpoint}}
\put(150.42,237.32){\usebox{\plotpoint}}
\multiput(160,255)(10.792,17.729){2}{\usebox{\plotpoint}}
\put(181.75,291.11){\usebox{\plotpoint}}
\put(192.60,308.80){\usebox{\plotpoint}}
\put(203.60,326.40){\usebox{\plotpoint}}
\multiput(214,344)(11.513,17.270){2}{\usebox{\plotpoint}}
\put(236.73,379.10){\usebox{\plotpoint}}
\put(248.25,396.36){\usebox{\plotpoint}}
\put(259.73,413.64){\usebox{\plotpoint}}
\put(270.89,431.13){\usebox{\plotpoint}}
\multiput(282,447)(11.720,17.130){2}{\usebox{\plotpoint}}
\put(306.40,482.28){\usebox{\plotpoint}}
\put(318.16,499.38){\usebox{\plotpoint}}
\put(330.00,516.43){\usebox{\plotpoint}}
\put(341.81,533.50){\usebox{\plotpoint}}
\put(353.76,550.46){\usebox{\plotpoint}}
\put(366.03,567.20){\usebox{\plotpoint}}
\put(378.22,584.01){\usebox{\plotpoint}}
\multiput(390,600)(12.152,16.826){2}{\usebox{\plotpoint}}
\put(414.98,634.26){\usebox{\plotpoint}}
\put(427.16,651.07){\usebox{\plotpoint}}
\put(439.76,667.55){\usebox{\plotpoint}}
\put(452.11,684.23){\usebox{\plotpoint}}
\put(464.62,700.79){\usebox{\plotpoint}}
\put(477.07,717.40){\usebox{\plotpoint}}
\put(489.66,733.88){\usebox{\plotpoint}}
\put(502.48,750.20){\usebox{\plotpoint}}
\put(514.80,766.89){\usebox{\plotpoint}}
\put(527.52,783.29){\usebox{\plotpoint}}
\put(540.15,799.76){\usebox{\plotpoint}}
\put(552.88,816.16){\usebox{\plotpoint}}
\multiput(565,832)(13.194,16.022){2}{\usebox{\plotpoint}}
\put(590.88,865.45){\usebox{\plotpoint}}
\put(603.98,881.55){\usebox{\plotpoint}}
\put(616.68,897.97){\usebox{\plotpoint}}
\put(629.77,914.07){\usebox{\plotpoint}}
\put(642.52,930.45){\usebox{\plotpoint}}
\put(655.55,946.60){\usebox{\plotpoint}}
\put(668.36,962.93){\usebox{\plotpoint}}
\put(681.33,979.12){\usebox{\plotpoint}}
\put(694.19,995.41){\usebox{\plotpoint}}
\put(707.12,1011.64){\usebox{\plotpoint}}
\put(720.03,1027.89){\usebox{\plotpoint}}
\put(732.90,1044.17){\usebox{\plotpoint}}
\put(746.06,1060.22){\usebox{\plotpoint}}
\put(759.19,1076.30){\usebox{\plotpoint}}
\put(772.19,1092.47){\usebox{\plotpoint}}
\put(785.11,1108.70){\usebox{\plotpoint}}
\put(798.49,1124.56){\usebox{\plotpoint}}
\put(811.24,1140.93){\usebox{\plotpoint}}
\put(824.41,1156.97){\usebox{\plotpoint}}
\put(837.52,1173.06){\usebox{\plotpoint}}
\put(850.70,1189.10){\usebox{\plotpoint}}
\put(863.81,1205.19){\usebox{\plotpoint}}
\put(876.99,1221.22){\usebox{\plotpoint}}
\put(890.09,1237.32){\usebox{\plotpoint}}
\put(903.28,1253.35){\usebox{\plotpoint}}
\multiput(916,1269)(13.194,16.022){2}{\usebox{\plotpoint}}
\put(942.66,1301.58){\usebox{\plotpoint}}
\put(955.85,1317.60){\usebox{\plotpoint}}
\put(968.95,1333.71){\usebox{\plotpoint}}
\put(982.57,1349.36){\usebox{\plotpoint}}
\put(995.29,1365.76){\usebox{\plotpoint}}
\put(1008.81,1381.50){\usebox{\plotpoint}}
\put(1021.99,1397.53){\usebox{\plotpoint}}
\put(1035.17,1413.56){\usebox{\plotpoint}}
\put(1048.28,1429.66){\usebox{\plotpoint}}
\put(1061.83,1445.38){\usebox{\plotpoint}}
\put(1075.05,1461.37){\usebox{\plotpoint}}
\put(1088.22,1477.41){\usebox{\plotpoint}}
\put(1101.34,1493.50){\usebox{\plotpoint}}
\put(1114.85,1509.25){\usebox{\plotpoint}}
\put(1128.11,1525.21){\usebox{\plotpoint}}
\put(1141.61,1540.98){\usebox{\plotpoint}}
\put(1154.56,1557.19){\usebox{\plotpoint}}
\put(1167.85,1573.11){\usebox{\plotpoint}}
\put(1181.16,1589.04){\usebox{\plotpoint}}
\put(1194.61,1604.84){\usebox{\plotpoint}}
\put(1207.93,1620.76){\usebox{\plotpoint}}
\put(1221.37,1636.56){\usebox{\plotpoint}}
\put(1234.41,1652.69){\usebox{\plotpoint}}
\put(1247.61,1668.70){\usebox{\plotpoint}}
\put(1260.97,1684.58){\usebox{\plotpoint}}
\put(1274.37,1700.42){\usebox{\plotpoint}}
\put(1287.74,1716.30){\usebox{\plotpoint}}
\put(1301.13,1732.15){\usebox{\plotpoint}}
\put(1314.51,1748.01){\usebox{\plotpoint}}
\put(1327.89,1763.88){\usebox{\plotpoint}}
\put(1341.28,1779.73){\usebox{\plotpoint}}
\multiput(1348,1788)(12.966,16.207){0}{\usebox{\plotpoint}}
\put(1352,1793){\usebox{\plotpoint}}
\end{picture}
\end	{center}
\caption{Continuum string tension versus number of colours
in $D=2+1$. Line is the fit in eqn~\ref{C7}.}
\label	{fig_plot_string3}
\end 	{figure}

I begin with the string tension, $\sigma$, since it turns out 
to be our most accurately calculated physical quantity.
We use smeared Polyakov loops 
\cite{CM-MT-Q},
to obtain $a^2\sigma$ for several values of the lattice spacing $a$.
We then extrapolate the lattice results, using the asymptotic
relation $\beta = 4/ag^2$, to obtain 
the continuum string tension in units of $g^2$:
\begin{equation}
{{\surd \sigma} \over g^2} = 
\lim_{\beta\to\infty} {\beta \over{2N_c}} a\surd\sigma    \label{C6}
\end{equation}
The results for $SU(2)$,$SU(3)$ and $SU(4)$ in $D=2+1$
\cite{MTSUN}
are shown in Table~\ref{n_string}
and are plotted in Fig.~\ref{fig_plot_string3}.
We immediately see that there is an
approximate linear rise with $N_c$ and we find that we
can obtain a good fit with
\begin{equation}
{{\surd \sigma} \over g^2} = 0.1974(12) N_c 
-{0.120(8) \over N_c}.         \label{C7}
\end{equation}
We obtain a similar behaviour with the light
glueball masses (see below).

Some observations.

$\bullet$ For large $N_c$, 
eqn~\ref{C7} 
tells us that 
$\surd\sigma \propto g^2 N_c$. That is to say, 
the overall mass scale of the theory, call
it $\mu$, is proportional to $g^2 N_c$. In other words, 
in units of the mass scale of the theory
\begin{equation}
g^2 \propto {\mu \over N_c}.          \label{C8}
\end{equation}
While this coincides with the usual expectation based on an 
analysis of Feynman diagrams, we note that here the
argument is fully non-perturbative.

$\bullet$ The string tension is non-zero for all $N_c$ 
and, in particular, for $N_c \to \infty$ (when expressed
in units of $g^2 N_c$ or the lightest glueball masses - see
below). This confirms the basic assumption that needs to
be made in 4 dimensions
in order to extract the usual phenomenology of the 
large-$N_c$ theory.

$\bullet$ In the pure gauge sector one expects (again
from an analysis of Feynman diagrams) 
\cite{N-C}
that the first correction to the large-$N_c$ limit will be
$O(1/N^2_c)$ relative to the leading term. The fit in
eqn~\ref{C7}
is indeed of this form. We note that if we try a fit 
with a $O(1/N_c)$ correction instead (which would be 
appropriate if we had quarks) then we obtain an unacceptably 
poor $\chi^2$ (corresponding to a confidence level of only 
$\sim 2\%$ in contrast to the $\sim 45\%$ we obtain for the
quadratic correction). We may regard this as providing some 
non-perturbative support for this diagram-based expectation.

$\bullet$ The coefficient of the correction term in
eqn~\ref{C7}
is comparable to that of the leading term, suggesting an
expansion in powers of $1/N_c$ that is rapidly convergent.
Indeed one has to go to $N_c=1$ before the correction term 
becomes comparable to the leading term. While the $SU(1)$
theory is completely trivial, we note that the $U(1)$
theory has a zero string tension (in the sense that
$\surd\sigma/g^2 = 0$ in the continuum limit).

Let me now turn from the string tension to the mass spectrum. 
Recall that since we are in $D=2+1$, states of opposite
parity are degenerate as long as $J \not = 0$.
This degeneracy is broken by lattice spacing and
finite volume corrections. I will present the
results separately for the $P=+$ and $P=-$ states
so as to provide an explicit check on the presence of any
such unwanted corrections.

\begin	{figure}[p]
\begin	{center}
\leavevmode
\input	{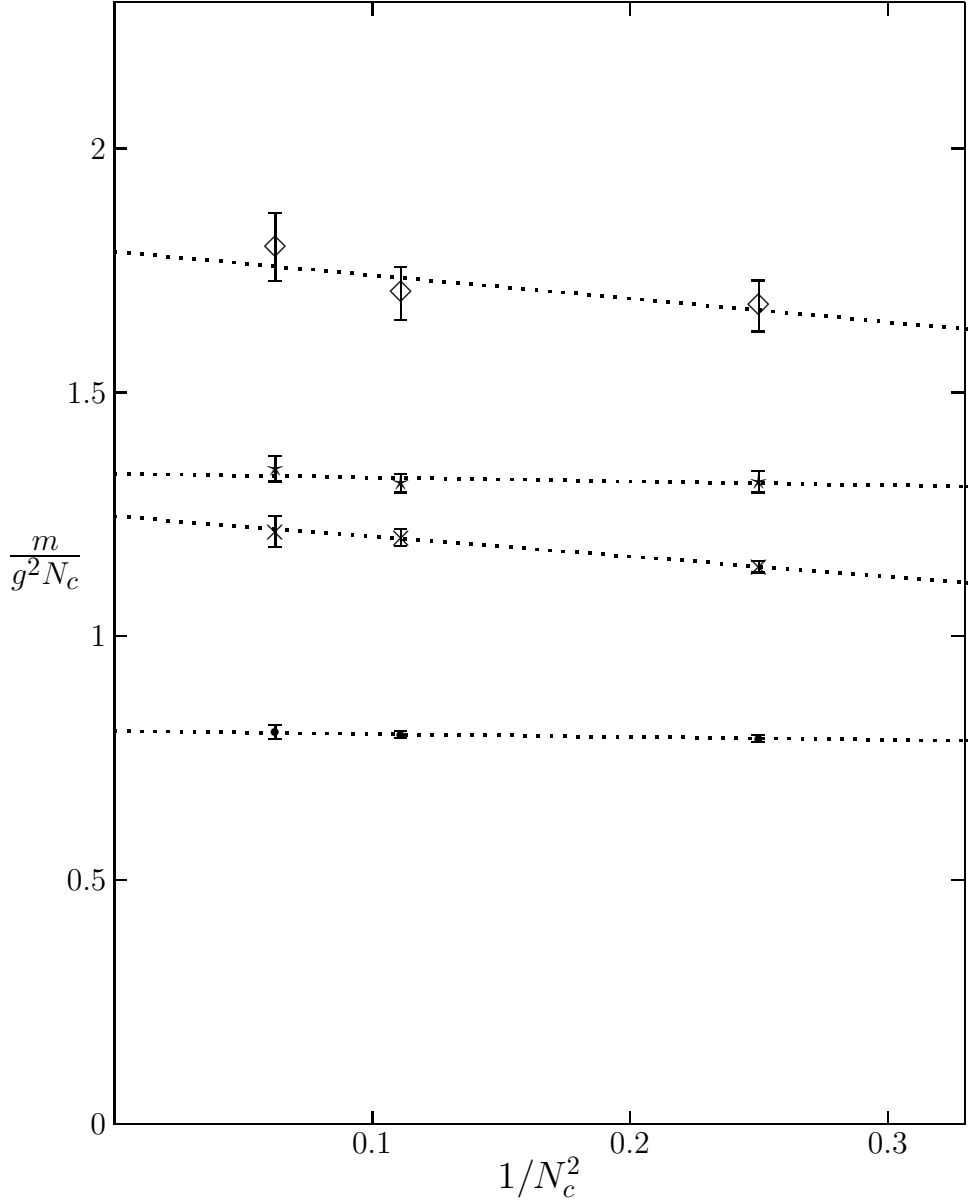}
\end	{center}
\caption{Some continuum glueball masses, in $D=3$, for 2,3,4
colours: $0^{++}$($\bullet$), $0^{++*}$($\times$), $2^{++}$($\star$), 
$0^{-+}$($\diamond$) and linear fits. }

\label	{fig_plot_glue3g}
\end 	{figure}

I begin with the $C=+$ spectrum since the $SU(2)$ spectrum
does not contain $C=-$ states. In this case we have
masses for three values of $N_c$, and so can check
how good is a fit of the kind in 
eqn~\ref{C7}.
In
Fig.~\ref{fig_plot_glue3g}.
I plot the ratio $m_G/g^2N_c$ against $1/N^2_c$ for a selection
of the lightest states, $G$. On this plot a fit of the form
in 
eqn~\ref{C7}.
will be a straight line and I show the best such fits.
As we can see, the data is
consistent with such a $1/N^2_c$ correction being
dominant for $N_c \geq 2$. However what is 
really striking is the lack of $any$ apparent $N_c$
dependence for the lightest $0^{++}$ and $2^{++}$
states. 

In Table~\ref{ng_plus} I present the results of fitting the
$C=+$ states to the form
\begin{equation}
{{m_G}\over{g^2 N_c}}
= R_{\infty} + {R_{slope} \over N^2_c}.          \label{C9}
\end{equation}
where $R_{\infty}={{m_G}\over{g^2 N_c}}{\Bigl/}_{N_c=\infty}$.
(Note that the errors on the slope and intercept are highly
correlated.) The confidence levels of the
fits are quite acceptable 
suggesting once again that for $N_c \geq 2$ a moderately 
sized correction of the form $\sim 1/N^2_c$ is all that
is needed. Note that since the variation with $N_c$ is small,
the exact form of the correction used will not have
a large impact on the extrapolation to $N_c=\infty$
(except in estimating the errors).

These calculations confirm my earlier claim that the physical 
mass scale at large $N_c$ is $g^2N_c$. So if we consider ratios
of $m_G$ to $\surd\sigma$ (as was explicitly
done in 
\cite{MTLAT96})
we will find that they have finite non-zero limits as
$N_c\to\infty$ : that is to say, the large-$N_c$ theory 
possesses linear confinement.

For the $C=-$ states we only have masses for 2 values of
$N_c$ and we cannot therefore check whether a fit
of the form in 
eqn~\ref{C9}
is statistically favoured or not. 
However given that such a fit has proved accurate for 
the $C=+$ masses and for the string tension down to
$N_c=2$ it seems entirely reasonable to assume that
it will be appropriate for $N_c \geq 3$ for the
$C=-$ masses. Assuming this we obtain the results
shown in Table~\ref{ng_minus} for the $N_c = \infty$ limit
and for the coefficient of the first correction.
The `lever arm' on this extrapolation is, of course, shorter 
than for the $C=+$ states and that leads to correspondingly 
larger errors. This will be dramatically improved once
I incorporate the $SU(5)$ spectrum into the analysis. 

The results in the Tables provide us not only with values
for the various mass ratios in the limit $N_c\to\infty$
but also, when inserted into 
eqn~\ref{C7},
predictions for $all$ values of $N_c$.
 
Finally, I should remark that I have also calculated the deconfining
temperature, $T_c$, for $SU(2)$
\cite{MTD3SU2T}
and for $SU(3)$
\cite{MTD3SU3}.
Extrapolating as in 
eqn~\ref{C7}, 
we find
\begin{equation}
{T_c \over {g^2 N_c}} = 0.1745(52)
+{0.079(23) \over N^2_c}.            \label{C10}
\end{equation}
Of course, extrapolating from $N_c=2,3$ is much less reliable
than extrapolating from $N_c=3,4$, and so this relation
should be treated with some caution.

\subsection {3+1 dimensions.}

\label	{4dim}

Our knowledge of 4 dimensional gauge theories is much less 
precise. As far as continuum properties are concerned, 
quantities that are known with reasonable accuracy include the 
string tension, the lightest scalar and tensor glueballs, the
deconfining temperature and the topological susceptibility. 
As in 3 dimensions, the $SU(2)$ and $SU(3)$ values are 
within $\sim 20\%$ of each other, which encourages us
to investigate the $SU(4)$ theory so as to see whether
we are indeed `close' to $N_c=\infty$. Of course these
$SU(4)$ calculations are much slower than in $D=2+1$ and the
results I will present here are of a very preliminary nature
\cite{MTSUN}.

I use the standard plaquette action, and so the
first potential hurdle is the presence of the well-known 
bulk transition that occurs as we increase $\beta$ from strong 
towards weak coupling. To locate this transition I have performed 
a scan on a $10^4$ lattice and found that it occurred
at $\beta = 10.4 \pm 0.1$. This corresponds to
a rather large value of the lattice spacing, $a$,
and so does not lie in the range of couplings within which 
we shall be working, i.e. $\beta=$10.7,10.9 and 11.1.

Our calculation consists of 4000,6000 and 3000 sweeps on 
$10^4$,$12^4$ and $16^4$ lattices at $\beta=$10.7, 10.9 and 
11.1 respectively. Every fifth sweep we calculated correlations 
of (smeared) gluonic loops and from these we extracted
the string tension and the masses of the lightest 
$0^{++}$ and $2^{++}$ particles, using standard techniques
\cite{CM-MT-Q,cm-mt}.
These are presented in Table~\ref{ngk_4d}. We also calculated
the topological susceptibility, $a^4\chi_t$.
The charge $Q$ was obtained using the 
cooling method discussed in the previous lecture.
These calculations were performed every 50 sweeps. Overall this
corresponds to rather small statistics and the errors are
therefore unlikely to be very reliable. 

We see from Table~\ref{ngk_4d} that the most accurate 
physical quantity in our calculations is the string tension, 
$\sigma$. Can we learn from it how $g^2$ varies with $N_c$, just 
as we did in $D=2+1$? We focus on a
particular embodiment of this question:
if we compare different $SU(N_c)$ theories at a value of
$a$ which is the same in physical units, i.e. for which
$a\surd\sigma$ is the same, does the bare coupling
vary as $1/N_c$, i.e. does $\beta \equiv 2N_c/g^2 \propto N^2_c$?
We perform this comparison for $\beta_4=10.9,11.1$. (For
convenience we shall label $\beta$ by the value
of $N_c$, i.e. we write it as $\beta_{N_c}$.)
To find the corresponding values of $\beta$ in
$SU(2)$ and $SU(3)$ we simply interpolate between the
values provided in (for example)
\cite{CM-MT-Q,cm-mt}.
Doing so we find that the values of $\beta$ corresponding
to $\beta_4=10.9,11.1$ are  $\beta_3 \simeq 5.972(18),6.071(24)$ 
and $\beta_2 \simeq 2.442(9),2.485(11)$ respectively. If we
simply scale $\beta_4$ by $N^2_c$ then what we would
expect to obtain is  $\beta_3 \simeq 6.131,6.244$ 
and $\beta_2 \simeq 2.725,2.775$ respectively. Superficially
the numbers look to be in the right ballpark, but
in fact the agreement is poor. For example 
$\beta_2=2.725$ and $\beta_2=2.442$ correspond to
values of $a\surd\sigma$ that differ by about a
factor of 3. 

This disagreement should not, however, be taken too
seriously, since it is well-known that the lattice
bare coupling is a very poor perturbative expansion parameter.
It is known that one can get a much better expansion
parameter if one uses instead the mean-field improved coupling,
$g^2_I$, obtained from $g^2$ by dividing
it by the average plaquette, $<{1\over N_c}TrU_p>$ 
\cite{Imp}.
Defining $\beta^I_{N_c} \equiv 2N_c/g^2_I(a)$ we find
that $\beta_4=10.9, 11.1$ correspond to $\beta^I_4=6.215, 6.474$
respectively. Scaling $\beta^I_4$ by $N^2_c$ we would
expect the equivalent $SU(3)$ and $SU(2)$ couplings to be
given by $\beta^I_3=3.496, 3.642$ and  $\beta^I_2=1.554, 1.619$.
What we actually find is that the equivalent couplings are
$\beta^I_3 \simeq 3.527(22), 3.649(28)$  and
$\beta^I_2=1.561(10), 1.613(12)$. The agreement is now
excellent. That is to say, if the $SU(N_c)$ mean-field 
improved bare-coupling is defined on a length scale that is
related to the physical length scale ($\surd\sigma$) by some
constant factor, then it varies as $g^2 \propto 1/N_c$.
This is, of course, the usual diagram-based expectation.

\begin	{figure}[p]
\begin	{center}
\leavevmode
\setlength{\unitlength}{0.240900pt}
\ifx\plotpoint\undefined\newsavebox{\plotpoint}\fi
\sbox{\plotpoint}{\rule[-0.200pt]{0.400pt}{0.400pt}}%
\begin{picture}(1500,1350)(0,0)
\font\gnuplot=cmr10 at 12pt
\gnuplot
\sbox{\plotpoint}{\rule[-0.200pt]{0.400pt}{0.400pt}}%
\put(120.0,31.0){\rule[-0.200pt]{4.818pt}{0.400pt}}
\put(108,31){\makebox(0,0)[r]{{$0$}}}
\put(1436.0,31.0){\rule[-0.200pt]{4.818pt}{0.400pt}}
\put(120.0,359.0){\rule[-0.200pt]{4.818pt}{0.400pt}}
\put(108,359){\makebox(0,0)[r]{{$2$}}}
\put(1436.0,359.0){\rule[-0.200pt]{4.818pt}{0.400pt}}
\put(120.0,687.0){\rule[-0.200pt]{4.818pt}{0.400pt}}
\put(108,687){\makebox(0,0)[r]{{$4$}}}
\put(1436.0,687.0){\rule[-0.200pt]{4.818pt}{0.400pt}}
\put(120.0,1015.0){\rule[-0.200pt]{4.818pt}{0.400pt}}
\put(108,1015){\makebox(0,0)[r]{{$6$}}}
\put(1436.0,1015.0){\rule[-0.200pt]{4.818pt}{0.400pt}}
\put(525.0,31.0){\rule[-0.200pt]{0.400pt}{4.818pt}}
\put(525,19){\makebox(0,0){\shortstack{\\ \\ \\ {$0.1$}}}}
\put(525.0,1323.0){\rule[-0.200pt]{0.400pt}{4.818pt}}
\put(930.0,31.0){\rule[-0.200pt]{0.400pt}{4.818pt}}
\put(930,19){\makebox(0,0){\shortstack{\\ \\ \\ {$0.2$}}}}
\put(930.0,1323.0){\rule[-0.200pt]{0.400pt}{4.818pt}}
\put(1335.0,31.0){\rule[-0.200pt]{0.400pt}{4.818pt}}
\put(1335,19){\makebox(0,0){\shortstack{\\ \\ \\ {$0.3$}}}}
\put(1335.0,1323.0){\rule[-0.200pt]{0.400pt}{4.818pt}}
\put(120.0,31.0){\rule[-0.200pt]{321.842pt}{0.400pt}}
\put(1456.0,31.0){\rule[-0.200pt]{0.400pt}{316.061pt}}
\put(120.0,1343.0){\rule[-0.200pt]{321.842pt}{0.400pt}}
\put(12,687){\makebox(0,0){{\Large{${m \over {\surd\sigma}}$}}}}
\put(788,-53){\makebox(0,0){{\large{$1/N^2_c$}}}}
\put(120.0,31.0){\rule[-0.200pt]{0.400pt}{316.061pt}}
\put(1132,666){\circle*{12}}
\put(570,628){\circle*{12}}
\put(379,585){\circle*{12}}
\put(367,684){\circle*{12}}
\put(1132.0,646.0){\rule[-0.200pt]{0.400pt}{9.395pt}}
\put(1122.0,646.0){\rule[-0.200pt]{4.818pt}{0.400pt}}
\put(1122.0,685.0){\rule[-0.200pt]{4.818pt}{0.400pt}}
\put(570.0,603.0){\rule[-0.200pt]{0.400pt}{12.045pt}}
\put(560.0,603.0){\rule[-0.200pt]{4.818pt}{0.400pt}}
\put(560.0,653.0){\rule[-0.200pt]{4.818pt}{0.400pt}}
\put(379.0,534.0){\rule[-0.200pt]{0.400pt}{24.572pt}}
\put(369.0,534.0){\rule[-0.200pt]{4.818pt}{0.400pt}}
\put(369.0,636.0){\rule[-0.200pt]{4.818pt}{0.400pt}}
\put(367.0,635.0){\rule[-0.200pt]{0.400pt}{23.608pt}}
\put(357.0,635.0){\rule[-0.200pt]{4.818pt}{0.400pt}}
\put(357.0,733.0){\rule[-0.200pt]{4.818pt}{0.400pt}}
\put(1132,954){\circle{12}}
\put(570,874){\circle{12}}
\put(379,854){\circle{12}}
\put(367,922){\circle{12}}
\put(1132.0,936.0){\rule[-0.200pt]{0.400pt}{8.672pt}}
\put(1122.0,936.0){\rule[-0.200pt]{4.818pt}{0.400pt}}
\put(1122.0,972.0){\rule[-0.200pt]{4.818pt}{0.400pt}}
\put(570.0,836.0){\rule[-0.200pt]{0.400pt}{18.308pt}}
\put(560.0,836.0){\rule[-0.200pt]{4.818pt}{0.400pt}}
\put(560.0,912.0){\rule[-0.200pt]{4.818pt}{0.400pt}}
\put(379.0,787.0){\rule[-0.200pt]{0.400pt}{32.521pt}}
\put(369.0,787.0){\rule[-0.200pt]{4.818pt}{0.400pt}}
\put(369.0,922.0){\rule[-0.200pt]{4.818pt}{0.400pt}}
\put(367.0,838.0){\rule[-0.200pt]{0.400pt}{40.230pt}}
\put(357.0,838.0){\rule[-0.200pt]{4.818pt}{0.400pt}}
\put(357.0,1005.0){\rule[-0.200pt]{4.818pt}{0.400pt}}
\end{picture}
\end	{center}
\caption{Lightest scalar ($\bullet$) and tensor ($\circ$)
 glueball masses in $D=4$. Continuum values for $N_c=2,3$
 and lattice values ($\beta=10.9$ and $\beta=11.1$)
 for $N_c=4$.}
\label	{fig_plot_glue4}
\end 	{figure}
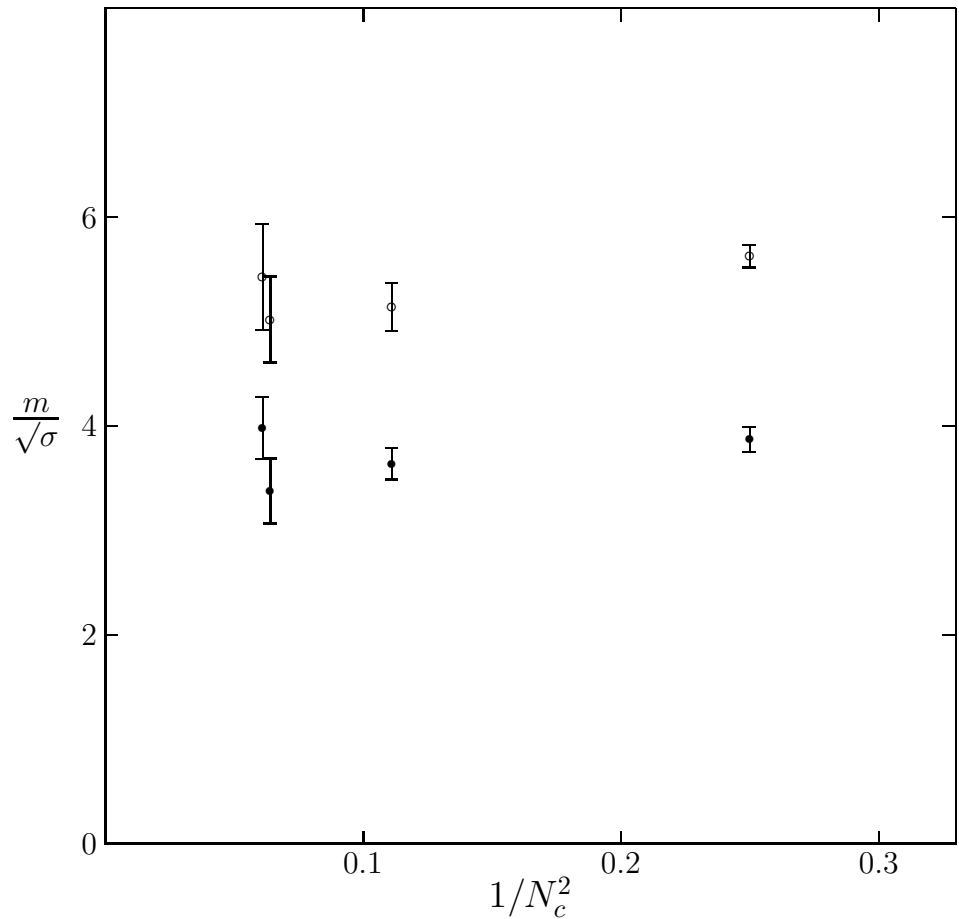

In Fig.~\ref{fig_plot_glue4} I plot the scalar and tensor glueball
masses, in units of $\surd\sigma$, as a function of $N_c$.
For $N_c=2,3$ we have used the continuum values. For $N_c=4$
the calculations are not precise enough to permit an extrapolation
to the continuum limit and so we simply present the values 
that we obtained at $\beta=10.9$ and 11.1. (We do not use the 
$\beta=10.7$ values since they have large errors and there is the
danger that the scalar mass may be reduced by its proximity to
the critical point at the end of the bulk transition line.)
Although the $N_c=4$ errors are quite large, it certainly seems that
there is little variation with $N_c$ for $N_c\geq 2$ and
any dependence appears to be consistent with being given by a simple
$1/N_c^2$ correction. The fact that these mass ratios appear
to have finite non-zero limits, implies that the large-$N_c$
theory is confining. 

\begin	{figure}[p]
\begin	{center}
\leavevmode
\setlength{\unitlength}{0.240900pt}
\ifx\plotpoint\undefined\newsavebox{\plotpoint}\fi
\sbox{\plotpoint}{\rule[-0.200pt]{0.400pt}{0.400pt}}%
\begin{picture}(1500,1350)(0,0)
\font\gnuplot=cmr10 at 12pt
\gnuplot
\sbox{\plotpoint}{\rule[-0.200pt]{0.400pt}{0.400pt}}%
\put(120.0,31.0){\rule[-0.200pt]{4.818pt}{0.400pt}}
\put(108,31){\makebox(0,0)[r]{{$0$}}}
\put(1436.0,31.0){\rule[-0.200pt]{4.818pt}{0.400pt}}
\put(120.0,381.0){\rule[-0.200pt]{4.818pt}{0.400pt}}
\put(108,381){\makebox(0,0)[r]{{$0.2$}}}
\put(1436.0,381.0){\rule[-0.200pt]{4.818pt}{0.400pt}}
\put(120.0,731.0){\rule[-0.200pt]{4.818pt}{0.400pt}}
\put(108,731){\makebox(0,0)[r]{{$0.4$}}}
\put(1436.0,731.0){\rule[-0.200pt]{4.818pt}{0.400pt}}
\put(120.0,1081.0){\rule[-0.200pt]{4.818pt}{0.400pt}}
\put(108,1081){\makebox(0,0)[r]{{$0.6$}}}
\put(1436.0,1081.0){\rule[-0.200pt]{4.818pt}{0.400pt}}
\put(525.0,31.0){\rule[-0.200pt]{0.400pt}{4.818pt}}
\put(525,19){\makebox(0,0){\shortstack{\\ \\ \\ {$0.1$}}}}
\put(525.0,1323.0){\rule[-0.200pt]{0.400pt}{4.818pt}}
\put(930.0,31.0){\rule[-0.200pt]{0.400pt}{4.818pt}}
\put(930,19){\makebox(0,0){\shortstack{\\ \\ \\ {$0.2$}}}}
\put(930.0,1323.0){\rule[-0.200pt]{0.400pt}{4.818pt}}
\put(1335.0,31.0){\rule[-0.200pt]{0.400pt}{4.818pt}}
\put(1335,19){\makebox(0,0){\shortstack{\\ \\ \\ {$0.3$}}}}
\put(1335.0,1323.0){\rule[-0.200pt]{0.400pt}{4.818pt}}
\put(120.0,31.0){\rule[-0.200pt]{321.842pt}{0.400pt}}
\put(1456.0,31.0){\rule[-0.200pt]{0.400pt}{316.061pt}}
\put(120.0,1343.0){\rule[-0.200pt]{321.842pt}{0.400pt}}
\put(-48,783){\makebox(0,0){{\Large{${{\chi_t^{1\over 4}} \over {\surd\sigma}}$}}}}
\put(788,-89){\makebox(0,0){{\large{$1/N^2_c$}}}}
\put(120.0,31.0){\rule[-0.200pt]{0.400pt}{316.061pt}}
\put(1132,853){\circle*{12}}
\put(570,795){\circle*{12}}
\put(373,743){\circle*{12}}
\put(373,568){\circle*{12}}
\put(1132.0,820.0){\rule[-0.200pt]{0.400pt}{15.899pt}}
\put(1122.0,820.0){\rule[-0.200pt]{4.818pt}{0.400pt}}
\put(1122.0,886.0){\rule[-0.200pt]{4.818pt}{0.400pt}}
\put(570.0,752.0){\rule[-0.200pt]{0.400pt}{20.958pt}}
\put(560.0,752.0){\rule[-0.200pt]{4.818pt}{0.400pt}}
\put(560.0,839.0){\rule[-0.200pt]{4.818pt}{0.400pt}}
\put(373.0,692.0){\rule[-0.200pt]{0.400pt}{24.572pt}}
\put(363.0,692.0){\rule[-0.200pt]{4.818pt}{0.400pt}}
\put(363.0,794.0){\rule[-0.200pt]{4.818pt}{0.400pt}}
\put(373.0,488.0){\rule[-0.200pt]{0.400pt}{38.785pt}}
\put(363.0,488.0){\rule[-0.200pt]{4.818pt}{0.400pt}}
\put(363.0,649.0){\rule[-0.200pt]{4.818pt}{0.400pt}}
\end{picture}
\end	{center}
\vskip 0.15in
\caption{The topological susceptibility: continuum values
 for $N_c=2,3$ and lattice values ($\beta=10.9$ and $\beta=11.1$) 
 for $N_c=4$.}
%
\label	{fig_plot_top4}
\end 	{figure}
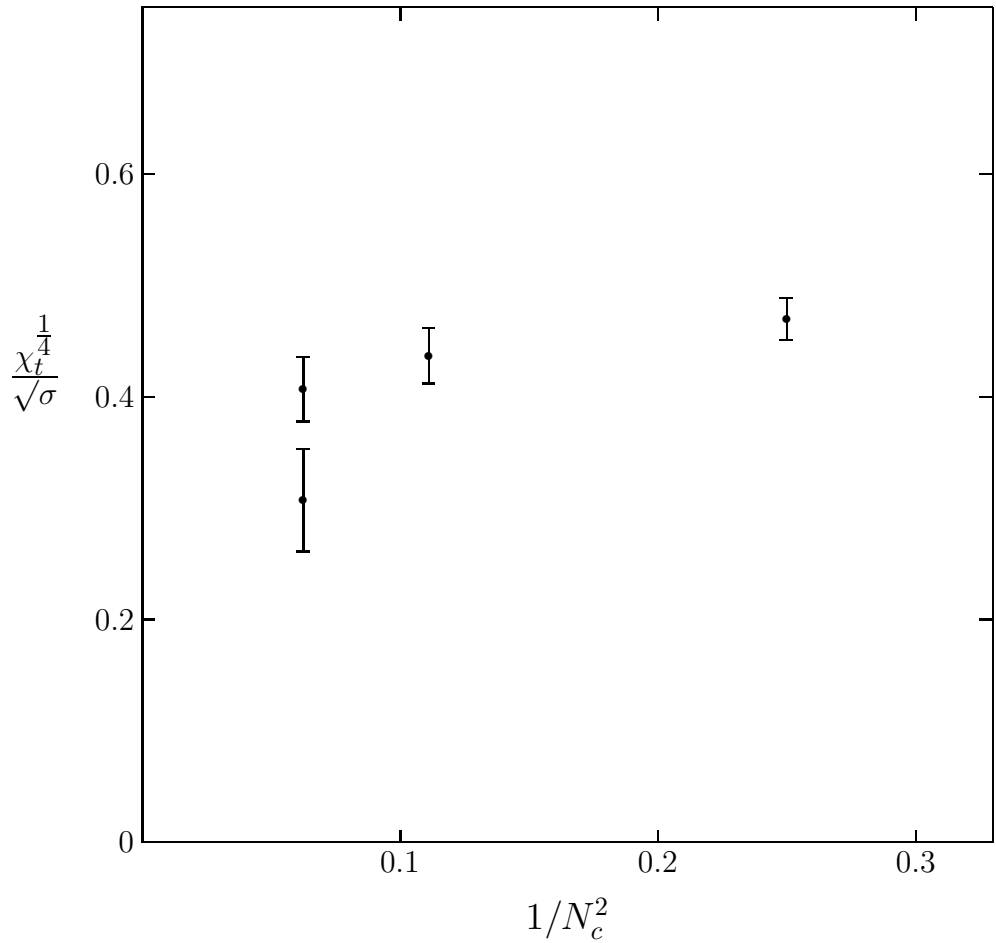

As mentioned earlier we have also calculated the topological
susceptibility. In Fig.~\ref{fig_plot_top4} we plot 
the dimensionless ratio $\chi_t^{1/4}/\surd\sigma$ as a
function of $N_c$. Once again the $N_c$=2 and 3 values
are continuum extrapolations of lattice values,
while in the case of $SU(4)$ we simply display the lattice
values obtained at $\beta$=10.9 and 11.1.
As we remarked earlier one expects, semiclassically, very 
few small instantons for $SU(4)$ and this is confirmed in our cooling
calculations. This has the advantage that the 
lattice ambiguities that arise when instantons 
are not much larger than $a$
are reduced as compared to $SU(3)$, and dramatically reduced
as compared to $SU(2)$. This implies that the interesting 
large-$N_c$ physics of topology (and the related meson physics)
should be straightforward to study.

\subsection {Conclusions.}

\label		{conc}

I have shown you the mass spectra and string tensions of 
gauge theories with $N_c=2,3,4$ in 3 dimensions. We saw
that there is only a small variation with $N_c$
and that this can be accurately described by a modest $O(1/N_c^2)$ 
correction. That is to say, such theories are close to their 
$N_c=\infty$ limit for all values of $N_c \geq 2$. We also saw that
the large-$N_c$ theory is confining and that $g^2 \propto 1/N_c$
when expressed in physical units. This confirms, in a fully
non-perturbative way,  expectations arrived at from analyses
of Feynman diagrams. It simultaneously provides a unified understanding of 
all our $SU(N_c)$ theories in terms of just the one theory, 
$SU(\infty)$, with modest corrections to it. In practical terms
this means that, from the parameters in our Tables, we know the 
corresponding masses for $all$ values of $N_c$.

Our calculation in 4 dimensions, while quite preliminary, 
does suggest that the situation may be much the same there.
This is obviously something that needs to be done much better.
Of particular interest are topological fluctuations at large $N_c$:
the instanton size density, $D(\rho)$, loses its small $\rho$
tail and lattice topology should become unambiguous. 
At the same time it would be interesting to investigate the
behaviour of hadrons composed of quarks: for example
the $\eta^\prime$ mass should vanish as $N_c \to \infty$. 
There are many problems of real theoretical interest here.
I hope some of you will choose to get involved.

\vskip 1.5in

{\Large{\bf Acknowledgements}}

My thanks to Pierre van Baal for inviting me to lecture 
at this School and for his efforts in making it such a success.
I would also like to thank him, David Olive and Peter West
for inviting me to the very stimulating Workshop at the 
Isaac Newton Institute that preceeded the School and, finally,
the Institute itself for providing such a splendid research
environment. 


%
%
\begin 	{table}[p]
\begin	{center}
\begin	{tabular}
{|c| r@{.}l@{ (}r@{) }|}
\hline
\multicolumn{1}{|c|}{$N_c$} & 
\multicolumn{3}{|c|}{${{\surd\sigma}/{g^2}}$} \\
\hline
 2 & 0&3350 &  15 \\
 3 & 0&5530 &  20 \\
 4 & 0&7564 &  45 \\
\hline
\end	{tabular}
\end	{center}
\caption{The $D=2+1$ $SU(N_c)$ confining string tension.}
\label	{n_string}
\end 	{table}

\begin 	{table}[p]
\begin	{center}
\begin	{tabular}
{|c| r@{.}l@{ (}r@{) }| r@{.}l@{ (}r@{) }|c|}
\hline
\multicolumn{1}{|c|}{$\beta$} &
\multicolumn{3}{|c|}{$R_{\infty}$} & 
\multicolumn{3}{|c|}{$R_{slope}$} \\
\hline
 $\surd\sigma$ & 0&1974&  12 & 	-0&12 &  1 \\
 $0^{++}$      & 0&805 &  13 & 	-0&06 &  8 \\
 $0^{++\star}$ & 1&245 &  27 & 	-0&41 & 14 \\
 $0^{-+}$      & 1&788 &  88 & 	-0&48 & 56 \\
 $2^{++}$      & 1&333 &  29 & 	-0&08 & 18 \\
 $2^{-+}$      & 1&340 &  40 & 	-0&01 & 24 \\
 $1^{++}$      & 1&946 &  75 & 	-0&59 & 47 \\
 $1^{-+}$      & 1&919 & 115 & 	-0&18 & 75 \\
\hline
\end	{tabular}
\end	{center}
\caption{States with $C=+$ in $D=2+1$ : $R_{\infty} \equiv 
{\displaystyle\lim_{N_c\to\infty}{{m_G}\over{g^2N_c}}}$
and $R_{slope}$ is the coefficient of the $1/N^2_c$ 
correction in eqn~\ref{C9}.}
\label	{ng_plus}
\end 	{table}
\begin 	{table}[p]
\begin	{center}
\begin	{tabular}
{|c| r@{.}l@{ (}r@{) }| r@{.}l@{ (}r@{.}l@{) }|}
\hline
\multicolumn{1}{|c|}{$G$} & 
\multicolumn{3}{|c|}{$R_{\infty}$} & 
\multicolumn{4}{|c|}{$R_{slope}$}\\
\hline
 $0^{--}$      & 1&18 &  6 & 	 0&1 &  0&6 \\
 $0^{--\star}$ & 1&47 & 10 & 	 0&3 &  1&1 \\
 $0^{+-}$      & 1&98 & 28 & 	-0&4 &  2&7 \\
 $2^{--}$      & 1&52 & 14 & 	 0&9 &  1&4 \\
 $2^{+-}$      & 1&58 & 13 & 	-0&4 &  1&3 \\
 $1^{--}$      & 1&85 & 15 & 	-0&3 &  1&5 \\
 $1^{+-}$      & 1&78 & 23 & 	 1&3 &  2&3 \\
\hline
\end	{tabular}
\end	{center}
\caption{As in Table 2 but for states with $C=-$.}
\label	{ng_minus}
\end 	{table}
\begin 	{table}[p]
\begin	{center}
\begin	{tabular}
{|c| r@{.}l@{ (}r@{) }|r@{.}l@{ (}r@{) }|r@{.}l@{ (}r@{) }|}
\hline
\multicolumn{1}{|c|}{$\beta$} &
\multicolumn{3}{|c|}{$a\surd\sigma$} & 
\multicolumn{3}{|c|}{$am_{0^{++}}$} & 
\multicolumn{3}{|c|}{$am_{2^{++}}$} \\
\hline
 10.7  &  0&296 & 14 & 	0&98 & 17 & 1&78 & 34 \\
 10.9  &  0&229 &  7 & 	0&77 &  8 & 1&20 & 10 \\
 11.1  &  0&196 &  7 & 	0&78 &  6 & 1&08 & 10 \\
\hline
\end	{tabular}
\end	{center}
\caption{$SU(4)$ in 4 dimensions; masses calculated at
the values of $\beta$ shown.}
\label	{ngk_4d}
\end 	{table}

\newpage

\newpage

\newpage

\newpage

\newpage

\newpage

\begin	{figure}[p]
\begin	{center}
\leavevmode
\input	{plot_qsu3}
\end	{center}
\vskip 0.15in
\caption{The SU(3) topological susceptibility: from all peaks ($\bullet$)
and from peaks with $Q(x_{peak}) \leq 1/16\pi^2$ ($\times$). Sample
extrapolations to a common continuum limit are shown.}  
\label	{fig-QSU3}
\end 	{figure}

\newpage

\begin	{figure}[p]
\begin	{center}
\leavevmode
\input	{plot_qsu2}
\end	{center}
\vskip 0.15in
\caption{The SU(2) topological susceptibility: from all peaks ($\bullet$)
and from peaks with $Q(x_{peak}) \leq 1/16\pi^2$ ($\times$). Sample
extrapolations to a common continuum limit are shown.}  
\label	{fig-QSU2}
\end 	{figure}

\newpage

\newpage

\newpage

\newpage

\newpage

\newpage

\newpage

\vskip 1.2in

\begin{thebibliography}{99}




\bibitem{GF11-decay} 
J.Sexton et al., Phys Rev. Letters, 75 (1995) 4563.

\bibitem{ukqcd} 
G.Bali et al, Phys. Lett. B309 (1993) 378.

\bibitem{GF11-G1} 
H. Chen et al, Nucl. Phys. Proc. Suppl. B34 (1994) 357.


\bibitem{fec-mt}
F. E. Close and M. Teper, inpreparation. 

\bibitem{book-C} 
M. Creutz, Quarks, Gluons and Lattices (CUP 1983).

\bibitem{book-MM}
I. Montvay and G. Munster, Quantum Fields on a Lattice (CUP 1994).
 
\bibitem{Luscher} 
M. Luscher, Comm. Math. Phys. 104 (1986) 177; 1983 Cargese Lectures.

\bibitem{Sym} 
K. Symanzik, Nucl. Phys. B226 (1983) 187.

\bibitem{book-Perkins}
D. Perkins, Introduction to High Energy Physics (Addison-Wesley 1987).

\bibitem{cm-mt} 
C. Michael and M. Teper, Nucl. Phys. B314 (1989) 347.

\bibitem{deF-G} 
Ph.de Forcrand et al, Phys. Lett. B152 (1985) 107.

\bibitem{string-tension}
Ph.de Forcrand et al, Phys. Lett. B160 (1985) 137;
P. Bacilieri et al,  Phys. Lett. B205 (1988) 535;
S. Perantonis and C. Michael, Nucl. Phys. B347 (1990) 854;
G. Bali and K. Schilling,  Phys. Rev. D46 (1992) 2636;
Phys. Rev. D47 (1993) 661;
S. Booth et al, Nucl. Phys. B294 (1992) 385;
C. Allton et al, Nucl. Phys. B407 (1993) 331;
H. Wittig, Nucl. Phys. Proc. Suppl. B42 (1995) 288.

\bibitem{P-M}
C. Morningstar and M. Peardon, Phys. Rev. D56 (1997) 4043.
 
\bibitem{GF11-H} 
F.Butler et al, Nucl. Phys. B430 (1994) 179.

\bibitem{UKQCD-H} 
R. Kenway et al, Nucl. Phys. Proc. Suppl. B53 (1997) 206;
H. Shanahan et al, Phys. Rev. D55 (1997) 1548;
S. Ryan, Ph.D. Thesis, Edinburgh 1996.

\bibitem{GF11-ss} 
W. Lee and D. Weingarten, Nucl. Phys. Proc. Suppl. B53 (1997) 236.

\bibitem{CM-ss}
P. Lacock et al,  Phys. Rev. D54 (1996) 6997.

\bibitem{book-Close} 
F. E. Close, Introduction to Quarks and Partons (Academic Press 1979).

\bibitem{Isgur}
S. Godfrey and N. Isgur, Phys. Rev. D32 (1985) 189. 

\bibitem{PDT}
Particle Data Group, Phys. Rev. D54 (1996) 1. 

\bibitem{Amsler-Close}
C. Amsler and F. Close, Phys. Rev. D53 (1996) 295.

\bibitem{GF11-G3} 
D. Weingarten, Nucl. Phys. Proc. Suppl. B53 (1997) 232.

\bibitem{Genov} 
M. Genovese, Phys. Rev. D46 (1992) 5204.

\bibitem{Jaffe}
R. Jaffe, Phys. Rev. D15 (1977) 267.

 

\bibitem{U1-Cole} 
S. Coleman, Aspects of Symmetry, Ch.7 (CUP 1985). 

\bibitem{U1-H} 
G. 't Hooft, Physics Reports 142 (1986) 357.

\bibitem{Banks-Casher}
T. Banks and A. Casher, Nucl. Phys. B169 (1980) 103. 

\bibitem{Shuryak} 
T. Schaefer and E. Shuryak, hep-ph/9610451.

\bibitem{SH-MT} 
S. Hands and M. Teper, Nucl. Phys. B347 (1990) 819.

\bibitem{dowrick-teper} 
N. Dowrick and M. Teper, Nucl. Phys. Proc. Suppl. B42 (1995) 237.

\bibitem{diV} 
P. di Vecchia et al, Nucl. Phys. B192 (1981) 392.

\bibitem{PISA} 
M. Campositrini et al, Phys. Lett. B212 (1988) 206.

\bibitem{cool-MT} 
M. Teper, Phys. Lett. B162 (1985) 357.

\bibitem{U1-Wit} 
E. Witten, Nucl. Phys. B156 (1979) 269.

\bibitem{U1-Ven} 
G. Veneziano, Nucl. Phys. B159 (1979) 213.

\bibitem{JH-MT}
J. Hoek et al, Nucl. Phys. B288 (1987) 589. 

\bibitem{MT-SU3} 
M. Teper, Phys. Lett. B202 (1988) 553.

\bibitem{CM-MT-Q} 
C. Michael and M. Teper, Nucl. Phys. B305 (1988) 453.

\bibitem{DP-MT} 
D. Pugh and M. Teper, Phys. Lett. B218 (1989) 326; B224 (1989) 159.

\bibitem{unpub-QSU2} 
M. Teper, unpublished.

\bibitem{deF-QSU2} 
P. de Forcrand et al, Nucl. Phys. B499 (1997) 409.

\bibitem{DeG-SU2} 
T. DeGrand, A. Hasenfratz and T. Kovacs,  hep-lat/9705009 

\bibitem{CM-PS} 
C. Michael and P. Spencer,  Phys. Rev. D52 (1995) 4691. 

\bibitem{Brower} 
R. Brower et al,  Nucl. Phys. Proc. Suppl. 53 (1997) 547.

\bibitem{DS-MT} 
D. Smith et al, hep-lat/9709128; in preparation.




\bibitem{N-H} 
G. 't Hooft, Nucl. Phys. B72 (1974) 461.

\bibitem{N-W} 
E. Witten, Nucl. Phys. B160 (1979) 57.

\bibitem{N-C} 
S. Coleman, 1979 Erice Lectures.

\bibitem{N-Das} 
S.R. Das, Rev. Mod. Phys. 59 (1987) 235.

\bibitem{N-EK} 
T. Eguchi and H. Kawai, Phys. Rev. Lett. 48 (1982) 1063.

\bibitem{MTD3SU2G} 
M. Teper, Phys. Lett. B289 (1992) 115.

\bibitem{MTD3SU2K} 
M. Teper, Phys. Lett. B311 (1993) 223.

\bibitem{MTD3SU3} 
M. Teper, in preparation.

\bibitem{ISG-PAT} 
N. Isgur and J. Paton, Phys. Rev. D31 (1985) 2910.

\bibitem{TM-MT} 
T. Moretto and M. Teper, hep-lat/9312035.

\bibitem{RJ-MT} 
R. Johnson and M. Teper, hep-lat/9709083

\bibitem{FA-SD} 
F. Antonuccio and S. Dalley, Nucl. Phys. B461 (1996) 275.

\bibitem{MTSUN} 
M. Teper, Phys. Lett. B397 (1997) 223.

\bibitem{MTLAT96} 
M. Teper, Nucl. Phys. (Proc. Suppl.) 53 (1997 715.

\bibitem{MTD3SU2T} 
M. Teper, Phys. Lett. B313 (1993) 417.

\bibitem{Imp} 
G. Parisi, in {\it High Energy Physics} - 1980(AIP 1981);
G. Lepage and P. Mackenzie, Phys. Rev. D48 (1993) 2250.


%
\end{thebibliography}
\end{document}